\documentclass[11pt,a4paper]{article}

\usepackage{url}
\usepackage{jcappub}
\usepackage{bm}
\usepackage{amsfonts}
\usepackage{subfigure}
\usepackage{graphicx}

\newcommand{\be}{\begin{equation}}
\newcommand{\ee}{\end{equation}}
\newcommand{\bea}{\begin{eqnarray}}
\newcommand{\eea}{\end{eqnarray}}

\newcommand{\bk}{{\bf k}}
\newcommand{\bl}{{\bf l}}

\newcommand{\bq}{{\bf q}}
\newcommand{\bp}{{\bf p}}

\newcommand{\beq}{\begin{equation}}
\newcommand{\eeq}{\end{equation}}

\begin{document}

\title{Improved axion emissivity from a supernova via  nucleon-nucleon  bremsstrahlung}

\author[a,b]{Pierluca Carenza,}
\author[c]{Tobias Fischer,}
\author[d]{Maurizio Giannotti,}
\author[e]{Gang Guo,}
\author[e,f]{Gabriel Mart\'inez-Pinedo,}
\author[a,b]{Alessandro Mirizzi}

\affiliation[a]{Dipartimento Interateneo di Fisica ``Michelangelo Merlin'', Via Amendola 173, 70126 Bari, Italy}
\affiliation[b]{Istituto Nazionale di Fisica Nucleare - Sezione di Bari, Via Amendola 173, 70126 Bari, Italy}
\affiliation[c]{Institute for Theoretical Physics, University of Wroc\l{}aw, Pl. M. Borna 9, 50-204 Wroc\l{}aw, Poland}
\affiliation[d]{Physical Sciences, Barry University, 11300 NE 2nd Ave., Miami Shores, FL 33161, USA}
\affiliation[e]{GSI Helmholtzzentrum f\"ur Schwerionenforschung,
  Planckstra{\ss}e~1, 64291 Darmstadt, Germany}
\affiliation[f]{Institut
  f{\"u}r Kernphysik (Theoriezentrum), Technische Universit{\"a}t
  Darmstadt, Schlossgartenstra{\ss}e 2, 64298 Darmstadt, Germany}

\emailAdd{pierluca.carenza@ba.infn.it, tobias.fischer@ift.uni.wroc.pl, MGiannotti@barry.edu, gangg23@gmail.com, alessandro.mirizzi@ba.infn.it}

\abstract{
The most efficient axion production mechanism in a supernova (SN) core is the nucleon-nucleon bremsstrahlung.
 This process has been often modeled  at the level of the vacuum one-pion exchange (OPE) approximation.
  Starting from this naive recipe, we revise the calculation including systematically different effects, 
  namely  a non-vanishing mass for the exchanged pion, the contribution from the two-pions exchange, effective in-medium nucleon masses and 
 multiple nucleon scatterings. Moreover, we allow for an arbitrary degree of nucleon degeneracy. 
A self consistent treatment of the axion emission rate including all these effects is currently missing. The aim of this work is to provide such an analysis.
Furthermore, we demonstrate that the OPE potential with all the previous corrections
 gives rise to similar results as the on-shell $T$-matrix, and is therefore well justified for our and similar studies. 
We find that the axion emissivity is reduced by over an order of magnitude 
  with respect to the basic  OPE  calculation, after all these effects are accounted for. 
The implications for the axion mass bound and the impact for the next generation experimental axion searches is also discussed. 
 }
\maketitle

\section{Introduction}

One of the most puzzling and long-standing problems in particle physics is related to the absence of an expected CP violation in the strong interactions: {\em the strong CP problem}. 
The most elegant solution to this problem is based on the Peccei-Quinn (PQ) mechanism~\cite{Peccei:1977hh,Peccei:1977ur,Weinberg:1977ma,Wilczek:1977pj}, in which the Standard Model is enlarged with an additional global U(1)$_A$ symmetry, known as the PQ symmetry. 
The axion is the  Nambu-Goldstone boson of the PQ symmetry,
a low-mass pseudoscalar particle with properties similar to those of neutral pions.
The axion mass is given by~\cite{diCortona:2015ldu,Borsanyi:2016ksw}
\begin{equation}
m_a = \frac{5.7 \,\ \textrm{eV}}{f_a/10^6 \,\ \textrm{GeV}} \,\ ,
\end{equation}
where $f_a$ is the axion decay constant or PQ scale.
The axion interactions with photons, electrons, and hadrons are also controlled by the PQ constant  and scale as  $f_a^{-1}$.
Therefore, the PQ scale controls the axion phenomenology and is constrained by different experiments and astrophysical arguments
that involve interactions with photons, electrons, and hadrons (see~\cite{Tanabashi:2018oca,Irastorza:2018dyq,Giannotti:2017hny} for recent reviews).

Among the astrophysical systems, core-collapse supernovae
constitute a valuable laboratory to constrain the axion properties~\cite{Raffelt:2006cw}.
The most relevant process of axion emission in a SN core is the nucleon--nucleon (N--N) axion bremsstrahlung,  which involves the 
 axion-nucleon coupling~\cite{Turner:1987by}.
This interaction strength has been constrained by two different arguments based on the observed neutrino signal of supernova
 1987A~\cite{Burrows:1988ah,Burrows:1990pk,Raffelt:1987yt,Raffelt:1990yz}. 
It has been proposed that axion nucleon couplings which predict an axion energy loss greater or equal than the neutrino luminosity should be excluded as they would spoil the observed neutrino signal from SN 1987A.

According to the current results, which we are going to revise in the present work, axions with mass $m_a \lesssim 16$~meV, corresponding to $f_a \gtrsim 4 \times 10^8$~GeV, interact so weakly that the changes in the neutrino emission   of SN 1987A, would be unobservable~\cite{Raffelt:2006cw}. 
Analogously, very strongly interacting axions, with $f_a \lesssim 2 \times10^6$~GeV, would  guarantee a standard SN neutrino signal. 
In this case, axions would be so efficiently trapped in the SN core that their luminosity would be suppressed below any observable level. 
As discussed in~\cite{Engel:1990zd}, however, the region with slightly lower axion decay constant, $f_a \lesssim 3\times10^5$~GeV, should be excluded since axions would have caused too many events in the Kamiokande water Cherenkov detector that observed the SN 1987A neutrino signal. 
As a result, it was argued that the window $3 \times 10^5$~GeV $\lesssim  f_a \lesssim 2 \times10^6$~GeV, dubbed  ``hadronic axion window'', could not be excluded
by the SN argument
 alone~\cite{Moroi:1998qs}. 
  In particular, for hadronic axion models one has to invoke cosmological mass bounds to close
this window~\cite{Hannestad:2005df,Giusarma:2014zza}, corresponding to axion masses of few eV.

A renewed interest towards the  axion from SNe arose recently, partially because the SN 1987A  bound overlaps with the sensitivity range
of the planned experiments, in particular the International Axion Observatory (IAXO), which  searches for solar axions~\cite{Armengaud:2019uso}.
IAXO is sensitive to generic axion-like particles coupled to photons and has the potential to probe the QCD axion region up to masses $m_a \gtrsim 10^{-2}$~eV. 
Therefore, it is interesting to assess how robust the SN 1987A bound is, in relation to the 
IAXO potential.

In addition to the impact on the current experimental potential, the axion mass region allowed by the SN argument has other interesting phenomenological implications. 
In particular, it has been recently shown that for such values of the  axion mass it would be possible to detect peculiar modifications
in the neutrino signal from a future galactic SN~\cite{Fischer:2016cyd}. 
Additionally, a dedicated study~\cite{Raffelt:2011ft} predicted a non-negligible diffuse axion background from past core-collapse supernovae.
Finally, the existence of eV axions in the hadronic axion window is certainly an appealing possiblity, as planned experiments such as IAXO can probe those masses. 
In the case of hadronic axion models, most notably the KSVZ axions~\cite{Kim:1979if,Shifman:1979if}, 
the strong astrophysical constraints on the axion electron coupling~\cite{Viaux:2013lha,Straniero:2018fbv} provide only a weak bound on the PQ constant, since the coupling to electrons is suppressed by loop effects. 
Therefore, one has to rely on the weaker horizontal branch bound~\cite{Ayala:2014pea,Straniero:2015nvc}, $ g_{a\gamma}\lesssim 0.65\times 10^{-10} $~GeV$ ^{-1} $, which translates into 
$ f_a\geq 1.8 \,C_{\gamma}\times 10^{7}$~GeV, where $ C_{\gamma} $ is a model dependent constant. 
Thus, a value of $ C_{\gamma}\sim 0.1 $, contemplated by recent studies~\cite{DiLuzio:2016sbl}, would allow a PQ constant in the hadronic window.
Moreover, the cosmological constraint on the hadronic axion window can also be relaxed in nonstandard low-temperature-reheating cosmological scenarios, 
in which the thermal axion relic abundances is suppressed and cosmological limits are significantly loosened~\cite{Grin:2007yg}. 
It is thus relevant to investigate the true impact of the SN bound on the hadronic axion window since this is not related to the model-dependent strength of axion-photon coupling, and unaffected by cosmological arguments.

In order to have a realistic determination of axion impact
on the SN neutrino signal one needs an accurate determination of the axion production rate.
As stated above, the most relevant process of axion production in a SN core is the nucleon-nucleon (N–N) axion bremsstrahlung, which involves the axion-nucleon coupling. 
This process has been originally modeled  at the level of the vacuum one-pion exchange (OPE) 
approximation~\cite{Turner:1988bt,Carena:1988kr,Iwamoto:1984ir,Brinkmann:1988vi,Keil:1996ju}.
In literature it has been realized that different effects might influence and somehow reduce the emissivity with respect to the naive
OPE approximation. In particular, an attempt to account for these effects has been recently proposed in~\cite{Chang:2018rso}. 
There, different factors were introduced to schematically modify the OPE emission rate. 
As a results, it was found a sizable reduction of the axion emissivity.
However, that analysis ignored some relevant effects such as the modification of the nuclear mass in the dense plasma
	 and, more importantly, neglected the relation between the multiple nucleon scatterings and the other contributions to the emission rate. 
	As we shall see, this last effect compensates partially the contribution of the other corrections. 
	At any rate, motivated by the results in~\cite{Chang:2018rso} and because of the great importance of a quantitative assessment of the axion bremsstrahlung emission rate, we decided to take a closer look at the axion emissivity by including in a self-consistent calculation the different corrections to OPE.
Some of the effects we are including have, to the best of our knowledge, never been considered in other calculations of the axion emission rate.
More specifically, we are including the effect of a finite mass in the pion propagator, the contribution of the $\varrho$-meson exchange, which mimic the effects of a two-pions exchange~\cite{Ericson:1988wr}, 
the medium modification of the nucleon mass, and the impact of nucleon multiple scatterings. Moreover, 
we provide numerical evaluations of the axion emissivity valid for generic nucleon degeneracy.

The plan of our work is as follows.
In Section~2 we present the theoretical framework for the calculation of the axion emission rate, including  modifications beyond the naive OPE and we discuss the impact of the different effects on the axion emissivity. We also 
 calculate the axion luminosity based on our new emissivity in the case of free-streaming axions. 
Moreover, we demonstrate that the OPE potential, corrected by $\varrho$-meson exchange, gives results similar to those based on $T$-matrix, and therefore is well justified for our and similar studies.
  In Section~3 we discuss the impact of the corrections to  the axion emissivity on the axion mass bound and we compare with previous literature.
In Section~4 we calculate the axion opacity for trapped axions and calculate their luminosity in this case. 
Finally, in  Section~5 we summarize our  results and we conclude.

\section{Axion emissivity beyond the one-pion-exchange approximation}

\subsection{Axion-nucleon interaction and emissivity in naive one-pion-exchange}

The dominant axion production channel in a 
 SN core is the $N$--$N$  bremsstrahlung~\cite{Turner:1987by}:
\begin{eqnarray}
N_1 + N_2 \longrightarrow N_3 + N_4 + a\,,
\label{eq:brems}
\end{eqnarray}
 where $N_i $ are nucleons (protons or neutrons) and $ a $ is the axion field. 
The process \eqref{eq:brems} is induced by the axion-nucleon interaction described by the following 
Lagrangian term~\cite{Carena:1988kr},
\begin{equation}\label{eq:axion_N_coupling}
\mathcal{L}_{a N}=\sum_{i=p,n} \frac{g_{a i}}{2m_N}\,\overline N_i\gamma_\mu\gamma_5 N_i  \partial^\mu  a,
\end{equation}
with axion-nucleon couplings defined as follows,
\begin{equation}
g_{ai} =C_{ai} \frac{m_N}{f_a} = 1.65 \times 10^{-7} \left(\frac{m_a}{\textrm{eV}} \right)C_{ai}~,
\label{eq:coupl}
\end{equation}
where $f_a$ is the Peccei-Quinn energy scale and $ m_N$ is the nucleon mass (we assume $ m_n\simeq m_p $).
The coupling constants, $ C_{ai} $, for the benchmark axion models have been recently calculated~\cite{diCortona:2015ldu}.
Their most accurate values, including the uncertainties, are
\begin{eqnarray}
&& C_{ap}^{\rm KSVZ}=-0.47(3)\,, \qquad
C_{an}^{\rm KSVZ}=-0.02(3)\,, \label{eq:axion_couplings_KSVZ}
\\
&& C_{ap}^{\rm DFSZ}=-0.435\sin^2\beta+\left(-0.182\pm 0.025\right)\,, \nonumber \\
&& C_{an}^{\rm DFSZ}=0.414\sin^2\beta+\left(-0.16\pm 0.025\right)\,.
\label{eq:axion_couplings}
\end{eqnarray}
Here, $\tan\beta \equiv v_u/v_d$ represents the ratio of the two Higgs bosons in the DFSZ axion model~\cite{Dine:1981rt,Zhitnitsky:1980tq}.
It is theoretically constrained from both ends by the requirement of perturbative unitarity of the Yukawa couplings $ 0.28 < \tan\beta < 140$.
 
The nuclear axion bremsstrahlung rate is highly uncertain, mostly because of the lack of understanding of the nuclear interactions; approximations are commonly applied based on vacuum physics. 
 The matrix elements can only be calculated in specific frameworks, the most widely used being the OPE potential, which describes the two nucleon interaction with the exchange of a pion.
 In this approximation, the axion emission rate (in units of energy/time$\times$volume) can be calculated as~\cite{Brinkmann:1988vi} 
\begin{equation}
Q_a = \int 
 \frac{d^3p_a}{2\omega_a (2\pi)^3} \prod_{i=1,4} \frac{g_i d^3p_i}{2E_i(2\pi)^3}~\omega_a f_1 f_2 (1-f_3) (1-f_4)
 \sum_{\rm spins} S | {\mathcal M}|^2
\delta^4(p_1+p_2-p_3-p_4-p_a)  \,\ ,
\label{eq:emissivity}
\end{equation}
where the nucleon degeneracy factor $g_i=2$, $\omega_a$ is the axion energy, 
$S$  is the usual symmetry factor for identical
particles in the initial and final states,  and the matrix element
squared  $| {\mathcal M}|^2$ is summed over initial and final
spins.

Before proceeding, let's introduce the convenient notation $ g_{a}=m_{N} /f_{a} $ for the axion-nucleon couplings. 
This notation encapsulates the model dependent constants [Eq.~\eqref{eq:coupl}] which can be extracted as 
\begin{eqnarray}
C_{ap}&=& g_{ap}/ g_{a}  \,\ , \nonumber \\
C_{an}  &=& g_{an} /g_{a}  \, \ .
\end{eqnarray}

The matrix-squared element for the nucleon axion bremsstrahlung in the one-pion-exchange (OPE) can be conveniently written separating out the medium contribution ${\overline M}$~\cite{Brinkmann:1988vi}
{

\begin{equation}
S\times\sum|\mathcal{M}|^{2} =\frac{1}{4}\times\frac{g_{a}^{2}}{4m_{N}^{2}} \omega_{a}^{2}{\overline M}  \,\ ,
\label{eq:inimatrix2}
\end{equation}
with
\begin{equation}
 {\overline M}= 4\times\frac{256}{3}m_{N}^{4} \omega_{a}^{-2} \left(\frac{g_{A}}{2f_{\pi}}\right)^{4}\left(\mathcal{A}_{nn}+\mathcal{A}_{pp}+4 \mathcal{A}_{np}\right) \,\ ,
\label{eq:matrix}
\end{equation}
where   $g_{A}=1.26$ is the axial coupling, $f_{\pi}=92.4$ \textrm{MeV} is the pion decay constant~\footnote{The  coupling 
$({g_{A}}/{2f_{\pi}})^{4}$
is $\sim 20 \%$ smaller than
the $(f/m_\pi)^{4}$ coupling used in~\cite{Hannestad:1997gc}, with $f=1$.} 
and 
\begin{eqnarray}
\mathcal{A}_{NN}&=& C_{aN}^{2}
\left[\left(\frac{|\bk|^{2}}{|\bk|^{2}+m_{\pi}^{2}}\right)^{2}+\left(\frac{|\bl|^{2}}{|\bl|^{2}+m_{\pi}^{2}}\right)^{2}+(1-\xi)\left(\frac{|\bk|^{2}}{|\bk|^{2}+m_{\pi}^{2}}\right)\left(\frac{|\bl|^{2}}{|\bl|^{2}+m_{\pi}^{2}}\right)\right] \nonumber \\
\mathcal{A}_{np}&=&\left(C_{+}^{2}+C_{-}^{2}\right)\left(\frac{|\bk|^{2}}{|\bk|^{2}+m_{\pi}^{2}}\right)^{2}+\left(4C_{+}^{2}+2C_{-}^{2}\right)\left(\frac{|\bl|^{2}}{|\bl|^{2}+m_{\pi}^{2}}\right)^{2}+\nonumber \\
&-& 2\left[\left(C_{+}^{2}+C_{-}^{2}\right)-\left(3C_{+}^{2}+C_{-}^{2}\right)\frac{\xi}{3}\right]\left(\frac{|\bk|^{2}}{|\bk|^{2}+m_{\pi}^{2}}\right)\left(\frac{|\bl|^{2}}{|\bl|^{2}+m_{\pi}^{2}}\right) \,\ ,
\label{eq:amplitude}
\end{eqnarray}
with 
\begin{eqnarray}
\xi&=&3(\hat{\bk}\cdot\hat{\bl})^{2} \,\ ,\label{eq:csi}\\
C_{\pm}&=& \frac{1}{2}\left(C_{an}\pm C_{ap}\right) \,\ ,
\label{eq:CpCn}
\end{eqnarray}
while $\bk=\bp_{1}-\bp_{3}$ and  $\bl=\bp_{1}-\bp_{4}$ are the exchanged nucleus momenta.
In the above expression, the notation $\mathcal{A}_{NN}$ indicates either $\mathcal{A}_{nn} $ or  $\mathcal{A}_{pp} $.
Analogously $C_{aN} $ indicates $ C_{an,p} $.

We point out that the sign in the last exchange term (proportional to $|\bk|^{2}|\bl|^{2}$) of   $\mathcal{A}_{np}$  is opposite 
with respect to the one in Eq.~(B4) of~\cite{Raffelt:1993ix} (which is analogous to the one reported in
 the Appendix of~\cite{Brinkmann:1988vi}). This is probably due to the missing of an additional minus sign due to the exchange
 of $p$ and $n$, which have to be considered as ``identical'' particles under isospin symmetry (see, e.g., discussion in Sec.~10.2-10.3 of~\cite{Bjorken:1965sts}).
{One can also compare our sign with the corresponding sign in Eq.~(161) of~\cite{Yakovlev:2000jp}, for the case of neutrino emission. We checked that this change of sign in the exchange term would produce a relaxation of the axion mass bound by a factor $\lesssim 2$. Therefore, this is a non-negligible effect overlooked in previous literature.
 }

It has been  customary, in the previous literature, to neglect the pion mass $m_\pi$, so that the previous expressions in Eq.~(\ref{eq:amplitude}) become constants.
As we shall see, this approximation is justified at high temperatures, when the pion mass is subdominant with respect to the exchanged momenta $|\bk|^2 \sim |\bl|^2 \approx 3 m_N T $.
With this approximation, the integration over the nucleons momenta is considerably simplified and the most complete expression for the axion emissivity, valid for generic 
nucleon degeneracy, gives
\begin{eqnarray}
Q_a^{(0)} &\simeq& 64\left(\frac{f}{m_\pi}\right)^4
{m_N^{5/2}\,T^{13/2}}
\left\{
\left(1-\frac{\overline{\xi}}{3}\right)\,g_{ an}^2\,I(y_n,y_n) +
\left(1-\frac{\overline{\xi}}{3}\right)\,g_{ an}^2\,I(y_p,y_p)
\right.
\nonumber
\\
&&
\left.
+ \frac{4(3+2\overline{\xi})}{9}\left(\frac{g_{ an}^2+g_{ ap}^2}{2}\right)\,I(y_n,y_p) +
\frac{8(3+2\overline{\xi})}{9}\left(\frac{g_{ an}+g_{ ap}}{2}\right)^2\,I(y_n,y_p)
\right\}.
\label{eq:Qa}
\end{eqnarray}
%
The fitting functions $I(y_1,y_2)$ are given in Ref.~\cite{Brinkmann:1988vi}, with nucleon degeneracy $y_i=\mu_i^0/T$ and nucleon chemical potentials $\mu_i^0 = \mu_i-m_i$, where  bare nucleon masses are assumed. The factor $ \overline{\xi} $ parameterizes the degree of degeneracy, with $\overline{\xi}=0$ for degenerate matter and $\overline{\xi}=1.3078$ in the 
non-degenerate case~\cite{Raffelt:1993ix}.

\subsection{Nucleon structure function}

It is useful to express the axion emissivity $Q_a$ of Eq.~(\ref{eq:emissivity}) in terms of the structure function~\cite{Raffelt:1993ix,Hannestad:1997gc}
\begin{eqnarray}
S_{\sigma} &=& \frac{1}{n_B}
 \int  \prod_{i=1,4} \frac{d^3p_i}{2 m_N (2\pi)^3}~f_1 f_2 (1-f_3) (1-f_4)
{\overline M} \nonumber \\
 &\times&  \delta (E_1+E_2-E_3-E_4+\omega)\delta^3({\bf p}_1+{\bf p}_2-{\bf p}_3-{\bf p}_4)
  \,\ ,
\end{eqnarray}
where $n_B$ is the baryon density, 
and $\omega=-\omega_a$ for axion emission, i.e. is the transfer of energy
to the nuclear medium. The structure function obeys to the detailed balance $S_\sigma(-\omega)= S_{\sigma}(\omega)e^{-\omega/T}$.

One can write the axion emissivity in terms of the structure function as~\cite{Raffelt:1993ix}
\begin{equation}
Q_a = \frac{g_a^2}{16 \pi^2}\frac{n_B}{m_N^2} \int_{0}^{+\infty} d\omega e^{-\omega/T} \omega^4 S_\sigma(\omega) \,\ ,
\end{equation}
where $S_{\sigma}(\omega)$ reads
\begin{equation}
S_{\sigma}(\omega)= \frac{\Gamma_\sigma}{\omega^2} s(\omega/T) \,\ ,
\end{equation}
in terms of   the  nucleon ``spin fluctuation rate'' $\Gamma_\sigma$, while $s(x)$ is a dimensionless function. 
The factorization between $\Gamma_\sigma$ and $s(x)$ is not unique.
Following~\cite{Raffelt:1993ix,Hannestad:1997gc}
we  take $\Gamma_\sigma$ such that $s(0)=1$ for a medium of only one non-degenerate nucleon species and when 
the $\pi$ and the $\varrho$ mass exchange is neglected, leading to~\cite{Raffelt:1993ix,Hannestad:1997gc}
\begin{equation}
\Gamma_{\sigma}=4\pi^{-1.5}\rho\left(\frac{g_{A}}{2f_{\pi}}\right)^{4}T^{0.5}m_N^{0.5}=21.6 \,\ \textrm{MeV} \rho_{14}T_{\rm MeV}^{0.5}m_{938}^{0.5} \,\ ,
\label{eq:spinfluc}
\end{equation}
where $ \rho_{14}= \rho/10^{14} \textrm{g} \,\ \textrm{cm}^{-3}$, and $T_{\rm MeV}= T/1\,\ \textrm{MeV}$ and
$m_{938}= m_N/ 938 \,\ \textrm{MeV}$.

In this case, the explicit form of the $s(x)$ function for a medium composed of protons and neutrons is given in
Eq.~(\ref{eq:sfunction}) in the Appendix A.

Assuming that the nucleon spins evolve independently of each
other, i.e. ignoring possible spin-spin correlations, the full scattering kernel
must obey the normalization requirement, { that for a medium composed by a mixture of 
protons and neutrons reads}~\cite{Raffelt:1996di,Hannestad:1997gc}
\begin{equation}
\int_{-\infty}^{+\infty} \frac{d \omega}{2 \pi} S_{\sigma}(\omega) = \frac{1}{n_B}
\sum_{i=p,n} C_{ai}^2  \int \frac{2 d^3 {\bf p}}{(2\pi)^3}f_i (1-f_i) \,\ .
\label{eq:normal}
\end{equation}
In case of a single non-degenerate species, one can neglect the Pauli blocking factor 
$(1-f_i)$  obtaining that the normalization is one.

\subsection{Corrections to emissivity beyond naive one-pion-exchange}
\label{sec:correction}

Different improvements have been discussed in the literature, beyond the naive OPE. 
However, a reliable calculation is highly nontrivial. 
Difficulties include the description of an appropriate approximation for the nucleon-nucleon interaction potential, accounting for intermediate degree of nucleon
degeneracy (which is in general different for protons and neutrons), nucleon-nucleon correlations, and multiple nucleon scattering effects. 
We will not be able to resolve all of these issues. 
In particular, we will ignore NN correlations in this work.
However, we will include several improvements over the previous approximation.
More specifically, we are including the following corrections:
\subsubsection*{Nonzero pion mass in the pion propagator}
At a temperature $T$ the typical nucleon momentum is $(3m_NT)^{1/2}$ so that a typical
momentum exchange in a collision is of a similar magnitude. At $T = 10$~MeV this is about
170~MeV, only slightly larger than the pion mass of 135~MeV. This implies that the
pion mass cannot be ignored in the denominators in Eq.~(\ref{eq:amplitude}), while its impact
would decrease at higher temperature (see~\cite{Raffelt:1993ix,Hannestad:1997gc,Stoica:2009zh} for early investigations).
\subsubsection*{Two-pions exchange}
Two-pions exchange effects become important at distances below 
$2~\textrm{fm} \simeq  1.5~{m_\pi}^{-1}$. 
It has been proposed in~\cite{Ericson:1988wr} to evaluate the impact on the axion emissivity
 by mimicking the two-pions exchange contribution by a one--meson exchange, with effective mass
  $m_{\varrho} = 600$~MeV.
 One then modifies the propagator in Eq.~(\ref{eq:amplitude}) as
\begin{equation}
\left(\frac{|\bk|^{2}}{|\bk|^{2}+m_{\pi}^{2}}\right)^{2}
\to \left(\frac{|\bk|^{2}}{|\bk|^{2}+m_{\pi}^{2}} - C_\varrho \frac{|\bk|^{2}}{|\bk|^{2}+m_{\varrho}^{2}} \right)^{2} \,\ ,
\label{eq:rhomass}
\end{equation}
with $C_\varrho=1.67$. 
Taking  ${\bf q}^2 \simeq 3 T m_N$, at $T=10$~MeV one 
gets a $\sim 35 \%$ reduction ot the squared matrix element (see also~\cite{Hannestad:1997gc}). 
\begin{figure}[t!]
\vspace{0.cm}
\includegraphics[width=0.5\textwidth]{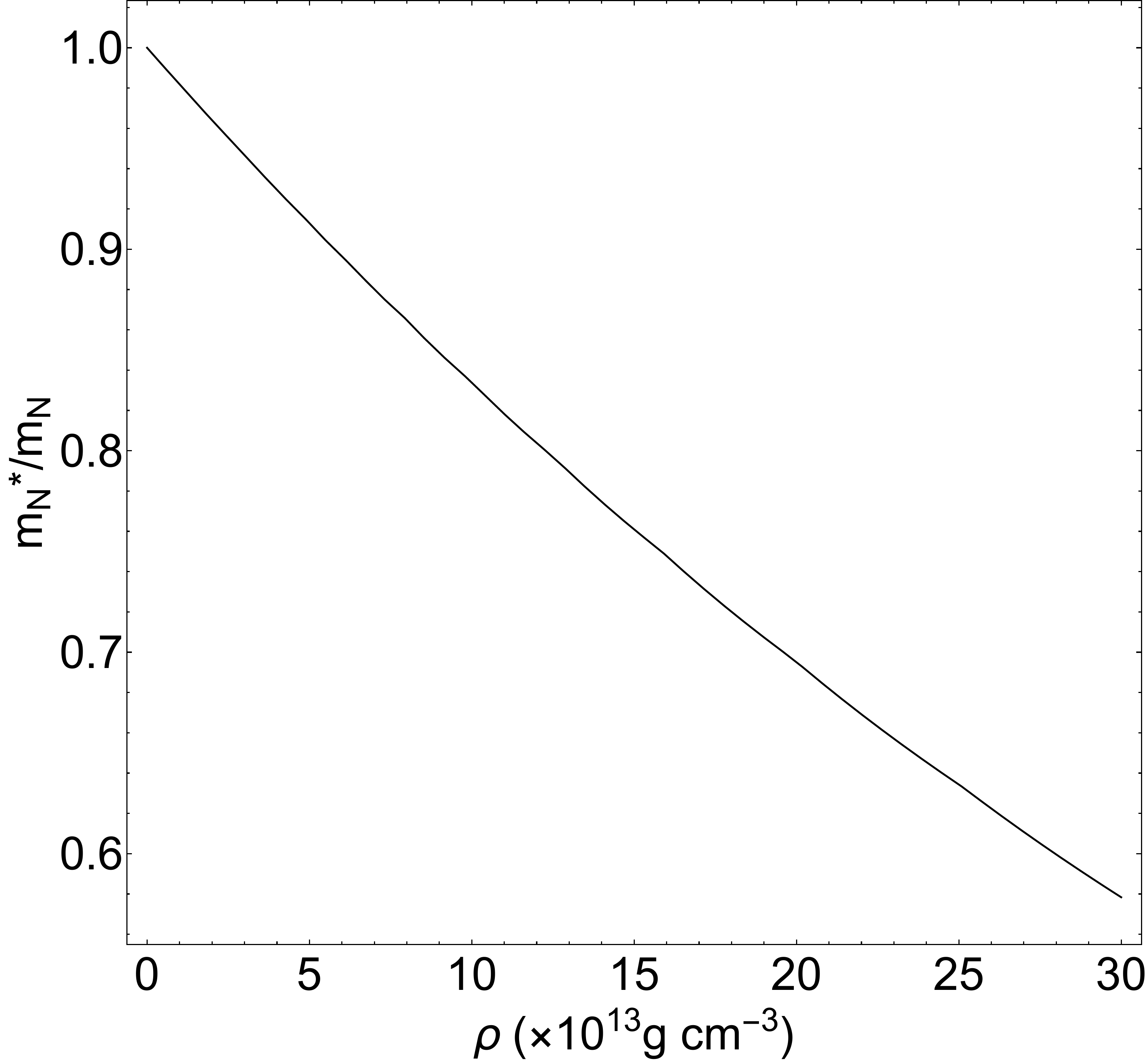}
\includegraphics[width=0.5\textwidth]{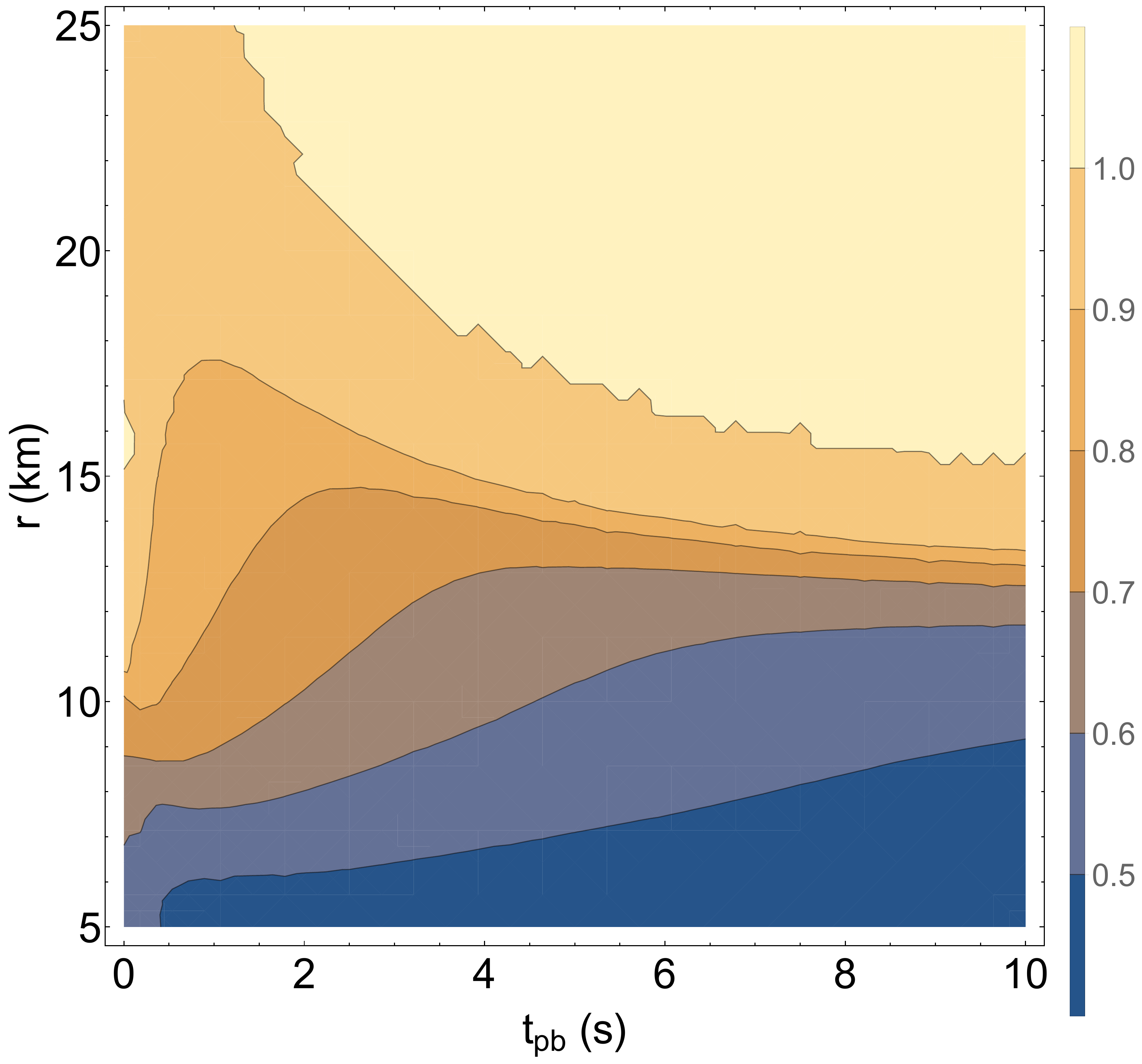}
\caption{Ratio of effective nucleon mass $m^{\ast}_N$ with respect to the vacuum value $ m_{N} $.
Left panel: $m^{\ast}_N/m_{N} $ as a function of $\rho$. Right panel: $m^{\ast}_N/m_{N} $ in the $t_{\rm pb}$-$r$ plane.}
\label{fig:mN}
\end{figure}
%
\subsubsection*{Effective nucleon mass}
In the relativistic mean-field treatment, the reduction of the nucleon masses due to the scalar interactions at high density leads to effective nucleon masses 
\begin{equation}
m^{\ast}_N= m_N+ \Sigma_S \,\ ,
\end{equation}

 which replace the vacuum mass $m_N$ in the Fermi-Dirac nucleon distributions Eq.~(\ref{eq:fddistrib})
 (see, e.g.~\cite{Hempel:2014ssa,Iwamoto:1984ir}) so that the non-relativistic nucleon energy can be approximated as~\cite{Hempel:2014ssa,MartinezPinedo:2012}
 \beq
 E_i \simeq m_N + \frac{{|{\bf p}_i|}^{2}}{2 m^{\ast}_N} +U_i \,\ ,
 \label{eq:energy}
 \eeq
 where  $U_i= \Sigma_S + \Sigma_V$ is the non-relativistic
mean-field potential that in the case of a relativistic model contains
contributions of the scalar ($\Sigma_S$) and vector ($ \Sigma_V$) self-energies (see
e.g.~\cite{Hempel:2014ssa}). 
 
 Figure \ref{fig:mN}  shows the ratio  of the effective nucleon mass $m^{\ast}_N$, based on the nuclear equation of state (EOS) given in~\cite{Typel:2009sy,Hempel:2009mc,Steiner:2012rk}, with respect to its vacuum value as a function of the density $\rho$ (left panel) and in the plane $t_{\rm pb}$-$r$ (right panel), where $ t_{\rm pb} $ is the post-bounce time and $ r $ the radial distance from the center of the SN.
 One realizes that for densities $\rho \lesssim 10^{14} \textrm{g} \,\ \textrm{cm}^{-3}$ at 
 $r \gtrsim 20$~km,  
 the effective nucleon mass  approaches the vacuum value.
 Conversely, at larger densities the correction becomes sizable and rather insensitive to the temperature.
 In particular, for $\rho \gtrsim 3 \times 10^{14} \textrm{g} \,\ \textrm{cm}^{-3}$ 
one should expect a $\sim 40 \%$ reduction in the nucleon mass.
In principle one should take into account also the density-dependence of $m_{\pi}$ and $m_{\varrho}$.
However, due to the large  uncertainties in this dependence we neglect this effect~\cite{Mayle:1989yx}.
%
\subsubsection*{Nucleon multiple scatterings}
As discussed in~\cite{Raffelt:1991pw,Raffelt:1993ix,Janka:1995ir,Raffelt:1996za,Sigl:1997ga}, many-body effects caused by multiple nucleon scatterings can also reduce the axion emissivity. 
 If nucleons constantly scatter with each other in the medium, their individual energies become uncertain so that the (energy dependent) classical structure function acquires a finite width.
As shown in the references cited above, 
 one can take the multiple scattering effects 
{by a phenomenological ansatz
 \begin{equation}
 S_{\sigma}(\omega)= \frac{\Gamma_\sigma}{\omega^2 + \Gamma^2} s(\omega/T)  \,\ . 
 \label{eq:Loren}
 \end{equation}
 The scattering kernel is obtained determining the $s(x)$ function from the bremsstrahlung calculation including the different
 approximations. Then the $\Gamma$ function in Eq.~(\ref{eq:Loren}) is determined in such a way that the normalization condition in
 Eq.~(\ref{eq:normal}) is fulfilled.
 One also has  to take into account that the structure function saturates for $\Gamma_\sigma/T \gtrsim 10$~\cite{Janka:1995ir}. 
  }


We give more details on how to implement the axion emissivity including all the previous corrections in Appendix A. 
{We remark that our calculation represents the state-of-the-art beyond OPE, including all corrections discussed separately
in the previous literature. Moreover, it is valid for arbitrary degeneracy. Therefore, it can be considered the generalization 
of the OPE calculation of Eq.~(\ref{eq:Qa}).}

\subsection{Impact of the different corrections}

\subsubsection*{SN reference model}
%
\begin{figure}[t!]
\vspace{0.cm}
\includegraphics[width=0.5\textwidth]{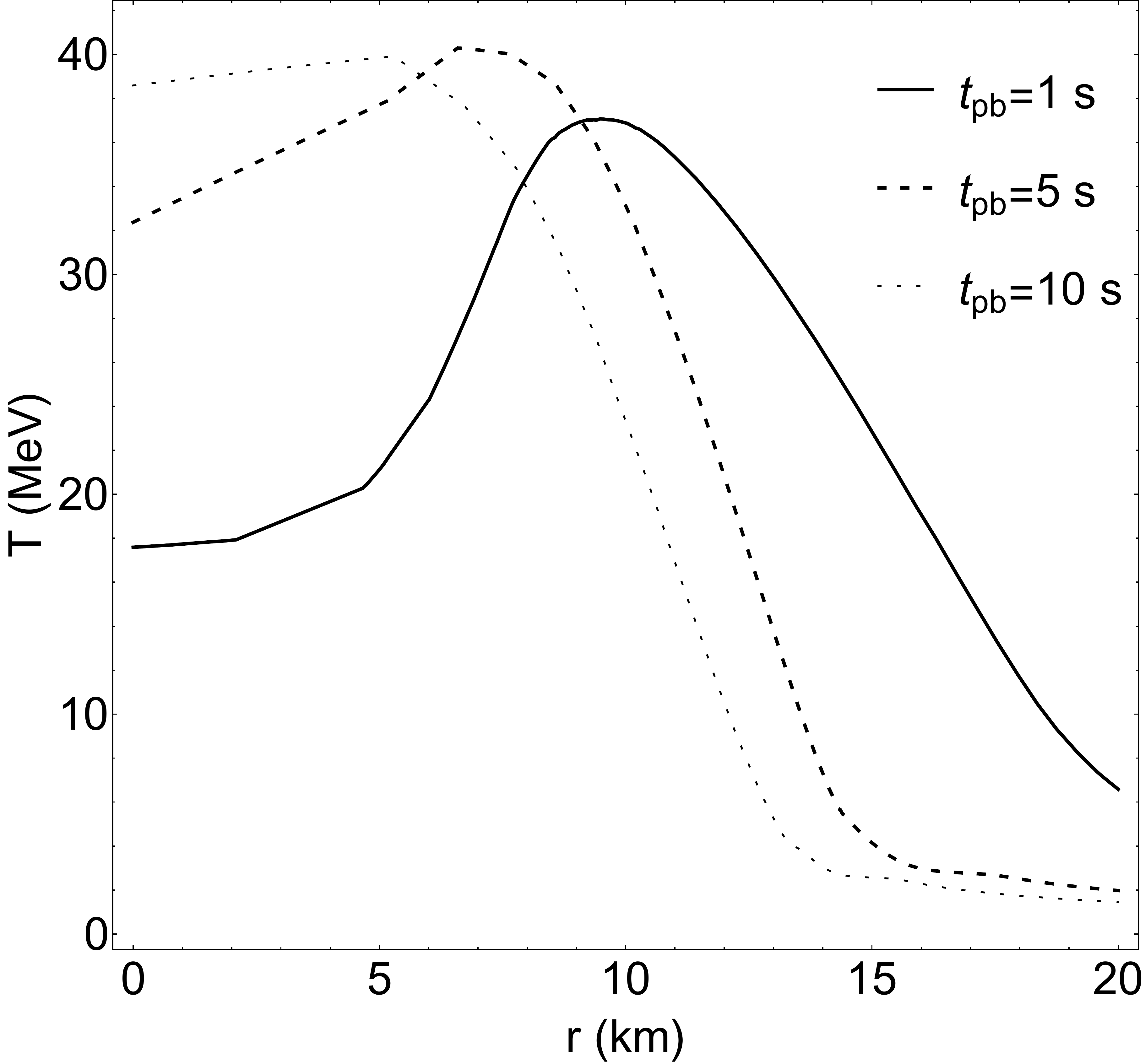}
\includegraphics[width=0.5\textwidth]{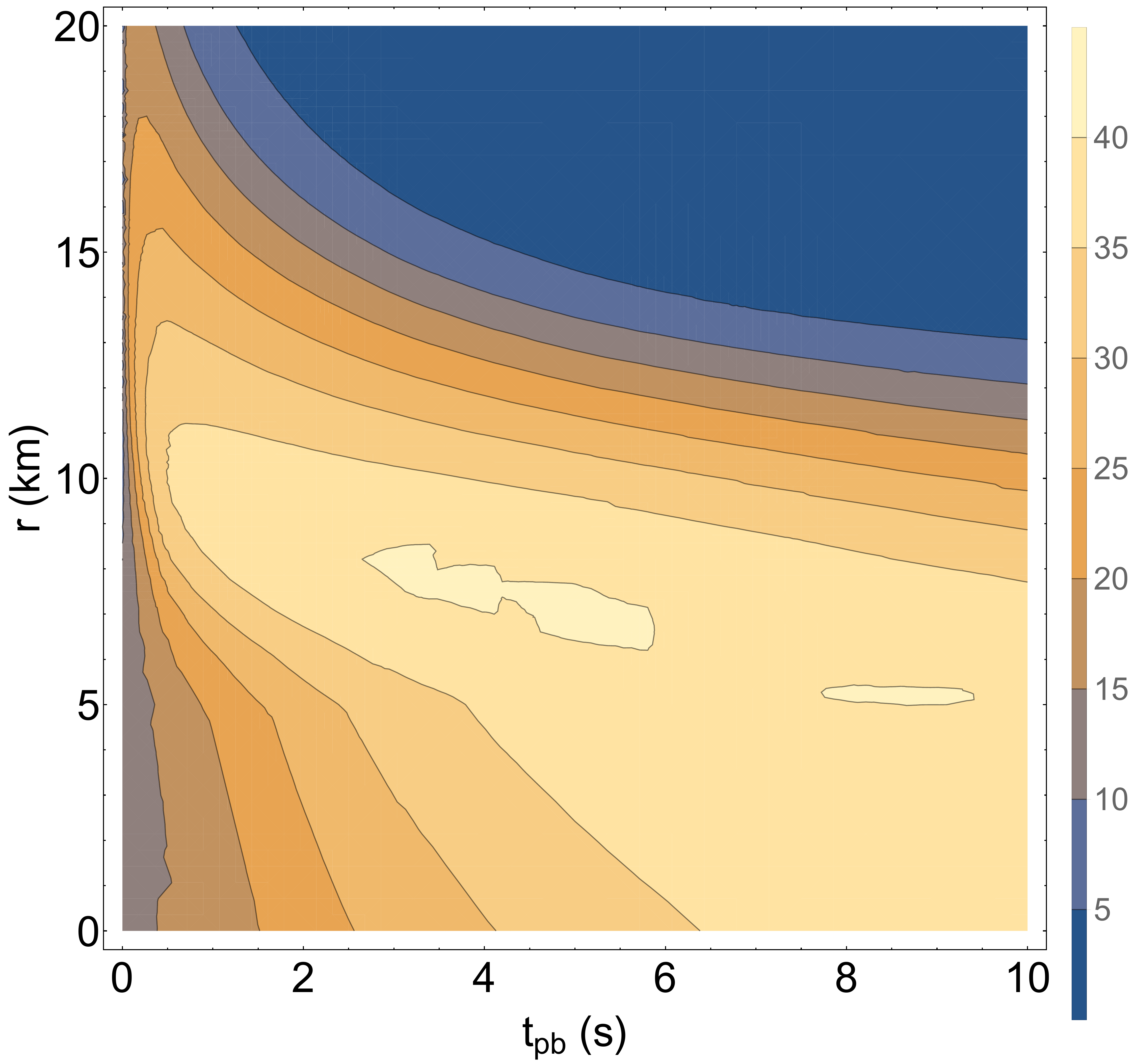}
\caption{Left panel: Radial evolution of the temperature $T$ at different post-bounce times $t_{\rm pb}$. Right panel:
$T$ behavior in the plane $t_{\rm pb}$-$r$.}
\label{fig:Temp}
\end{figure}
\begin{figure}[t!]
\vspace{0.cm}
\includegraphics[width=0.5\textwidth]{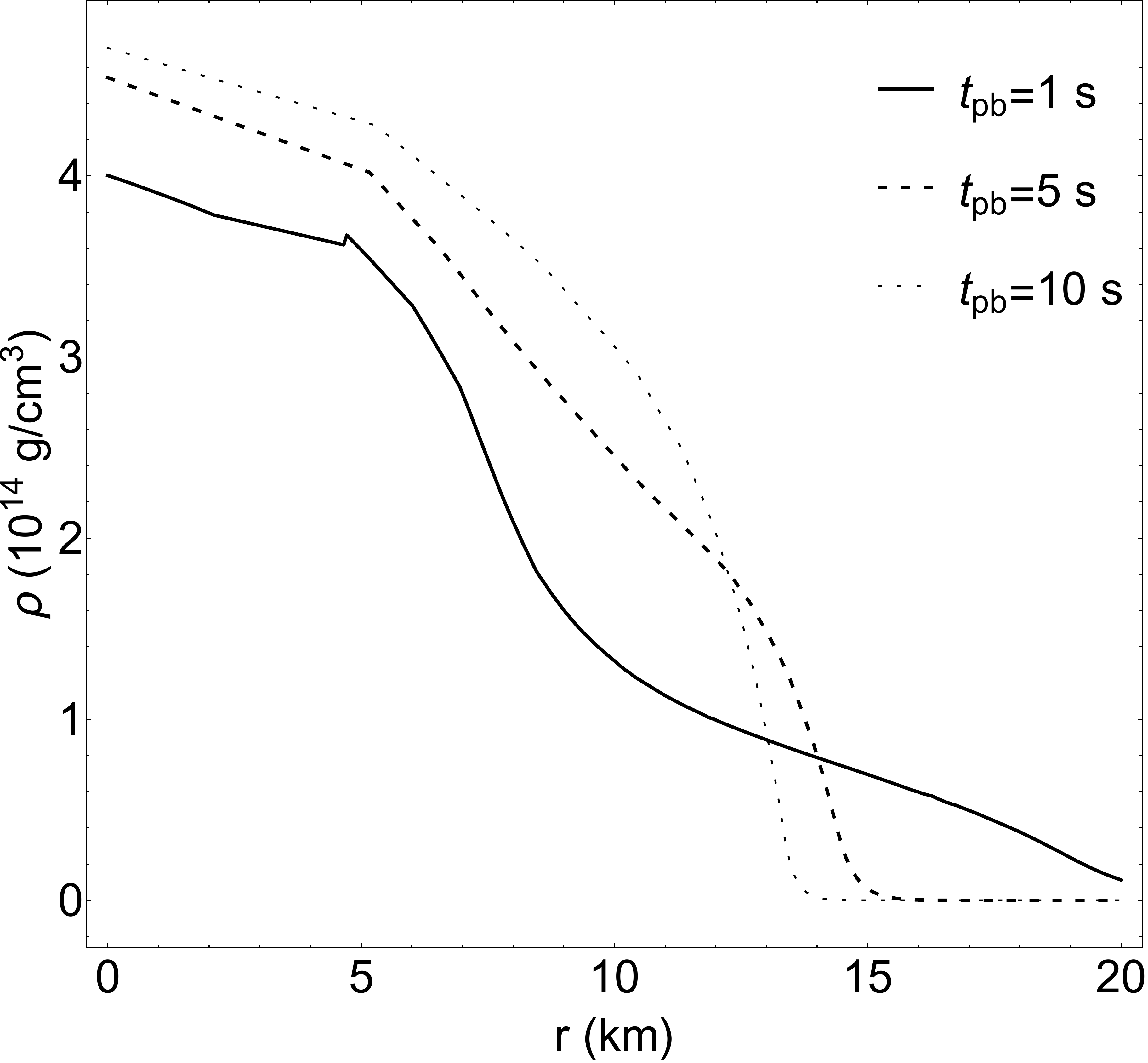}
\includegraphics[width=0.5\textwidth]{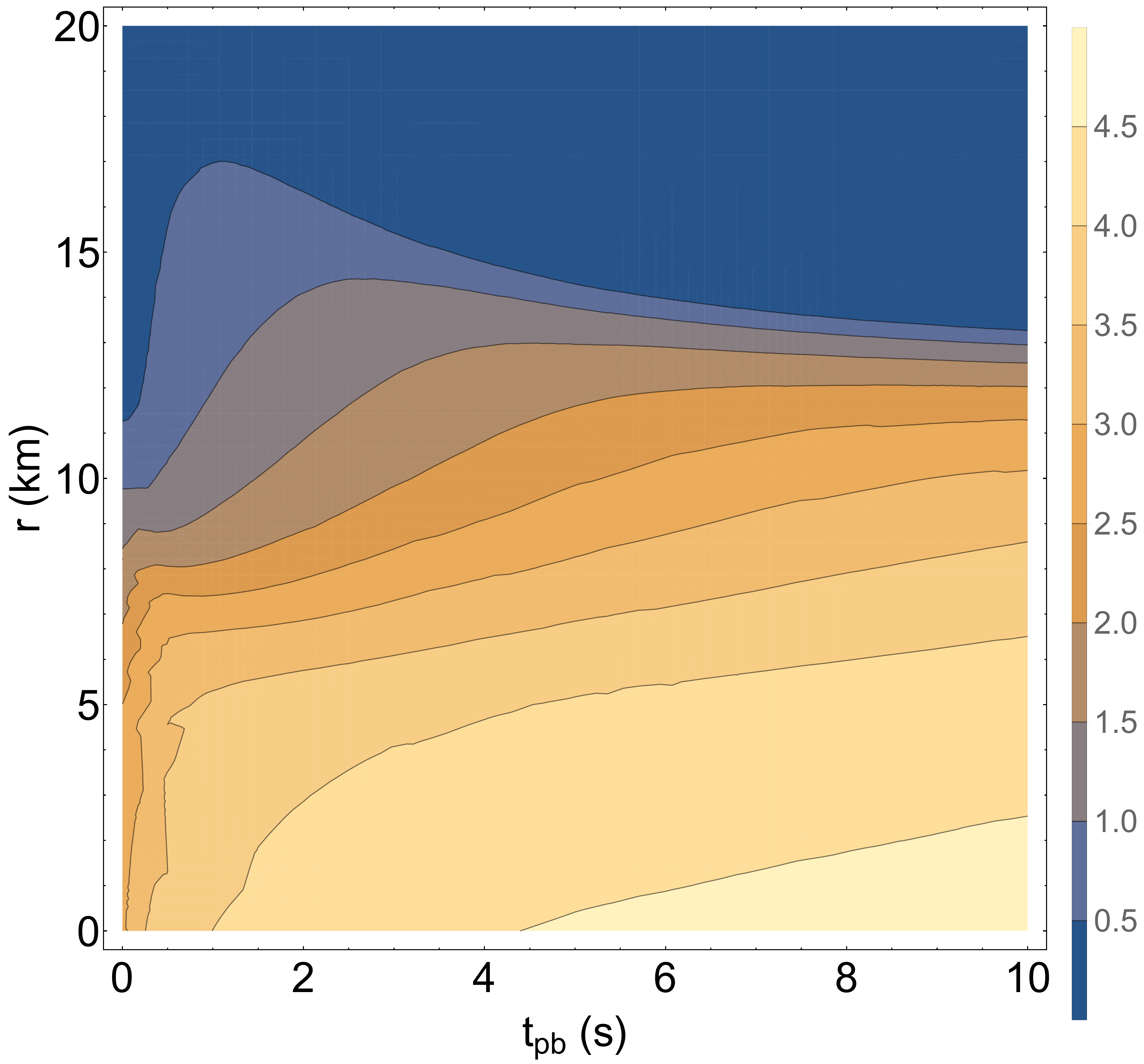}
\caption{Left panel: Radial evolution of the matter density $\rho$ at different post-bounce times $t_{\rm pb}$. Right panel:
$\rho$ behavior in the plane $t_{\rm pb}$-$r$.}
\label{fig:dens}
\end{figure}

In this Section we analyze the impact of the different corrections introduced beyond the naive OPE calculation of the axion emissivity. 
In order to connect  these results in relation to realistic SN models, 
we consider as SN reference a model with 18 $M_{\odot}$ progenitor 
simulated  in spherical symmetry with the  AGILE-BOLTZTRAN  code~\cite{Mezzacappa:1993gn,Liebendoerfer:2002xn}.
The matter density and temperature of our model  shown  in Fig.~\ref{fig:Temp} and \ref{fig:dens} 
 correspond to the simulation of the protoneutron star deleptonization of~\cite{Fischer:2016} without considering axions.
The left panels show the radial behavior of temperature and density at different post-bounce times
$t_{\rm pb}$, while the right panels 
present contours of $T$ and $\rho$ in the $t_{\rm pb}$ and radial coordinate $r$ plane. 
We can see that the temperature $T$ presents a peak of $\sim 40$~MeV 
at $r \simeq 10$~km. 
Given the steep temperature dependence of the axion emission rate, one expects the higher axion production around this peak temperature. One also realizes that at  larger 
$t_{\rm pb}$ the peak in the temperature  recedes, due to the cooling of the proto-neutron star.
Concerning the matter density, we realize that it exceeds the nuclear density ($\rho_0 \simeq 3 \times 10^{14}
\,\ \textrm{g} \,\  \textrm{cm}^{-3}$) in the deepest SN regions ($r\lesssim 10$~km). 
There the density increases with time due to the 
contraction of the proto-neutron star. 

In Fig.~\ref{fig:mucontour} we show  in the plane $t_{\rm pb}$-$r$  the contour plots for the degeneracy parameter
\beq
\eta_i = \frac{\mu_i-m_N}{T} \,\ ,
\label{eq:degen}
\eeq
where $\mu_i$ is the chemical potential,
for {interacting protons (upper left panel), interacting neutrons (upper right panel), free protons (lower left panel) and free neutrons (lower right panel).
We remark that in the case of interacting nucleons, due to Eq.~(\ref{eq:energy}) the {degeneracy parameter}  in Eq.~(\ref{eq:degen})  includes
also the non-relativistic
mean-field potential $U_i$ (see Appendix A) that in the case of a relativistic model contains
contributions of the scalar and vector self-energies~\cite{Hempel:2014ssa,MartinezPinedo:2012}, i.e.,~\footnote{{Note that in the previous version of this paper,  published on JCAP, we missed to include  the vector contribution, $\Sigma_V$, to the self-energy $U$ in
the calculation of the degeneracy parameter. This caused  inconsistencies in the calculation of the nuclear densities. 
Including the attractive $\Sigma_V$ contribution leads to a larger $ \eta_i $ (thus, to a slightly more degenerate plasma).}
 \begin{equation}
{\eta}_i = \frac{\mu_i -m_N - U_i}{T}\,\ .
\end{equation}
 }
 We realize that for typical SN conditions protons are {almost }  non-degenerate {$\eta_p< 0$}, while neutrons can be mildly degenerate
($\eta_n \gtrsim 1$). Therefore, the non-degenerate approximation often used in literature is not completely justified.
When needed,
we take as representative values of degeneracy for our estimation below 
{$\eta_p=-1$}  and $\eta_n=1$. Moreover, in the following for simplicity we assume $g_{ap}=g_{an}$. 
The generalization to different couplings is discussed in Section~\ref{sec:Axion mass bound and comparison with previous works}. 

\begin{figure}[t!]
\vspace{0.cm}
\includegraphics[width=0.5\textwidth]{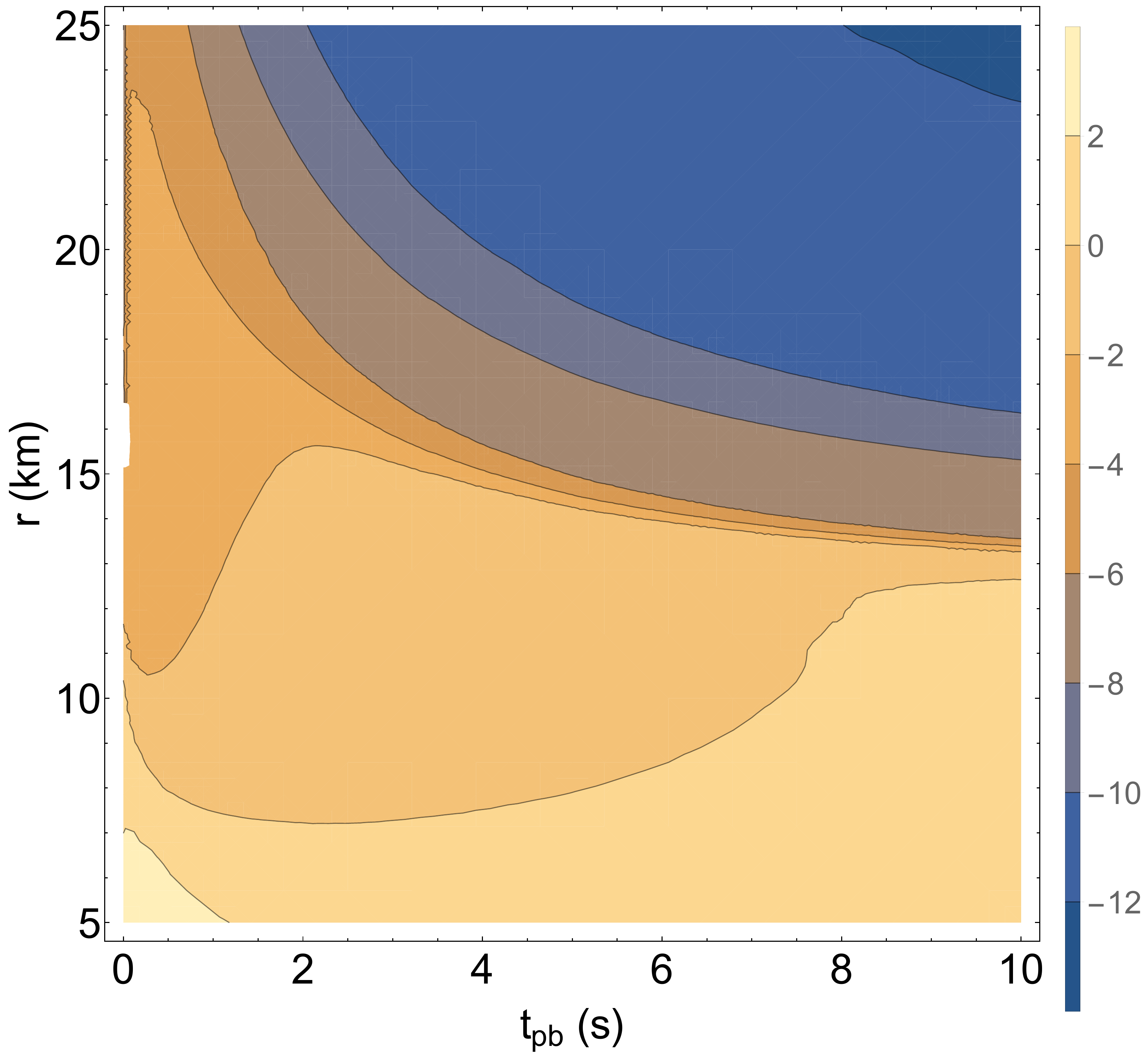}
\includegraphics[width=0.5\textwidth]{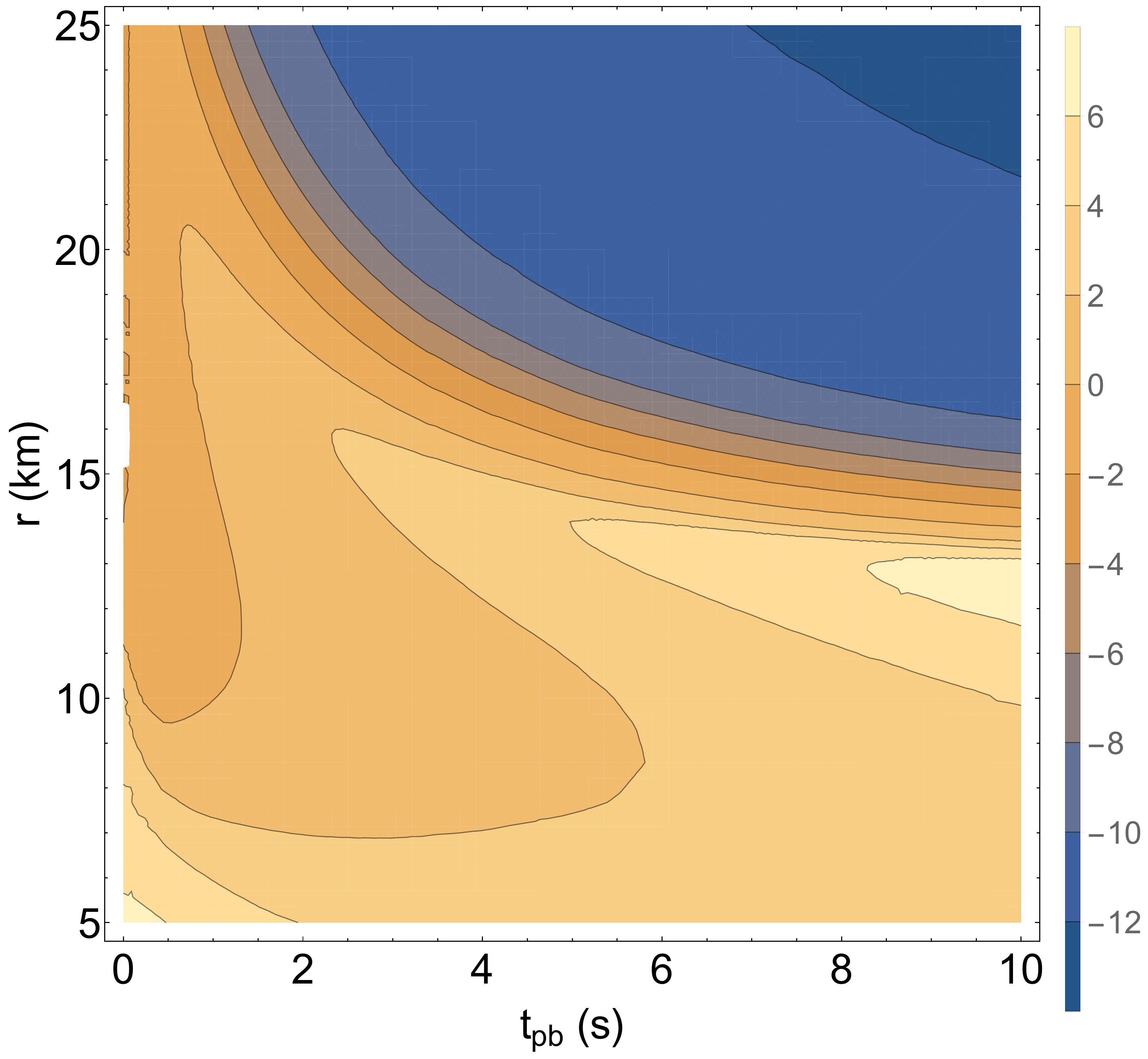}
\includegraphics[width=0.5\textwidth]{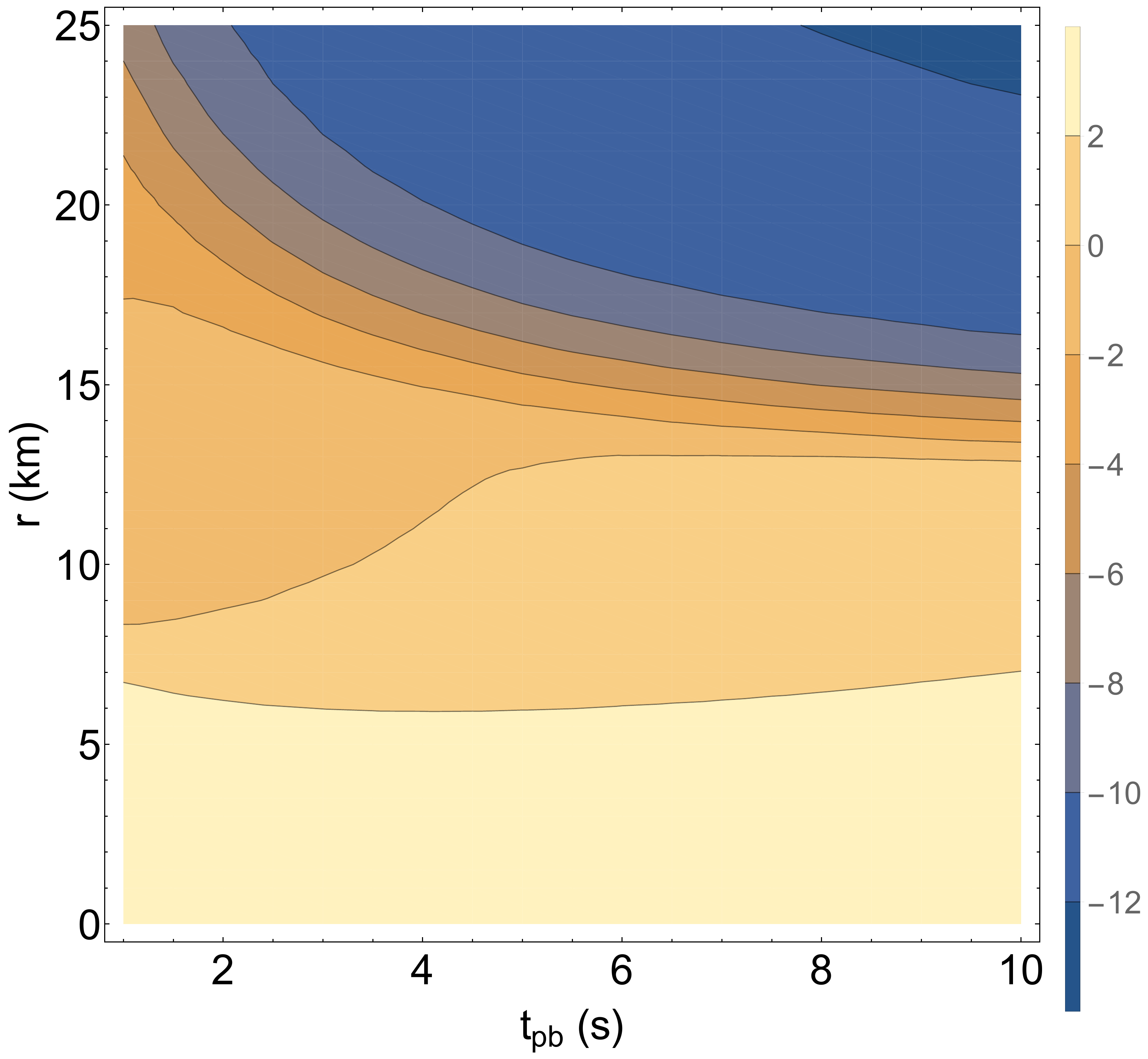}
\includegraphics[width=0.5\textwidth]{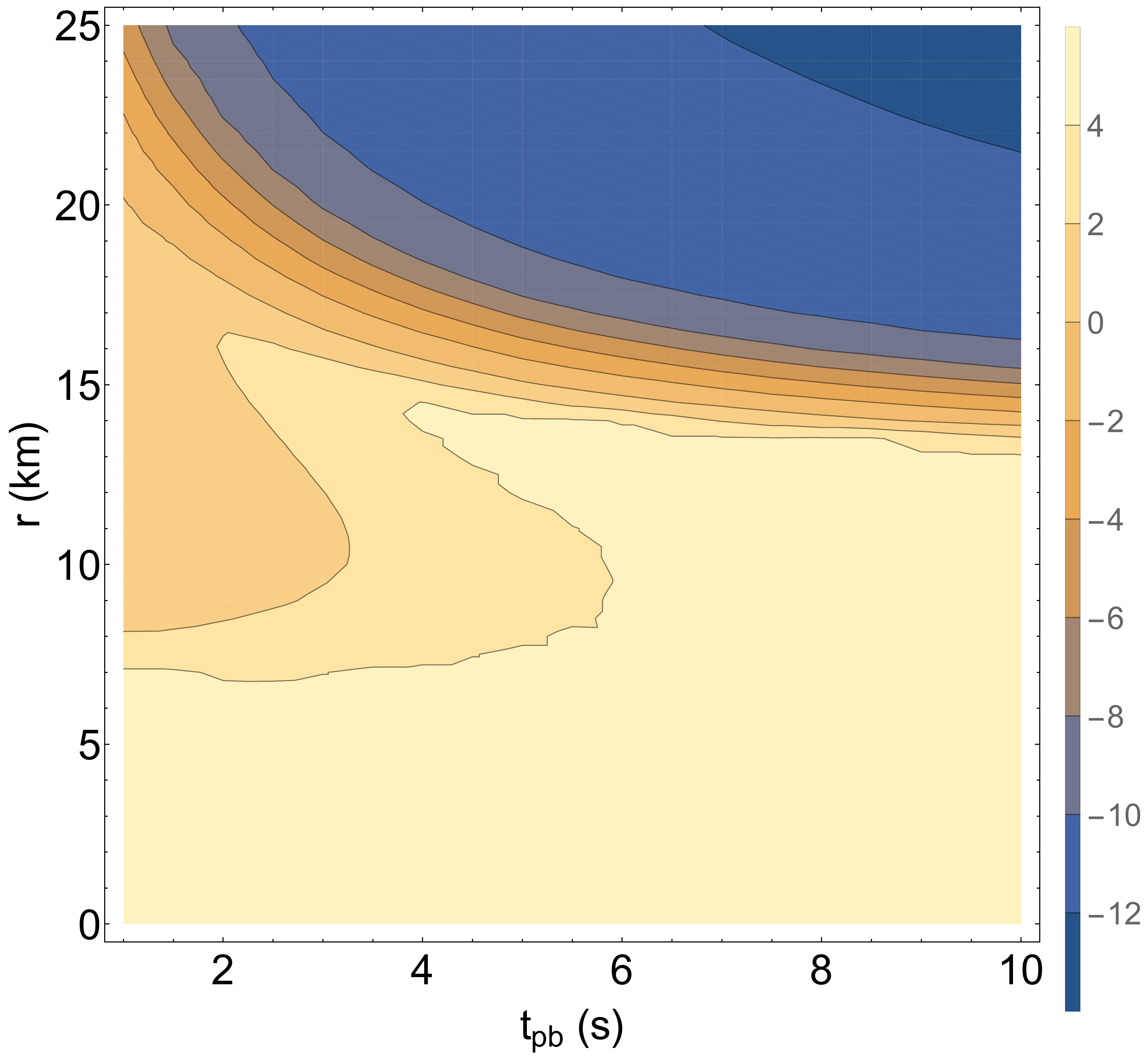}
\caption{{Contour plot of the {degeneracy parameter} $\eta$ in the plane $t_{\rm pb}$-$r$
for interacting protons (upper left panel), interacting neutrons (upper right panel), free protons (lower left panel) and free neutrons (lower right panel).}}
\label{fig:mucontour}
\end{figure}
%

\begin{figure}[t!]
\vspace{0.cm}
\includegraphics[width=0.5\textwidth]{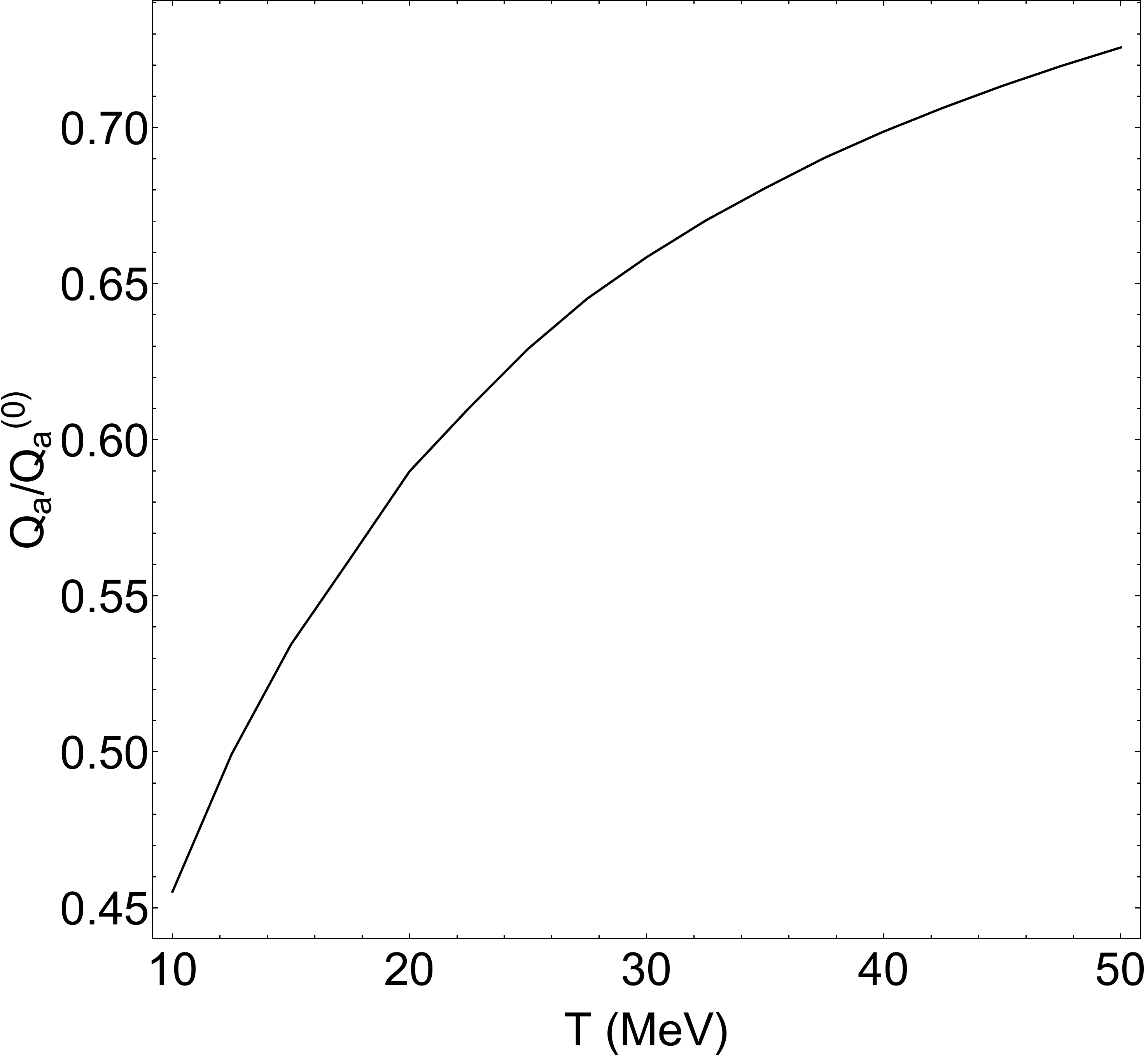}
\includegraphics[width=0.5\textwidth]{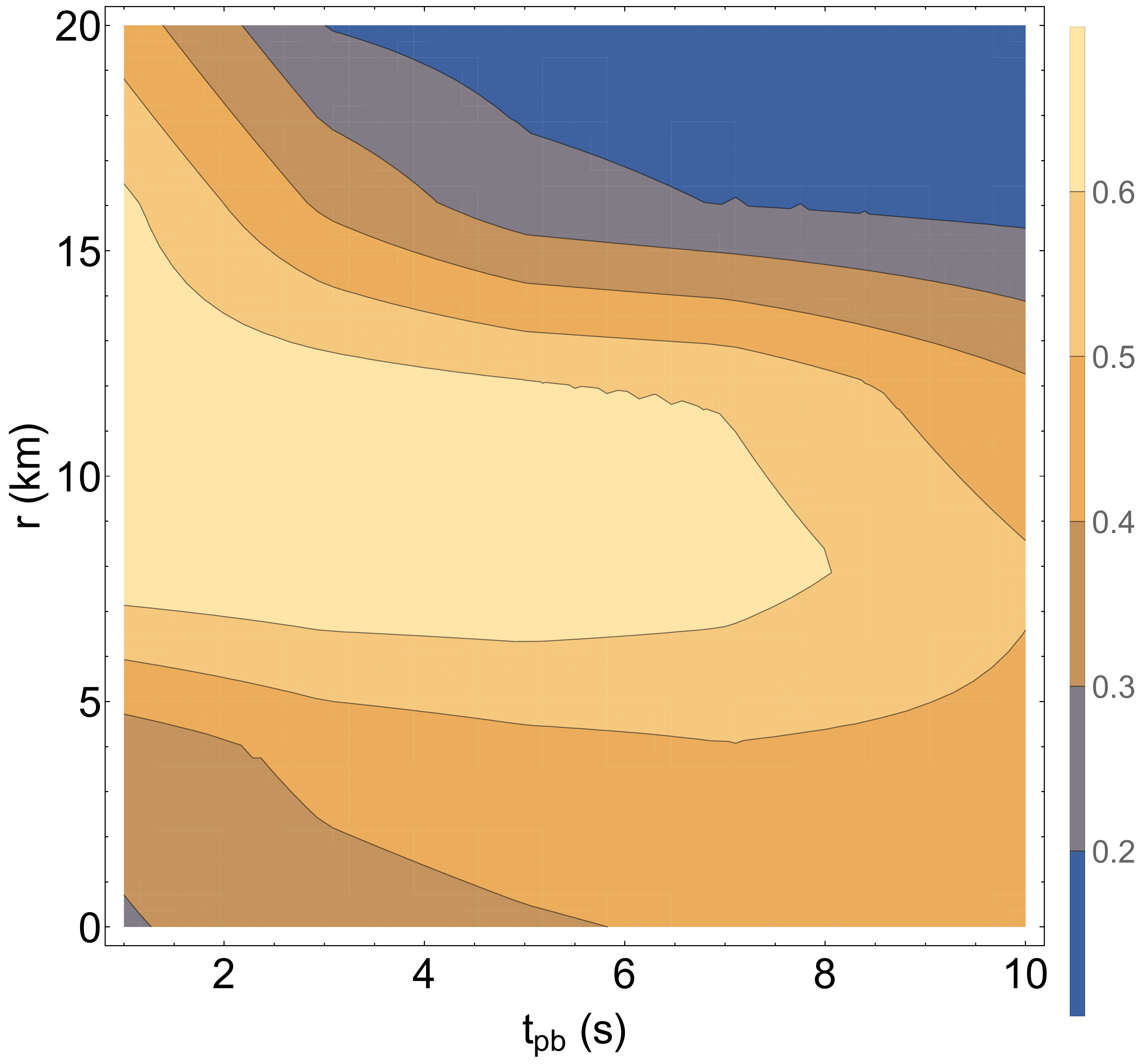}
\caption{Ratio of axion emissivity $Q_a/Q_a^{(0)}$ with pion mass correction.
Left panel: Ratio in function of $T$. Right panel: Plane $t_{\rm pb}$-$r$.}
\label{fig:mpi}
\end{figure}

\begin{figure}[t!]
\vspace{0.cm}
\includegraphics[width=0.5\textwidth]{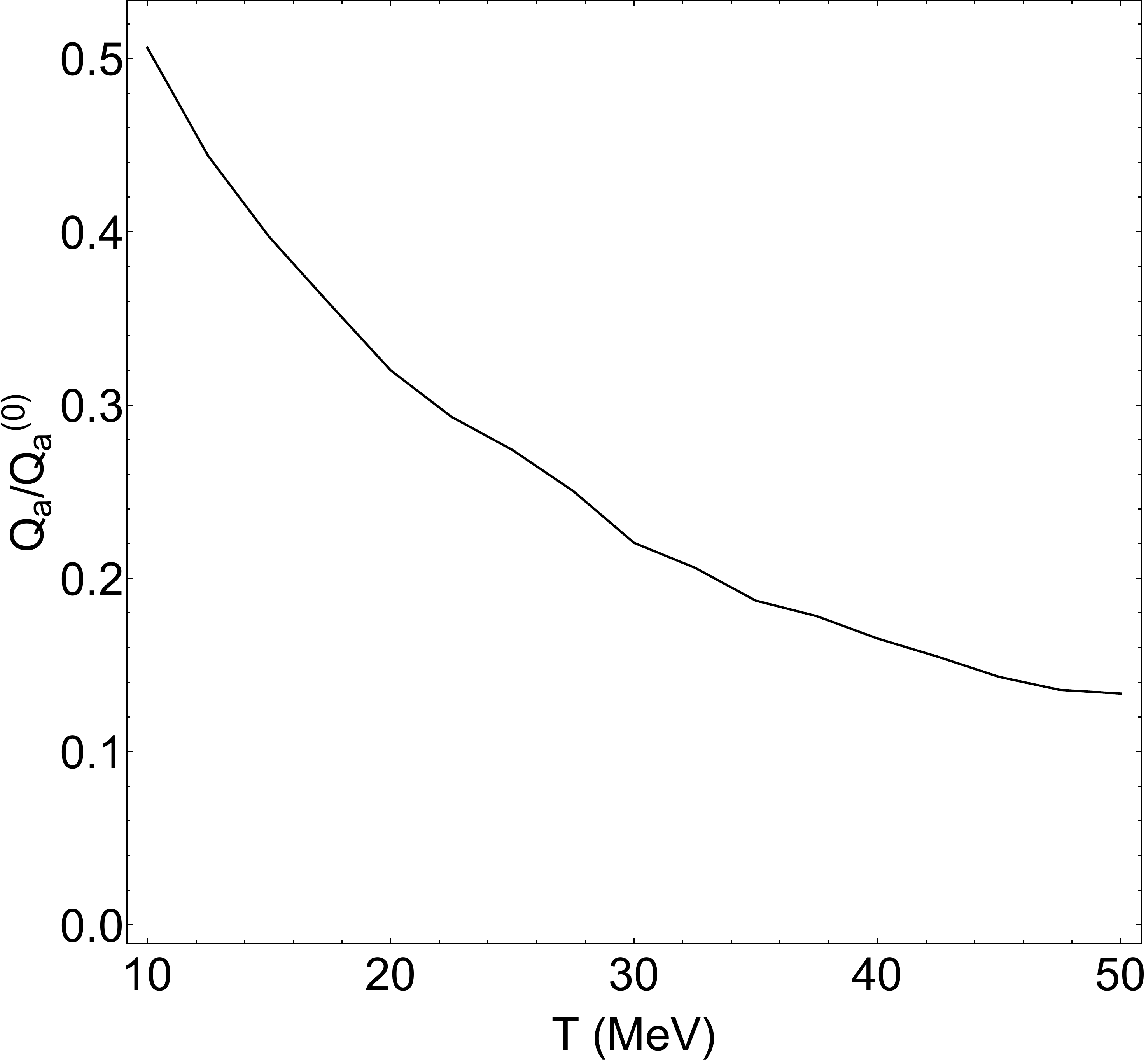}
\includegraphics[width=0.5\textwidth]{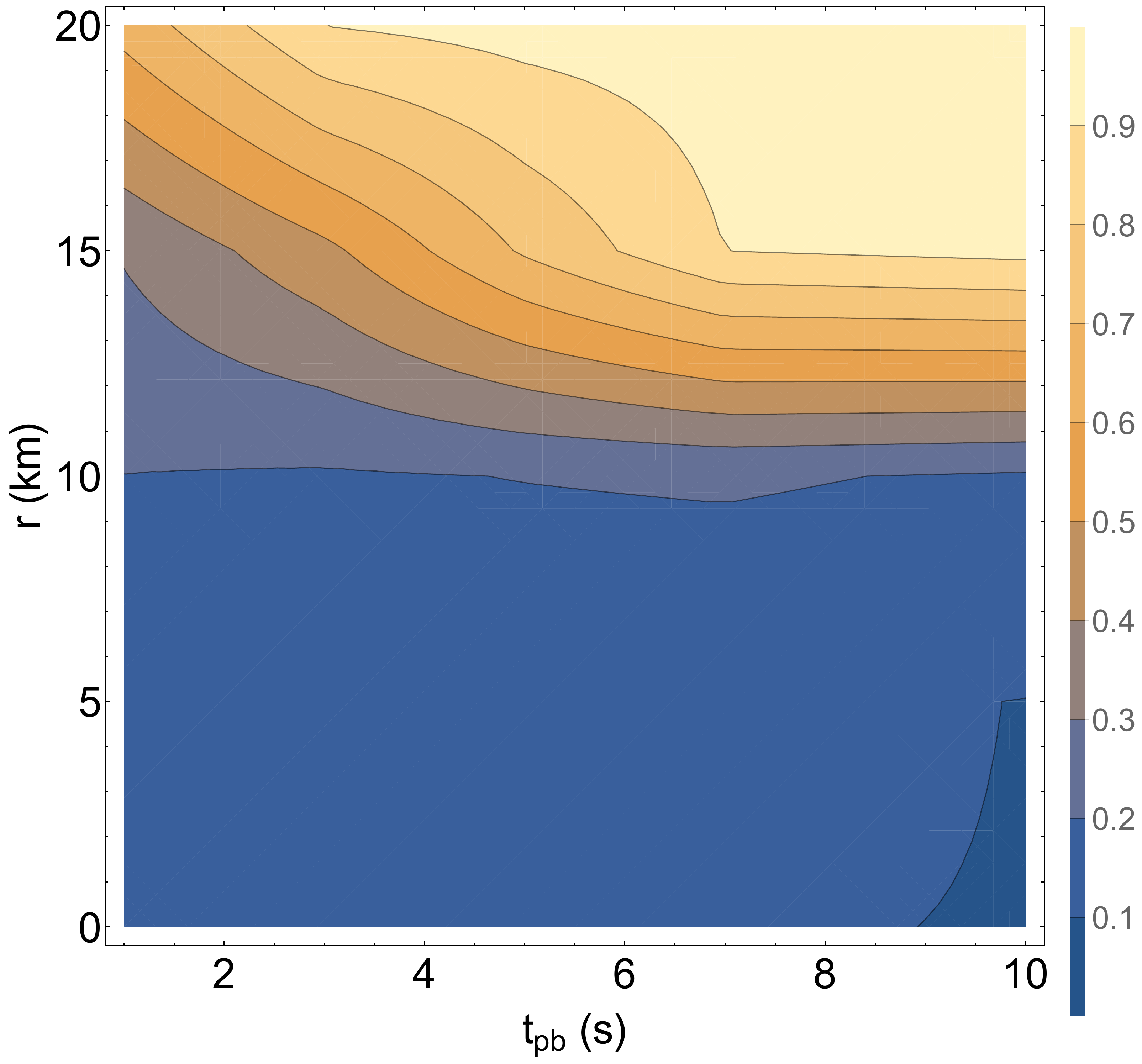}
\caption{Ratio of axion emissivity $Q_a/Q_a^{(0)}$ with  $\varrho$ mass correction
[see Eq.~(\ref{eq:rhomass})].
Left panel: Ratio in function of $T$. Right panel: Plane $t_{\rm pb}$-$r$.}
\label{fig:mrho}
\end{figure}

\begin{figure}[t!]
\vspace{0.cm}
\includegraphics[width=0.5\textwidth]{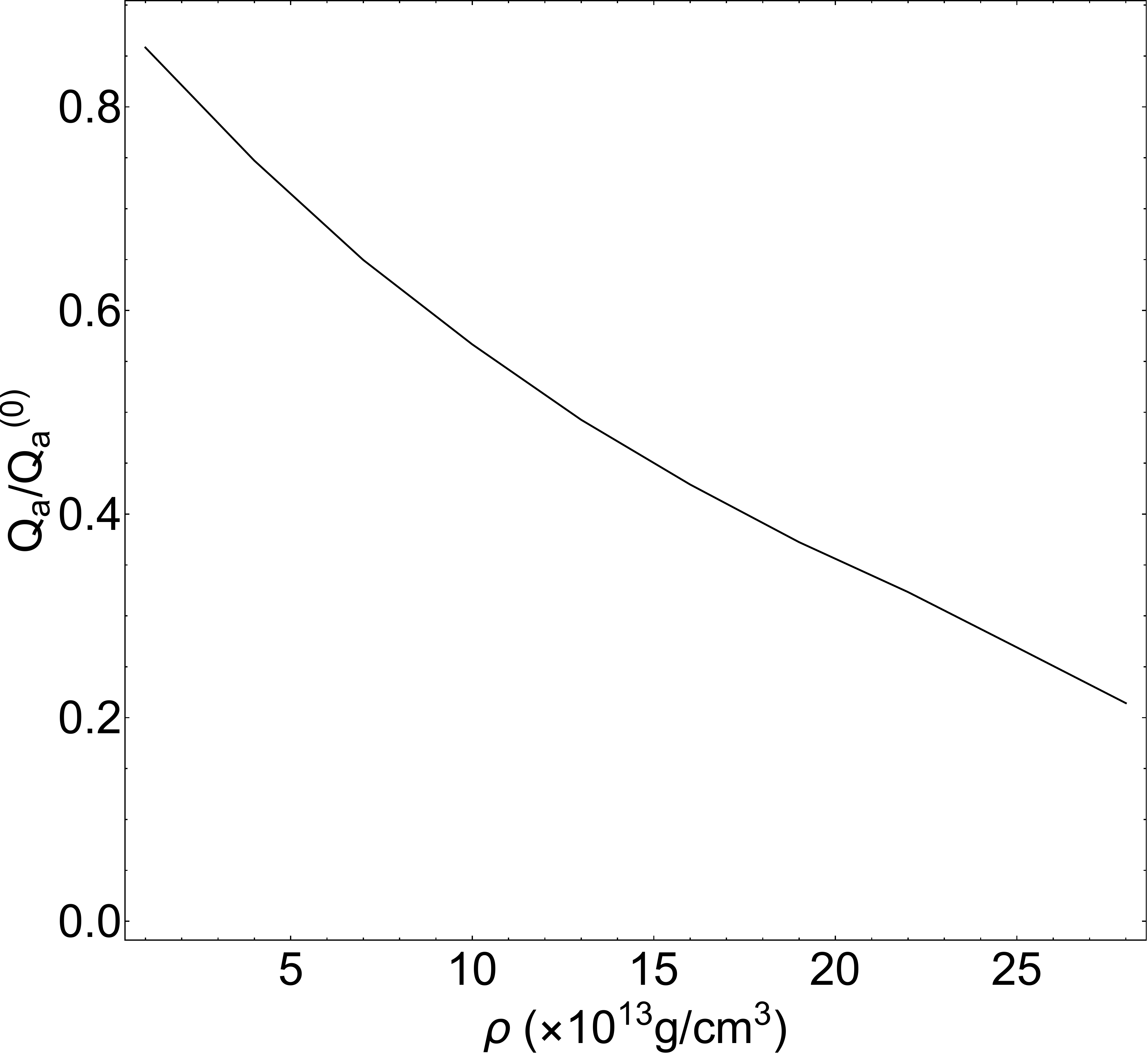}
\includegraphics[width=0.5\textwidth]{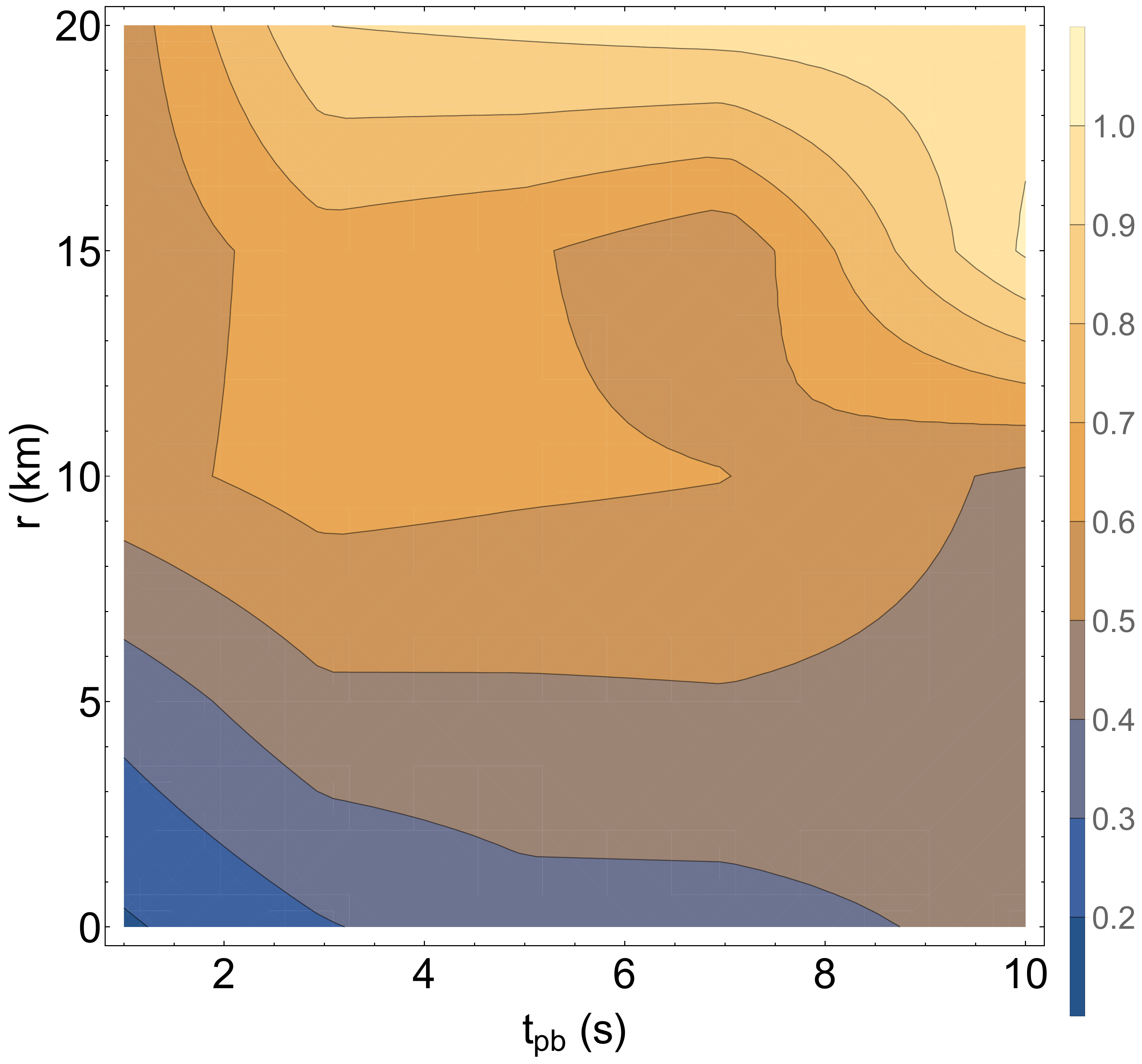}
\caption{Ratio of axion emissivity $Q_a/Q_a^{(0)}$  including the effective nucleon mass $m_N^{\ast}$.
Left panel: Ratio in function of the density $\rho$. Right panel: Plane $t_{\rm pb}$-$r$.}
\label{fig:numass}
\end{figure}

\subsubsection*{Nonzero pion mass in the pion propagator}

In order to investigate the impact of the different corrections on the emissivity we plot the ratio of the modified $Q_a$ with
respect to the naive OPE prescription $Q_a^{(0)}$  of Eq.~(\ref{eq:Qa}), 
where {in the latter case} for definitiveness,  we fix the value of $\overline{\xi}$ to the non-degenerate case {and we assumed
non-interacting nucleons, neglecting the corrections of the nucleon self-energy to the chemical potentials.}
In Fig.~\ref{fig:mpi} we consider the impact of the finite pion mass $m_\pi$ in the propagator. 
From the left panel, one realizes that the pion mass has a sizable effect for $T\lesssim 10$~MeV, when
the pion mass correction becomes larger than the thermal energy of the nucleon. In this case
the maximum reduction of the emissivity with respect to OPE is $\sim 50 \%$. As expected the 
correction  decreases with the temperature.
For typical SN conditions, one expects that the overall reduction of the emissivity 
at $r \lesssim 15$~km would be $\sim 40 \%$ with respect to the naive OPE for post-bounce times $t_{\rm pb} \lesssim 4$~s, becoming even larger than
$\sim 50 \%$ at later times.

\subsubsection*{Two-pions exchange}

In Fig.~\ref{fig:mrho} we show the impact of   $\varrho$-meson exchange in the propagator [see Eq.~(\ref{eq:rhomass})]. 
This effect suppresses the OPE rate in a 
sizable way up to $\gtrsim 80 \%$ for $T \gtrsim 30$~MeV (left panel).
Notice that the $\varrho$-meson exchange produces an opposite trend in the emissivity with respect to  a finite $m_{\pi}$. 
Indeed,  the rate suppression due to $\varrho$-exchange increases with the temperature. Therefore, it is more important in the deepest
regions of the star, i.e. $r\lesssim 10$~km (right panel).
Observe also that this suppression is rather stable in time. 

\subsubsection*{Effective nucleon mass {and chemical potential}}

In Fig.~\ref{fig:numass} we consider the medium correction of the nucleon mass $m_N^{\ast}$ and of {the chemical potential}.
As shown before, the effective nucleon mass decreases as  the density increases. As a consequence
we find a reduction of the emission rate $\gtrsim 60 \%$ for
$\rho \gtrsim 1.5 \times 10^{14} \,\ \textrm{g} \,\ \textrm{cm}^{-3}$ (left panel). This happens
in the inner SN core, i.e. $r\lesssim 10$~km (right panel).

\subsubsection*{Multiple nucleon scatterings}

In Fig.~\ref{fig:multisc} we consider the impact of the multiple nucleon scatterings. Since
the square of the spin-fluctuation rate entering the correction term, is 
quadratic in $\rho$ and linear in $1/T$ 
[cfr.  Eq.~\eqref{eq:spinfluc} and Eq.~\eqref{eq:Loren}]
one sees that the correction is larger at higher 
densities and lower temperatures. In particular, for  $\rho \gtrsim 1.5 \times 10^{14} \,\ \textrm{g} \,\ \textrm{cm}^{-3}$
one may reach a reduction of the OPE emissivity  greater than {$50 \%$}.
This  reduction of the emissivity with respect of OPE including multiple scattering effects
was already pointed out in early papers on the subject, e.g. in~\cite{Raffelt:1996za}.
However, we remark that the impact of the multiple spin fluctuations is reduced once the different corrections beyond OPE 
are included in the emissivity.
Indeed, as shown in Fig.~\ref{fig:s}, the $s(x)$ function (with $x=\omega/T$)
 in Eq.~\eqref{eq:Loren} taken at $t_{\rm pb}=1$~s and $r=10$~km,  decreases with respect to the OPE case once the effect of $\pi$ and $\varrho$
mass exchange are included. 
 As a result, due to the normalization condition of 
 Eq.~(\ref{eq:normal}),  the width of the Lorentzian function in Eq.~\eqref{eq:Loren} is also smaller. 
 To be more quantitative, let's define
the Lorentzian width $g \equiv\Gamma/ (\Gamma_\sigma/2)$ so that $g=1$ for a single non-degenerate nucleon 
species~\cite{Hannestad:1997gc}. 
In our case of mixed proton and neutron medium with intermediate degeneracy, one finds {$g=0.7$}, for OPE, 
{$g=0.55$} when the effects of a finite $\pi$ mass are included, and {$g=0.2$} when also the effects of the $\varrho$ exchange are taken into account.
Because of this fact, comparing the ratio $Q_a/Q_a^{(0)}$  including all the corrections except multiple scatterings (Fig.~\ref{fig:nomultisc}) 
with the one including also the multiple scatterings (Fig.~\ref{fig:all}) one realizes that the differences are minimal.

\subsubsection*{Including all the corrections}

From our analysis we find a sizable reduction of the OPE emissivity when all the OPE corrections are accounted for. 
This is shown in Fig.~\ref{fig:all}, where we see that the naive OPE emissivity may be reduced by more than one order of magnitude when all these corrections are accounted for. 
We also see that this reduction has a very mild dependence on the temperature, so that
it is rather flat in time as the star cools. 
Finally, in Fig.~\ref{fig:Qacomp} we compare
the axion emissivity $Q_a$ (in units $g_{an}^{2} \times 10^{52} \,\  \textrm{erg} \,\ \textrm{cm}^{-3} \textrm{s}^{-1}$) in the plane $t_{\rm pb}$-$r$ in the naive OPE approximation
(left panel) and  including all the corrections    (right panel). We consider the time window $t_{\rm pb} \in [1;2]$~s.
We realize that the bulk of the axion production 
occurs at low-radii $r \lesssim 10$~km, and there the rate is suppressed by more than  one order of magnitude with respect to the naive OPE, once all corrections are taken into consideration.

\begin{figure}[t!]
\vspace{0.cm}
\includegraphics[width=0.5\textwidth]{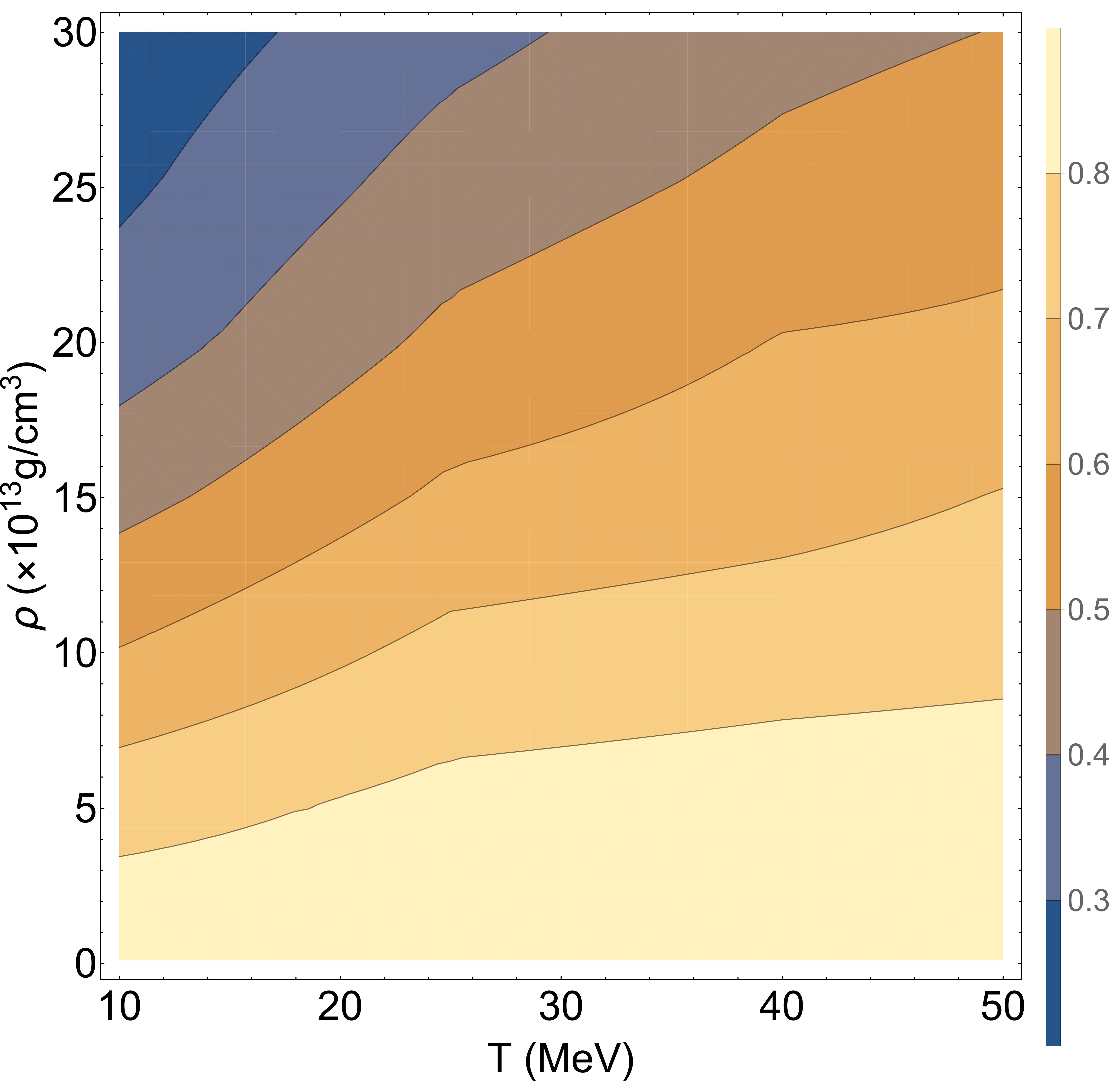}
\includegraphics[width=0.5\textwidth]{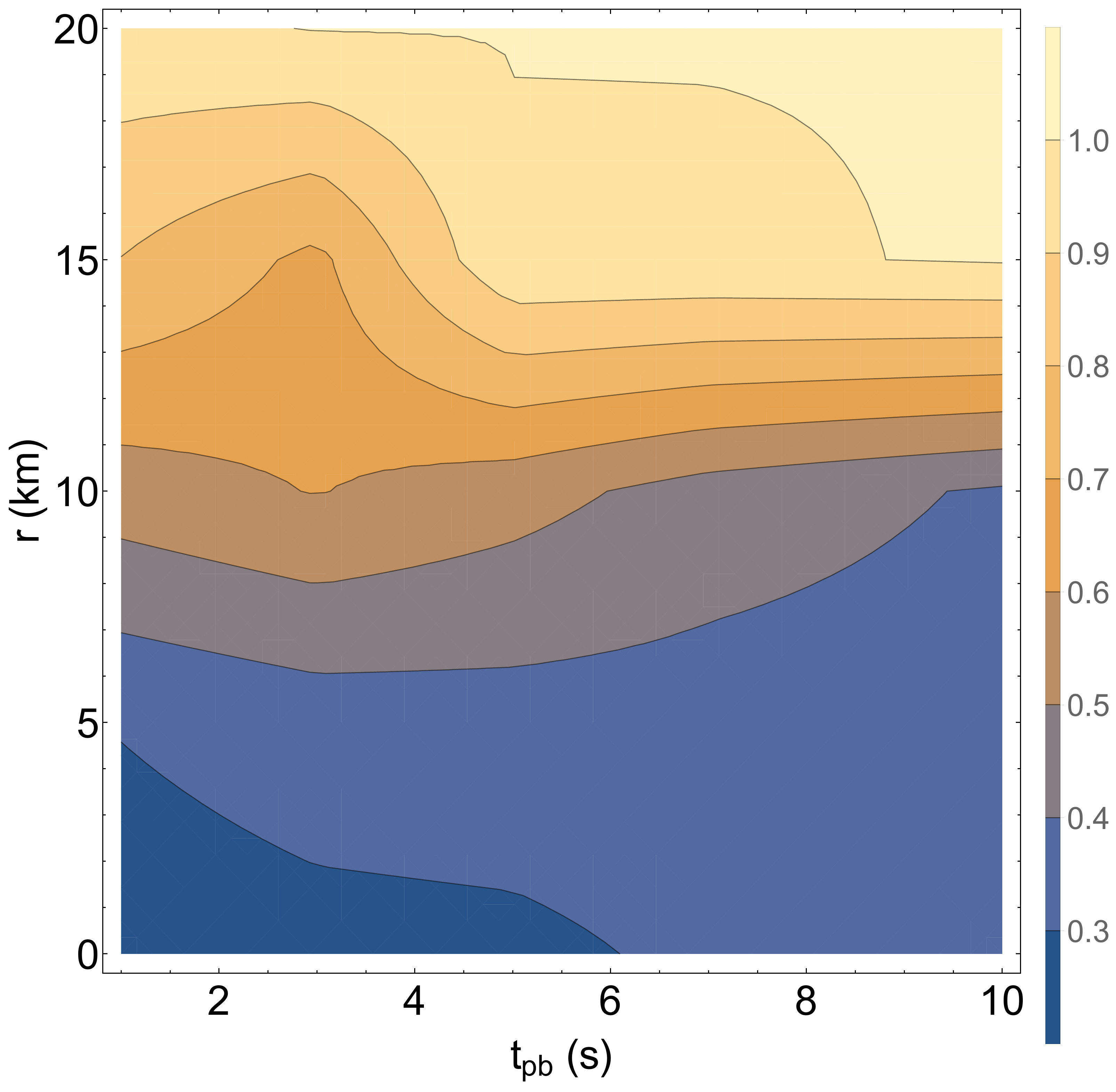}
\caption{Ratio of axion emissivity $Q_a/Q_a^{(0)}$ including multiple nucleon scatterings.
Left panel: Plane $T$-$\rho$. Right panel: Plane $t_{\rm pb}$-$r$.}
\label{fig:multisc}
\end{figure}

\begin{figure}[t!]
	\vspace{0.cm}
	\hspace{1.cm}
\includegraphics[width=0.8\textwidth]{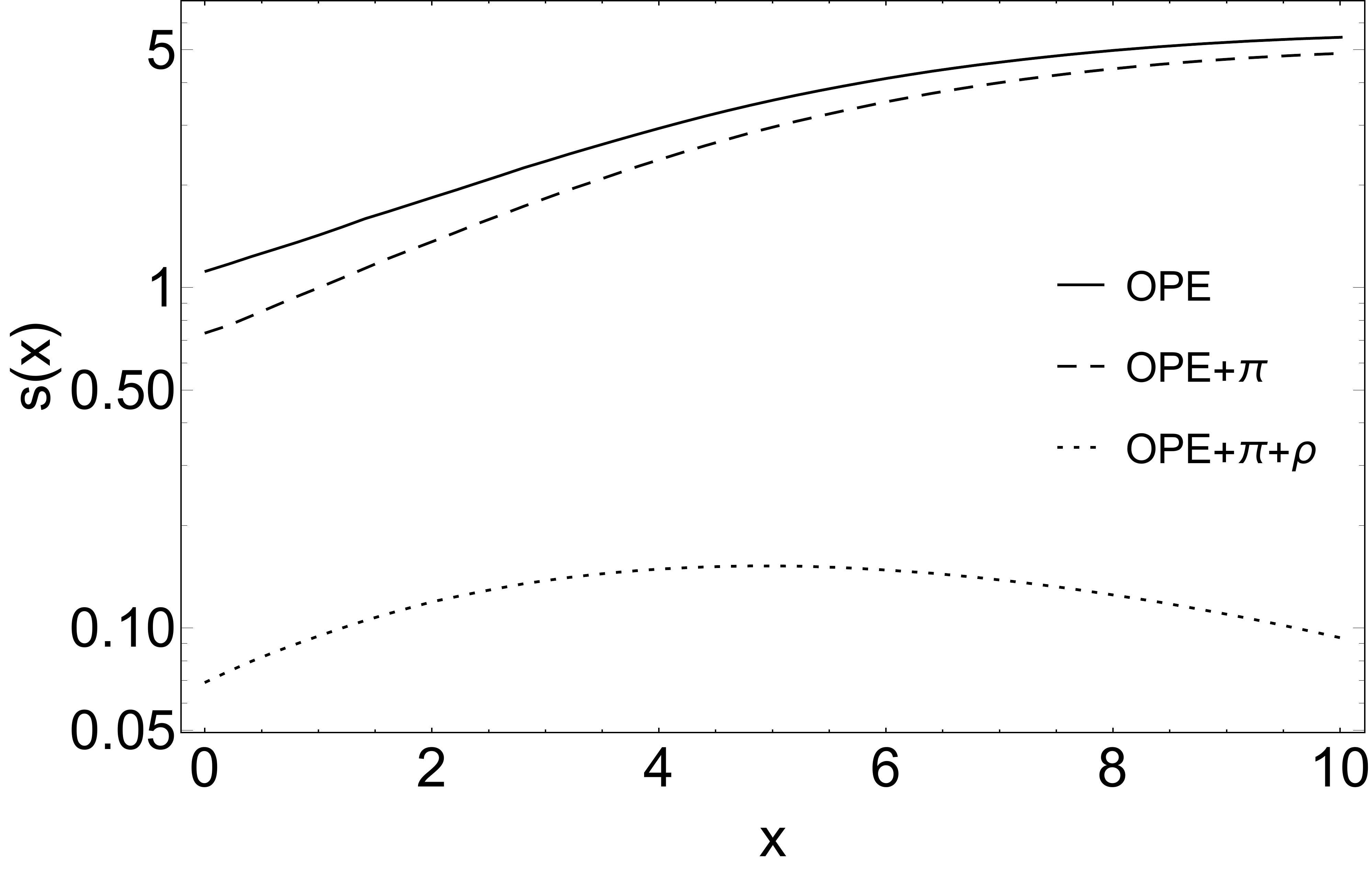}
\caption{Function $s(x)$ with $x=\omega/T$ defined in Eq.~(\ref{eq:sfunction}) for the SN model at $t_{\rm pb}=1$~s and $r=10$~km.{The dotted line refers to the full correct case, including the effective nucleon mass.}}
\label{fig:s}
\end{figure}

\begin{figure}[t!]
\vspace{0.cm}
\includegraphics[width=0.5\textwidth]{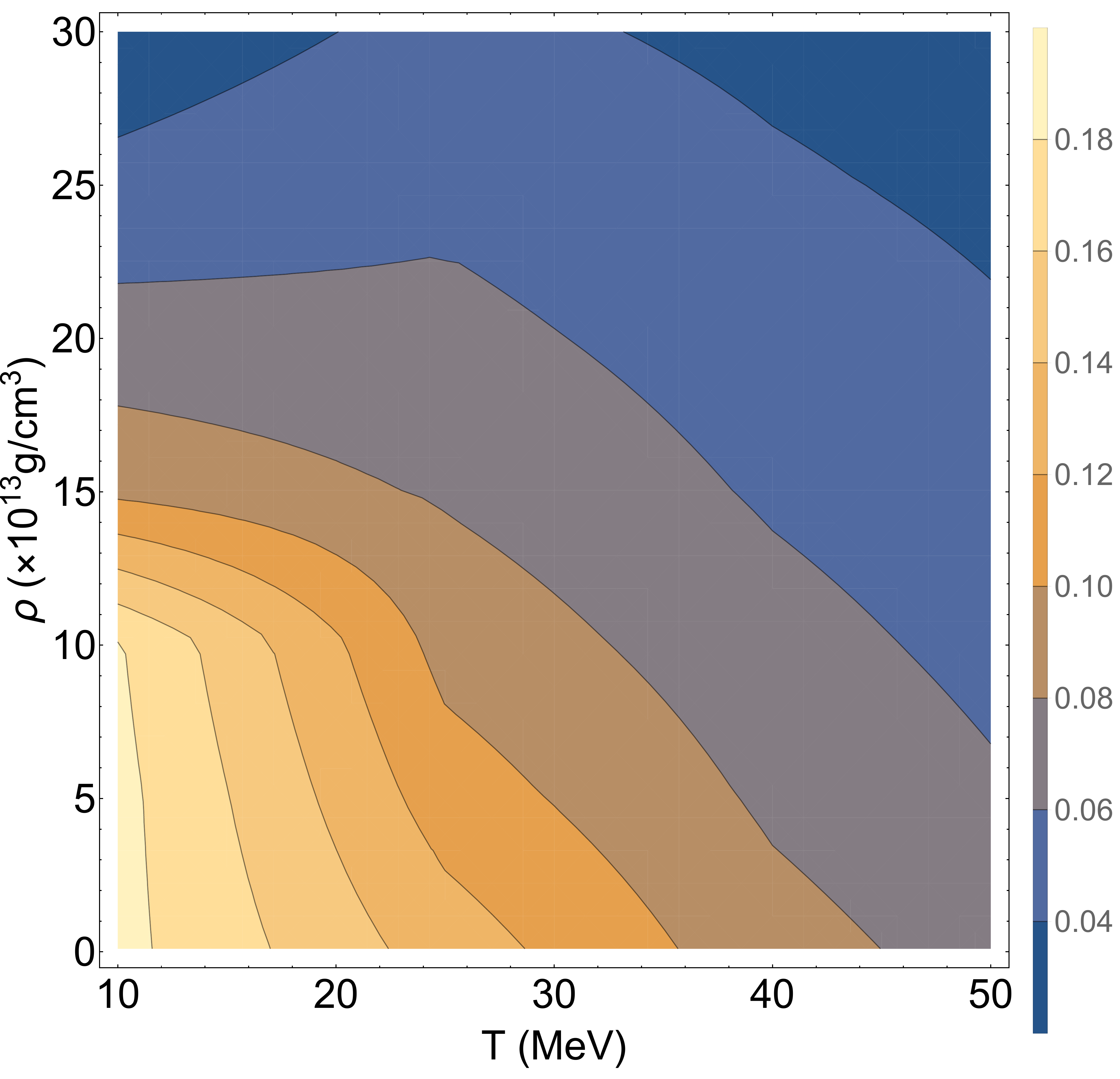}
\includegraphics[width=0.5\textwidth]{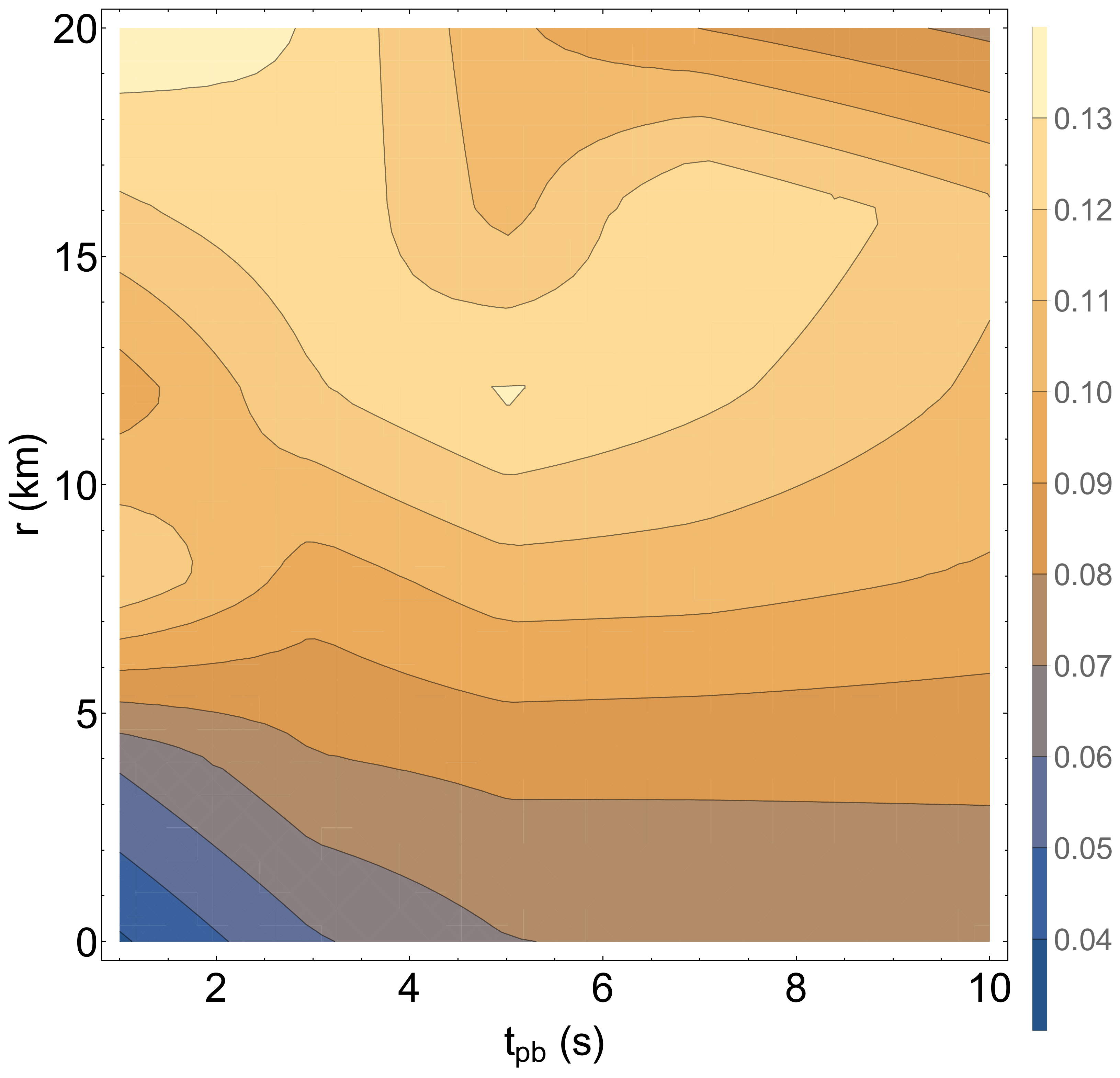}
\caption{Ratio of axion emissivity $Q_a/Q_a^{(0)}$  including all the corrections except multiple nucleon scatterings.
Left panel: Plane $T$-$\rho$. Right panel: Plane $t_{\rm pb}$-$r$.}
\label{fig:nomultisc}
\end{figure}

\begin{figure}[t!]
\vspace{0.cm}
\includegraphics[width=0.5\textwidth]{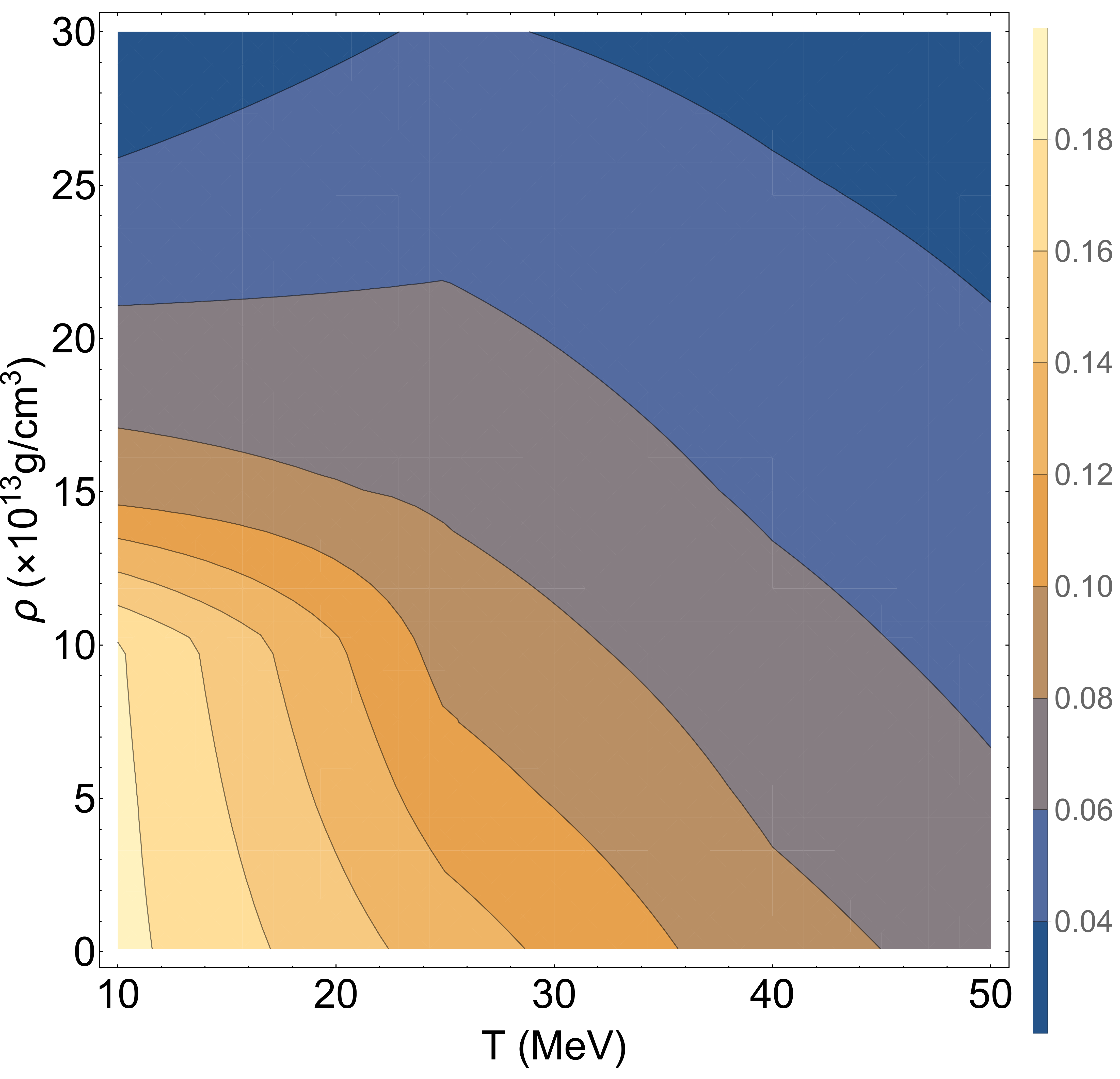}
\includegraphics[width=0.5\textwidth]{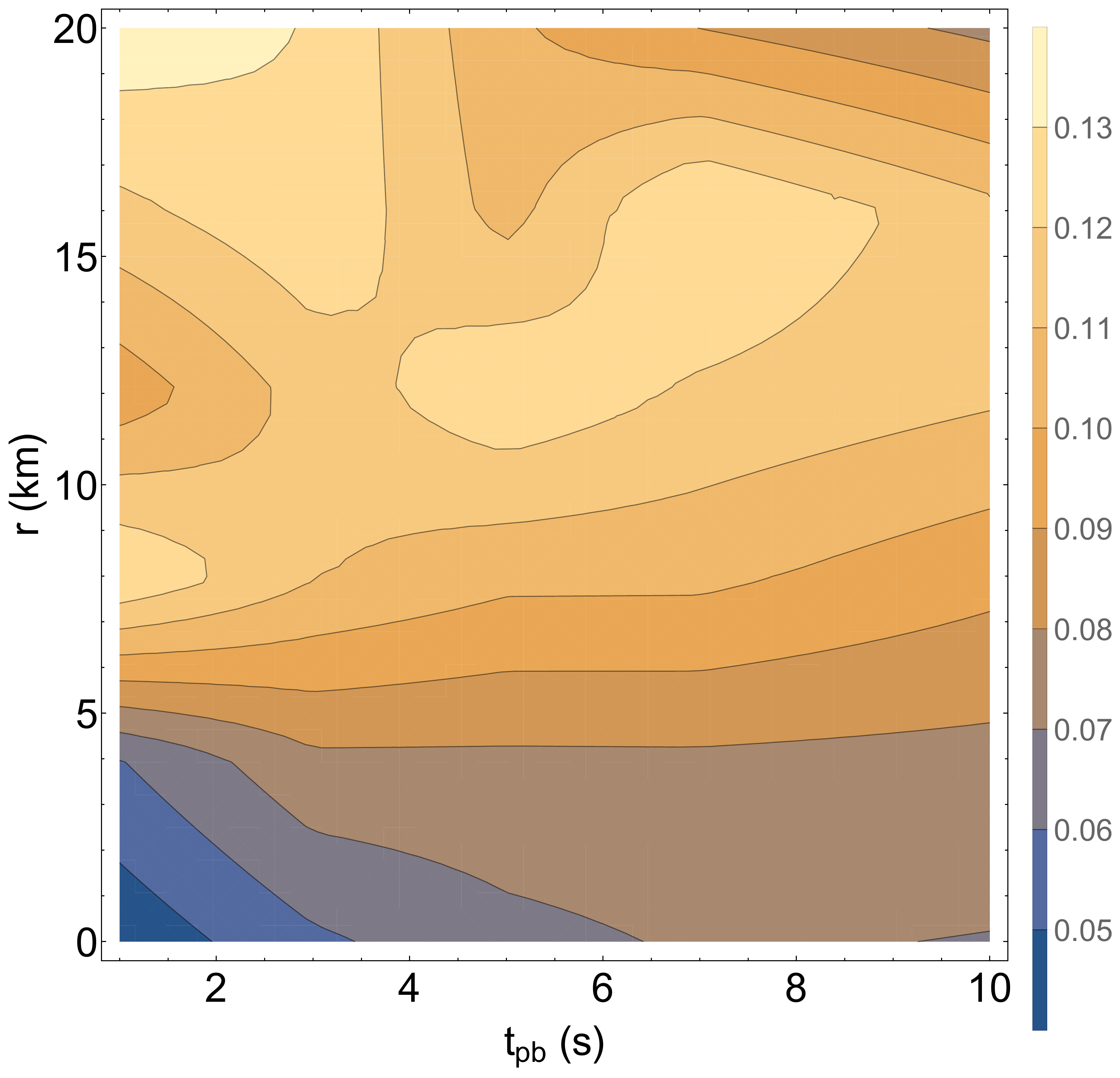}
\caption{Ratio of axion emissivity $Q_a/Q_a^{(0)}$  including all the corrections.
Left panel: Plane $T$-$\rho$. Right panel: Plane $t_{\rm pb}$-$r$.}
\label{fig:all}
\end{figure}

\begin{figure}[t!]
\vspace{0.cm}
\includegraphics[width=0.5\textwidth]{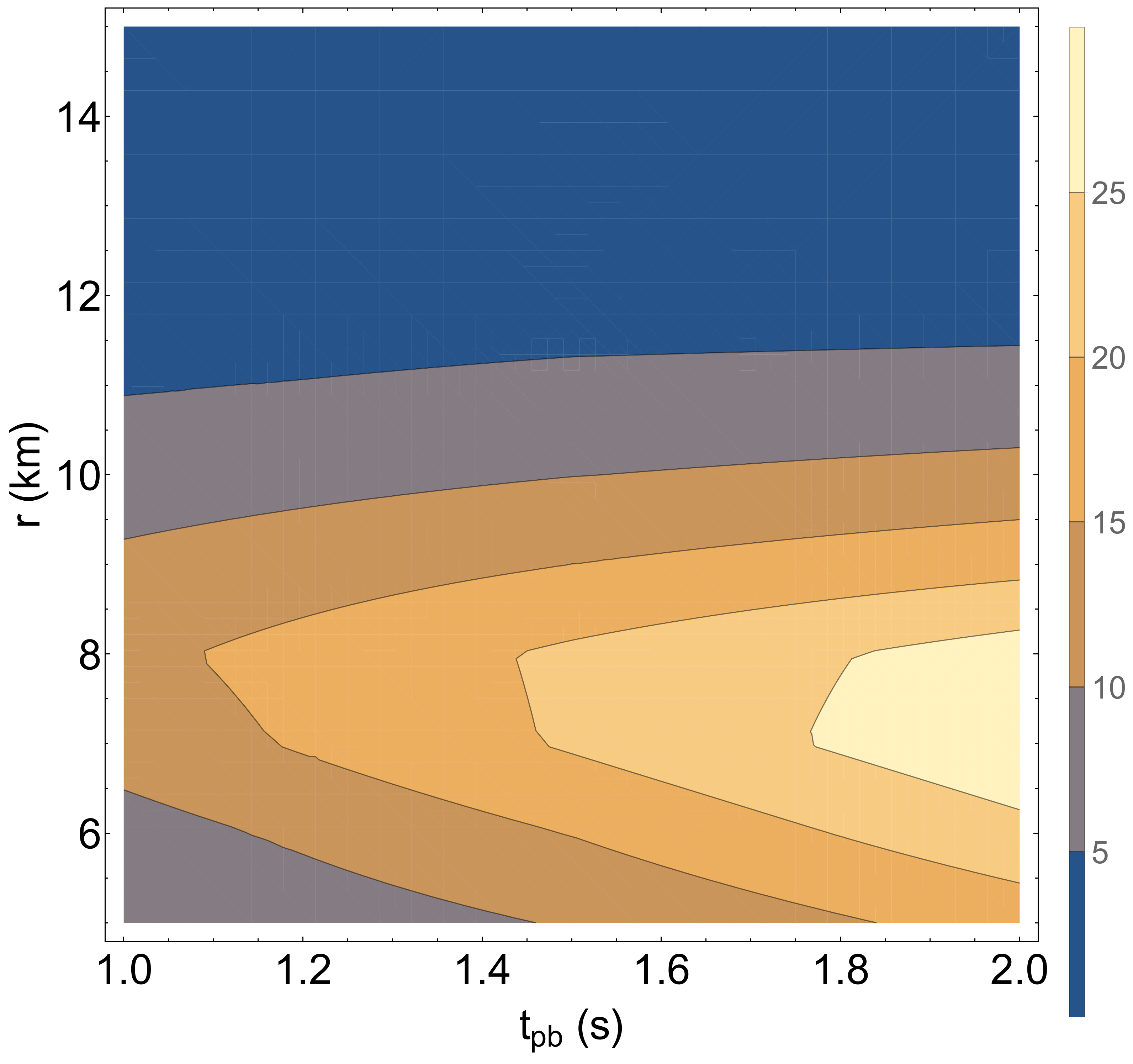}
\includegraphics[width=0.5\textwidth]{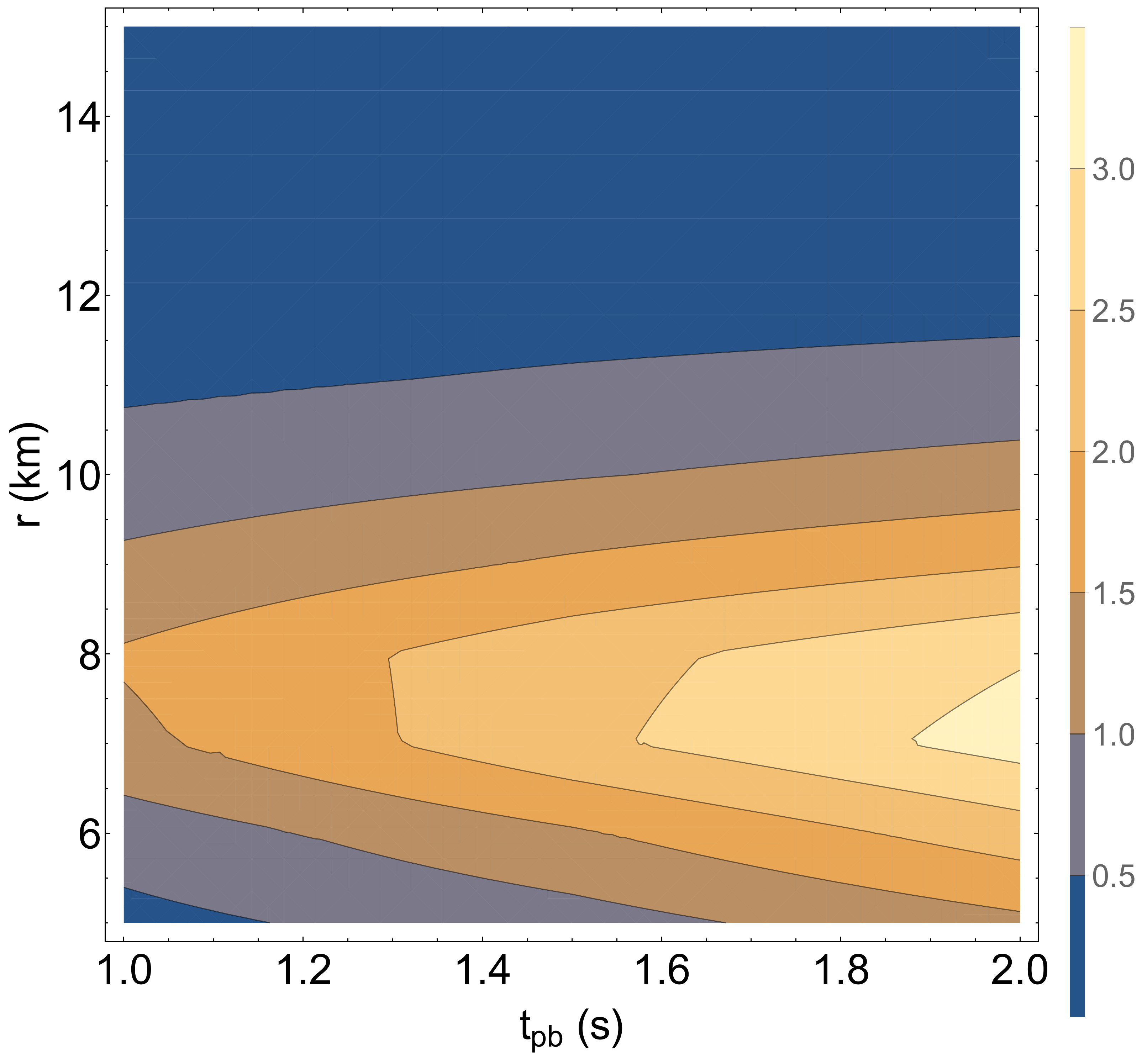}
\caption{Axion emissivity $Q_a$ (in units $g_{an}^{2} \times 10^{52}\,\ \textrm{erg} \,\ \textrm{cm}^{-3} \textrm{s}^{-1}$) in the plane $t_{\rm pb}$-$r$ for OPE
(left panel) and  including all the corrections    (right panel) for $t_{\rm pb} \in [1;2] $~s.}
\label{fig:Qacomp}
\end{figure}

\subsection{Justification of one-pion-exchange plus $\varrho$-meson exchange using  $T$-matrix results}

The rate of NN Bremsstrahlung depends sensitively on the dynamics of the two-nucleon system. An accurate description requires not only a reliable nuclear potential, but also a proper way to treat the non-perturbative property of the potential. Before presenting the bounds on axion for our improved treatments beyond OPE discussed above, we demonstrate here that the use of the OPE potential, corrected with the $\varrho$-meson exchange, gives rise to similar results as those obtained using the $T$-matrix, and is therefore well justified for our study.     

Axion as well as neutrino emissivities from NN Bremsstrahlung of neutron stars have been studied based on the on-shell $T$-matrix extracted from NN scattering data, and a significant difference has been found compared to the studies based on the OPE potential in the Born approximation \cite{Hanhart:2000ae}. Although the on-shell $T$-matrix result is only valid in the soft $\omega$-limit, it already signifies the necessity to use a more realistic nuclear potential to account for the short-range NN interaction and to go beyond the Born approximation. 
To compare with the results from using the OPE potential and the on-shell T-matrix from scattering data, studies based on modern chiral effective potentials have also been performed in the Born approximation for neutrino NN Bremsstrahlung in supernova matter \cite{Bacca:2008yr,Bacca:2011qd,Bartl:2014hoa,Bartl:2016iok,Guo:2019cvs}. The main lessons from these studies can be summarised as follows: 
\begin{itemize}
\item At low density/energy, the OPE potential results in similar rates as the chiral potential in the Born approximation, since the rates are dominated by the long-range part of the nuclear tensor force that can be well described by the OPE potential. 
Due to the non-perturbative and resonant properties of the nuclear force at low energy, the $T$-matrix gives rise to significantly higher rates. 
\item Around and above the nuclear saturation density, nuclear force becomes more perturbative and chiral effective potential in the Born approximation leads to similar rates as the $T$-matrix (see, e.g., Fig. 1 in \cite{Bartl:2014hoa}). The OPE potential, however, typically overestimates the rates by a factor of $\sim$ 5 as the short-range repulsive part of nuclear force becomes relevant at high density but is not included in the OPE potential.   
\end{itemize}
                              
Since axions are mainly emitted from high density regions in SN matter, we expect the OPE potential plus the $\varrho$-meson exchange  treated in the Born approximation can give similar rates as the $T$-matrix. To demonstrate this point, we follow the studies in \cite{Guo:2019cvs} to solve the half-off-shell $T$-matrix using the chiral effective potential of \cite{Entem:2017gor} and then use it to compute the axion emissivity. Unlike using the on-shell $T$-matrix extracted from scattering data, our studies based on the half-off-shell $T$-matrix are generally valid for axions with finite energy, i.e., $\omega > 0$. For illustration, we consider the case that $g_{ap}=g_{an}$, which is different from the neutrino case that the axial weak currents couple to neutron and proton differently with $C_A^n = -C_A^p$. We follow the similar formalisms shown in \cite{Guo:2019cvs} to obtain the related structure function for axion emission. For this specific case with $g_{an}=g_{ap}$, one only needs to change the operator $\bm{\mathcal{Y}}_u$ to $\bm{\mathcal{Y}}_u=\sum_r \bm{\sigma}^{(r)}_u$ in Eq. (B37) and modify the corresponding matrix element in Eq. (B42) in \cite{Guo:2019cvs}. 
   
Fig. \ref{fig:OPE_T} (left panel) compares the axion emissivities for $g_{ap}=g_{an}= 5 \times 10^{-10}$ as function of radius at $t_{\rm pb}=1$ s based on different nuclear potentials: naive OPE, OPE with nonzero pion mass, OPE with nonzero pion mass plus $\varrho$-meson exchange, and the vacuum half-off-shell $T$-matrix. Note that for all the cases studied in this figure, we have included the multiple scattering effects and used nucleon vacuum masses.
The right panel shows the ratio of the axion emissivity for the corrected OPE prescription  (including all corrections) with respect to the T-matrix calcuation.
We find a surprisingly good agreement between the results based on OPE plus $\varrho$-meson exchange and the $T$-matrix, especially for $r \lesssim 10$ km where most axions are emitted, where the modified OPE prescription underestimates by $\sim 20$\% the T-matrix result.  We checked that this agreement holds also for the other axion-nucleon coupling cases
and for later post-bounce times.    

$T$-matrix is affected by the Pauli blocking in nuclear medium, which can lead to different Bremsstrahlung rates as the vacuum $T$-matrix. It has been explored that the impact is typically within 20\% for hot and dense SN matter \cite{Guo:2019cvs}. Besides, three-body force becomes relevant at density above the nuclear saturation density and can affect the rates by (10--20)\% \cite{Bartl:2016oum}. Within an accuracy of (20--30)\%, our use of OPE potential plus $\varrho$-meson exchange correction should be well justified.

\begin{figure}[t!]
	\vspace{0.cm}
\includegraphics[width=0.5\textwidth]{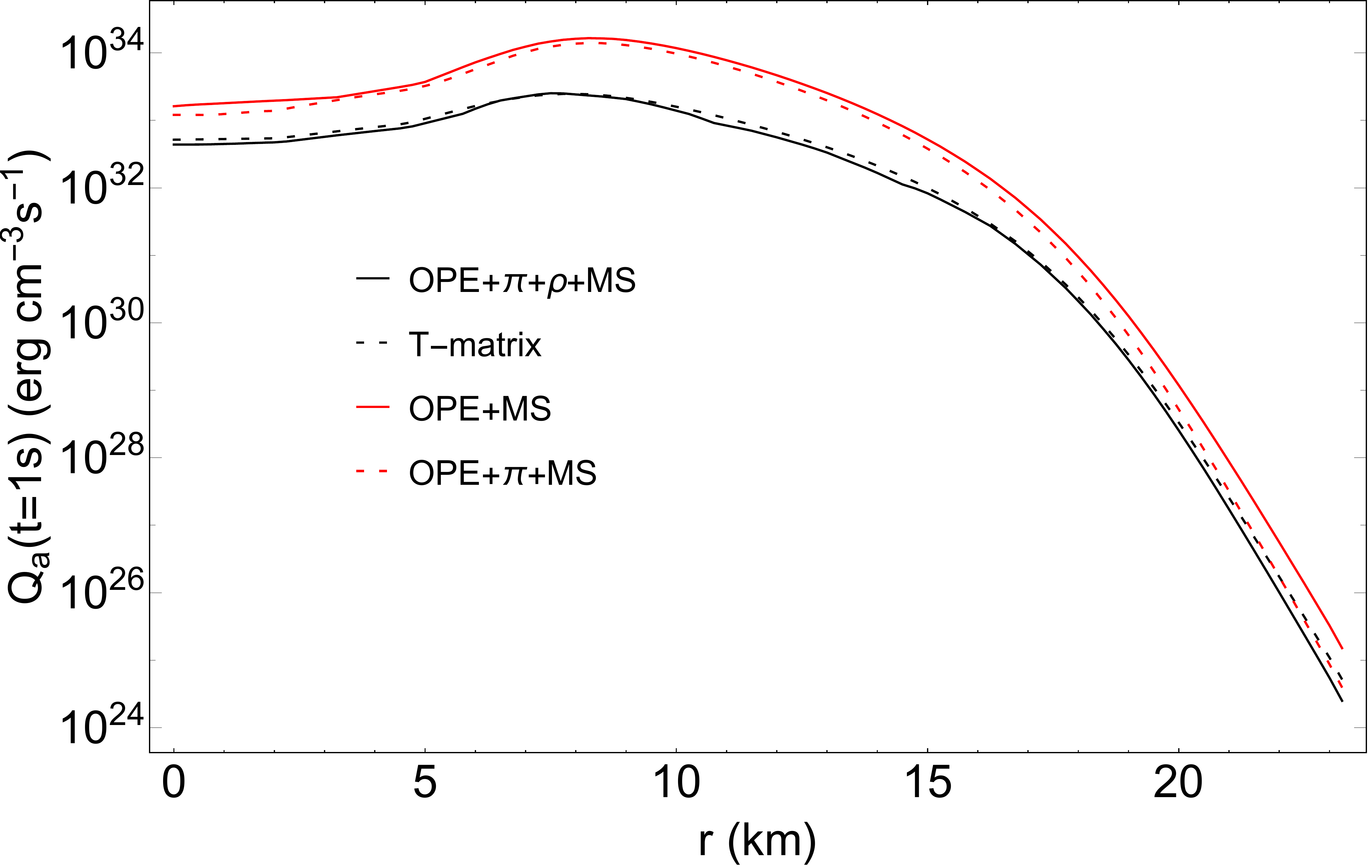}
\includegraphics[width=0.477\textwidth]{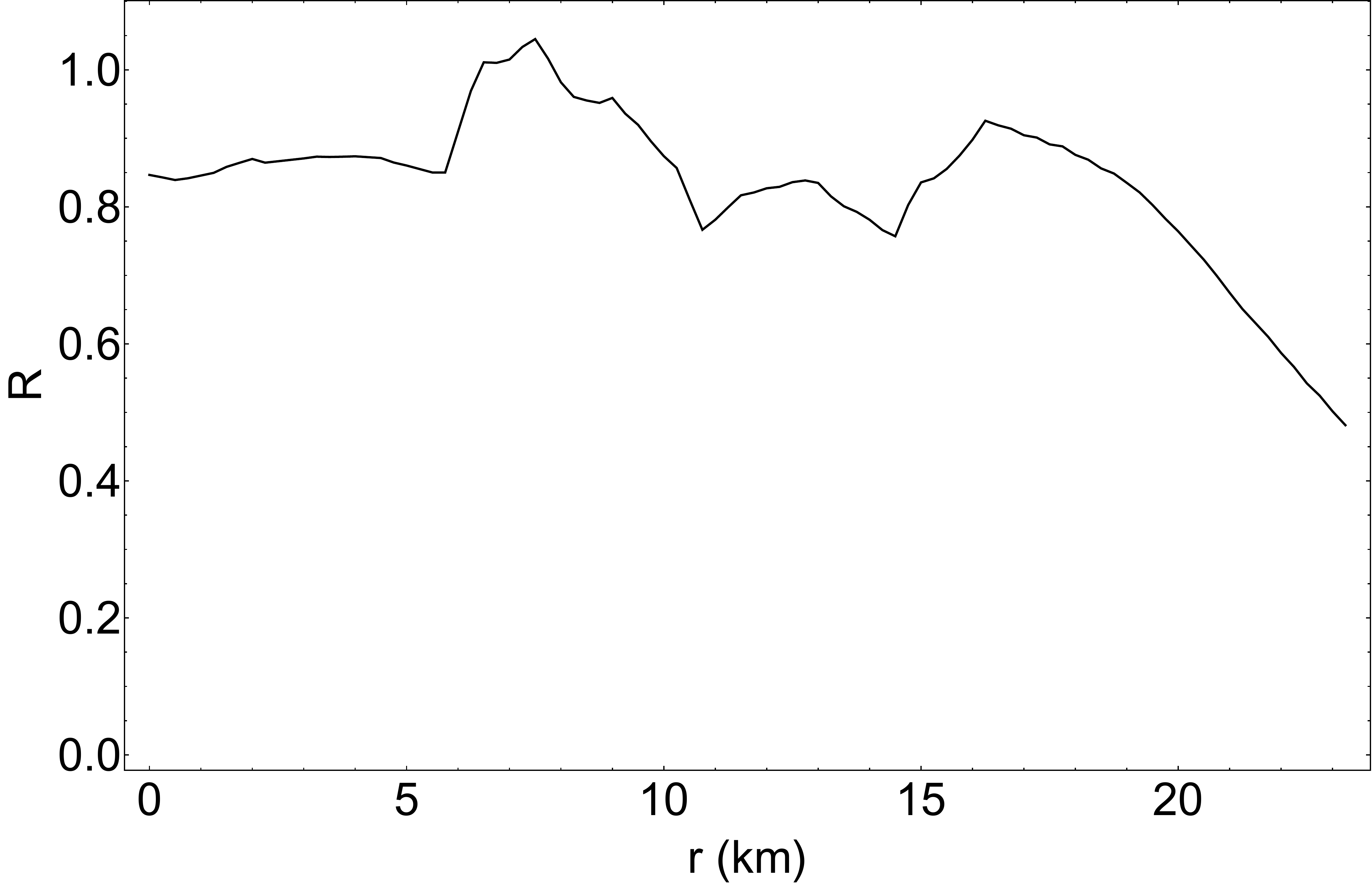}  
\caption{Left panel: Comparison of axion emissivities $Q_a$ for $g_{ap}=g_{an}= 5 \times 10^{-10}$
 as function of radius $r$ at $t_{\rm pb} = 1$ s based on different nuclear potentials {assuming free nucleons (i.e. no correction in  chemical potentials of the free nucleons}. Right panel: Ratio $R$ of
 the axion emissivity for the modified OPE prescription (including all corrections considered in the right panel) with respect to the T-matrix calcuation.}
\label{fig:OPE_T}
\end{figure}

\subsubsection*{Axion luminosity}

Let us finally calculate the axion luminosity, integrating the axion emissivity over the SN model~\cite{Fischer:2016cyd}
\begin{equation}
L_a = \int d r 4 \pi r^2 Q_a \,\ .
\end{equation}
In Figure \ref{fig:LaLnuf}, we compare 
the axion luminosity $L_a$ for    $g_{ap}=g_{an}= 5 \times 10^{-10}$  for $t_{\rm pb} > 1$~s,
in the case of  OPE (black continuous curve), and including 
the effective nucleon mass (black dot-dashed curve), a finite pion mass (black dotted  curve),  the exchange of the $\varrho$ meson (red dotted curve), and multiple nucleon scatterings (red continuous curve), compared with the $\bar\nu_e$ luminosity (black dashed curve).
We see that the inclusion of  all the corrections to OPE causes a reduction of the axion luminosity by roughly an order of magnitude, the major impacts coming from the effective nucleon mass $m_N^\ast$  and the exchange of the $\varrho$ meson. 
On the other hand, the effects of the non-zero pion mass and of multiple nucleon scatterings 
are subleading. 
Moreover, we see that for the chosen value of the axion-nucleon coupling the axion signal is larger than  the neutrino one.
This means that the axion feedback effect on the  neutrino signal cannot be ignored and should be self-consistently included in
SN simulations. We postpone this important task to a future work.

\begin{figure}[t!]
	\vspace{0.cm}
	\hspace{1.5cm}
\includegraphics[width=0.95\textwidth]{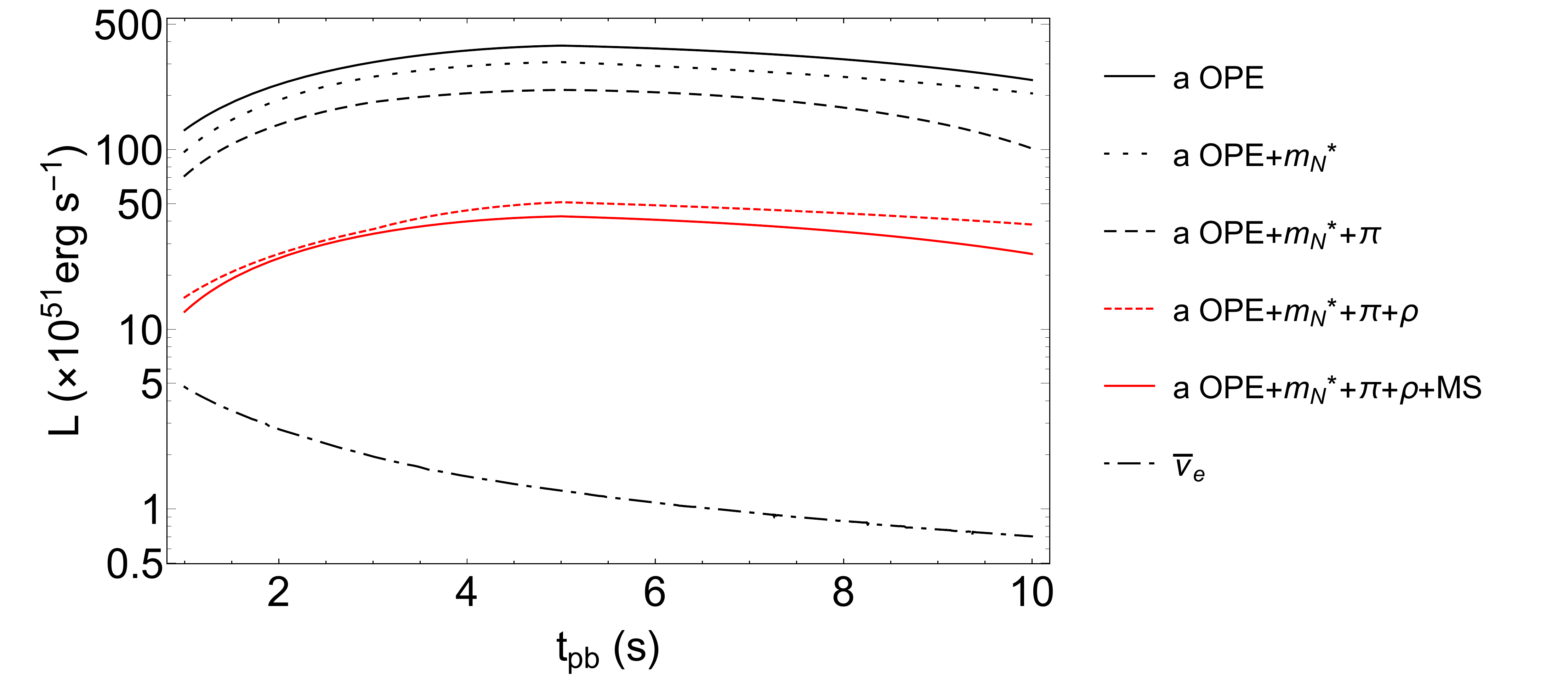}
\caption{Time evolution of the axion luminosity $L_a$ for $g_{an}=g_{ap}= 5 \times 10^{-10}$ for OPE (black continuous curve), and including 
		the effective nucleon mass (black dot-dashed curve), a finite pion mass (black dotted  curve),  the exchange of the $\varrho$ meson (red dotted curve), and multiple nucleon scatterings (red continuous curve), compared with the $\bar\nu_e$ luminosity (black dashed curve).}
\label{fig:LaLnuf}
\end{figure}

In Fig.~\ref{fig:LaLnu} we show 
 the ratio of the axion luminosity $L_a$ with respect to the total neutrino luminosity
  $L_{\nu}$ at $t_{\rm pb}=1$~s
as a function of $ g_{an}=g_{ap}$ for the complete axion emissivity (continuous curve) and for the naive OPE (dashed curve). 
The limit $L_a/L_\nu=1$ is shown as a horizontal dotted line.
We consider a relative early post-bounce time, since the feedback on the neutrino signal 
would be less important than for later times. 
Cases with $g_{ap}= g_{an} \lesssim 10^{-7}$ correspond to the free-streaming case. 
Note that in the free-streaming case the luminosity increases at larger couplings, since it scales as $g_{a}^2$.
As suggested in~\cite{Raffelt:1987yt}, axion-nucleon couplings that imply $L_a/L_{\nu} \gtrsim 1$ are excluded due to 
an excessive energy loss. For our SN model, in the case of the complete emission rate
the condition of equality between axion and neutrino luminosity is reached for { $g_{ap}=g_{an} \approx  6 \times 10^{-10}$
(to be compared with $g_{ap}=g_{an} \approx  2 \times 10^{-10}$ for OPE). } 
So, we expect that couplings larger than this value would cause an excessive energy loss and should be excluded. 
The limit on the axion-nucleon coupling (and on the axion mass) will be discussed the next Section.


\begin{figure}[t!]
	\vspace{0.cm}
	\hspace{1.5cm}
	\includegraphics[width=0.8\textwidth]{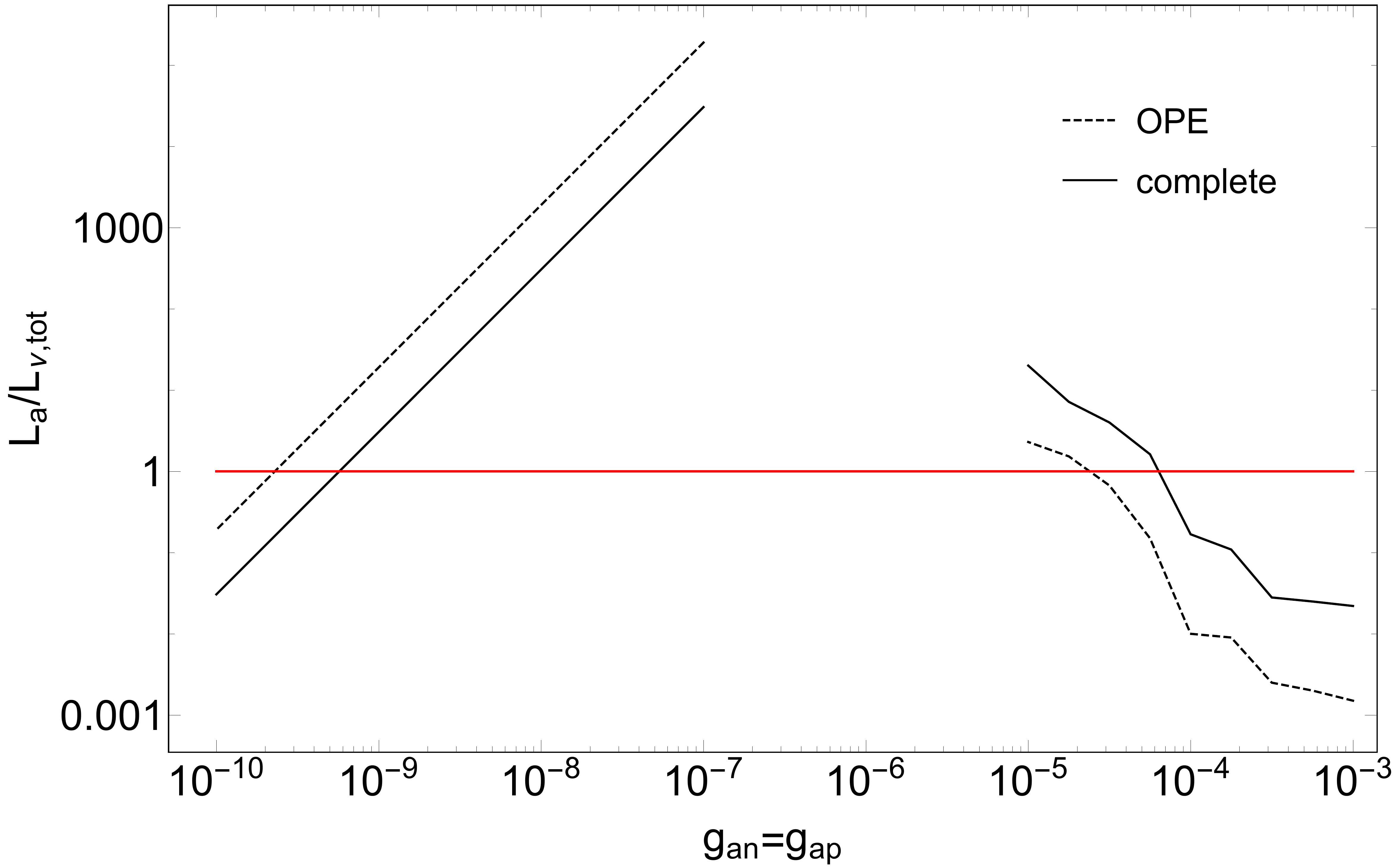}
	\caption{
				Ratio of the axion luminosity $L_a$ with respect to the total neutrino luminosity $L_{\nu}$ at $t_{\rm pb}=1$~s
		for the case of $g_{aN}\equiv g_{an}=g_{ap}$. Cases with $g_{aN} \gtrsim 10^{-6}$ corresponds to trapped axions, while
		for $g_{aN} \lesssim 10^{-7}$ one has free-streaming axions. 
		Note that in the intermediate range of couplings one finds a mixed regime, where axions are neither free-streaming out of the star, nor properly trapped. 
This regime is quite more challenging to evaluate numerically.
For this reason, in the plot  we have left this coupling range empty.
		Couplings with $L_a/L_{\nu} \gtrsim 1$ (horizontal
		dotted line) are excluded due to 
		an excessive energy loss.}
	\label{fig:LaLnu}
\end{figure}

\section{Axion mass bound and comparison with previous works}
\label{sec:Axion mass bound and comparison with previous works}
In order to assess the impact of our calculation on the axion mass bound based on emissivity from SN, here we compare our result with the previous 
literature. 
First of all, however, we should notice that previous works used different approximations and so there is a large range of variability in the quoted limits. 
For the sake of definitiveness we take as benchmark for 
our comparison, the result obtained in~\cite{Raffelt:2006cw} and quoted in the  latest edition of the Particle Data Group (PDG)~\cite{Tanabashi:2018oca}.
This bound was based on 
a simple criterium to estimate the impact of the axion emissivity from supernovae  proposed in~\cite{Raffelt:1990yz}.
Namely, 
the observation of neutrinos from the SN 1987A implies that the energy-loss rate $\varepsilon_a = Q_a/\rho$,  associated to exotic particles such as axions,
evaluated at $\rho = 3 \times 10^{14}$~g~cm$^{-3}$ and $T=30$~MeV, should satisfy 
\begin{equation}
\varepsilon_a \lesssim 1 \times 10^{19} \textrm{erg} \,\ \textrm{g}^{-1} \textrm{s}^{-1} \,\ .
\end{equation}
Assuming a SN with a core mass of $1 M_{\odot}= 2 \times 10^{33}$ g,
this gives an axion luminosity $L_a = \varepsilon_a M_{\odot} = 2 \times 10^{52}$~erg~s$^{-1}$. 
Assuming the non-degenerate limit of the axion emissivity amd setting setting $\xi=0$ in the squared-matrix element in Eq.~(\ref{eq:amplitude}),
and including only multi-scattering effects as a correction of the naive OPE, in~\cite{Raffelt:2006cw}
it was derived the mass bound $m_a \lesssim 16$~meV (corresponding to $f_a \gtrsim 4 \times 10^8$~GeV)
for the case of an hadronic axion model with $C_{an}=0$, $C_{ap}=-0.4$ and a proton fraction 
$Y_p=0.3$.
In order to compare with this result, in Table \ref{tab:massschmatic}
we present our bound on axion-nucleon coupling and on the mass for the schematic SN model described before,
assuming non-degenerate  emissivity and  for KVSZ hadronic axion model 
with $C_{ap}=-0.47$ [cfr. Eq.~(\ref{eq:axion_couplings_KSVZ})].
Note that for the case of OPE and multiple nucleon scatterings (OPE+MS) we find 
{ $m_a \lesssim 8$~meV }, which is  
{a factor two smaller than the}  value quoted in~\cite{Raffelt:2006cw}. 
{We realize that this difference should be attributed to different points. In particular, in~\cite{Raffelt:2006cw}
it is assumed that the medium is composed by a single nucleon species and that the total emissivity is simply
rescaled by $Y_p n_B$. However, from Eq.~(\ref{eq:sfunction}) it results that, assuming only the coupling with protons, the emissivity rate should scale as $Y_p^2$
and $Y_p Y_n$. In particular, in~\cite{Raffelt:2006cw} it was neglected the contribution to the emissivity due to the  $Y_p Y_n$ tem in Eq.~(\ref{eq:sfunction}), corresponding
to the $I(y_n,y_p)$ term in Eq.~(\ref{eq:Qa}). We checked that this latter term would dominate by a factor $\sim 4$ the axion emissivity, leading to mass bound stronger
by a factor $\sim2$.
Furthermore, we also find as width of the multiple-scattering Lorentzian function
[Eq.~(\ref{eq:Loren})] $g=0.7$ instead of $g=1$ found  in~\cite{Raffelt:2006cw}. As result the impact of this effect beyond OPE is reduced with respect 
to the previous calculation.}
We note that the ratio of the square of the coupling constant $g_{ap}^2$ in the case of OPE with respect to OPE+MS
 is {$\sim 0.45$}, which is consistent with the reduction of the axion emissivity discussed in the previous section.
 If we include the other corrections to the emissivity, the axion bound relaxes by a factor of {$ \sim 3 $ }with respect to OPE.
 The additional inclusion of the multiple nucleon scatterings changes this results by only {$\sim 10\%$}.

 In order to study the impact of the SN model on the axion mass bound, we consider the KSVZ  hadronic axion  model 
 and, rather than referring to the simplified SN model discussed above, we calculate the axion luminosity from our realistic SN reference model at
 $t_{\rm pb}=1$~s. We show these results in Table~\ref{tab:masssnreal}.  The differences with respect to the schematic SN model are less than a factor {$1.4$}. 
 This result is reassuring since it implies that the derivation of the axion mass
 bound is not strongly affected by the details of the SN model.
In this case, including all the corrections beyond OPE the relaxation of the axion mass bound is a factor { $\sim 3$}, leading to
{ $m_a \lesssim 15$~meV, comparable with the one obtained in the schematic calculation of~\cite{Raffelt:2006cw}}.


 \begin{table}[!t]
 \caption{Bounds on axion couplings and mass for KVSZ model in a schematic SN model with $\rho = 3 \times 10^{14}$~g~cm$^{-3}$ and $T=30$~MeV.}
\begin{center}
\begin{tabular}{llll}
\hline
$C_{ap}=-0.47\; ; C_{an}=0$ & 
$g_{ap}$ ($\times10^{-10}$)
& $m_{a}$ (meV)&$f_{a} (\times10^{8}$~GeV)\\
\hline
\hline
OPE&4&5&10.4 \\
OPE+MS&6&8&7.2\\
OPE+corr. (no MS) &8&10&5.5\\
OPE+corr.+MS& 9&11&5.1\\
\hline
\end{tabular}
\label{tab:massschmatic}
\end{center}
\end{table}


{
 \begin{table}[!t]
 \caption{Bounds on axion couplings and mass for KVSZ model in our SN model at  $t_{\rm pb}=1$~s.}
\begin{center}
\begin{tabular}{llll}
\hline
$C_{ap}=-0.47\; ; C_{an}=0$ & 
$g_{ap}$ ($\times10^{-10}$)
& $m_{a}$ (meV)&$f_{a} (\times10^{8}$~GeV)\\
\hline
\hline
OPE&4&5&10.4 \\
OPE+MS&5&6&9.7  \\
OPE+corr. (no MS) &11&14&4.2\\
OPE+corr.+MS&12&15&4.0 \\
\hline
\end{tabular}
\label{tab:masssnreal}
\end{center}
\end{table}
}

 We remark that  a  similar trend of  relaxation of the bound on  the axion-nucleon coupling  in the free-streaming regime has been recently presented in~\cite{Chang:2018rso}.  
	In their calculation, the authors of~\cite{Chang:2018rso} discuss some of the effects we also include, such as  the finite pion mass and 
the multiple nucleon scatterings. Moreover, they
attempted to go beyond OPE including also deviations coming from chiral perturbation theory~\cite{Hanhart:2000ae,Bartl:2014hoa,Bartl:2016iok}.
When we compare their \emph{single} correction beyond OPE with ours, we find a quantitative agreement. In particular, their correction coming 
from chiral perturbation theory (called $\gamma_h$ in their paper) has a similar behavior and strength as ours coming from the $\varrho$ exchange. 
 However,
their implementation of the \emph{overall} correction responsible for the emissivity suppression is  more schematic than ours, since
their different corrections are simply taken as  fudge factors on top of the naive {non-degenerate} OPE.
{However}, as we noticed above, the amplitude of the multiple scatterings effects 
(called $\gamma_f$ in their work) gets significantly reduced once other corrections are included, since they 
diminish the width of the Lorentzian function, Eq.~(\ref{eq:Loren}), fixed by the normalization
condition in Eq.~(\ref{eq:normal}).  {Therefore, their treatment leads to a suppression of the axion emissivity.}
{Indeed, their final result that may reach $m_a \lesssim 50$~meV for the KSVZ model 
is more than a factor 3 weaker than 
our finding.}

 In order to extend the previous analysis to generic axion models, we need to extract the axion luminosity dependence on the separate nucleon couplings, $ g_{an} $ and $ g_{ap} $.
 Considering our reference SN model at $t_{\rm pb}=1$~s we get
\begin{equation}
{
L_a \simeq 2.42\times 10^{70} \,\ \textrm{erg} \,\  \textrm{s}^{-1} (g_{an}^2+ 0.61 g_{ap}^2 + 0.53 g_{an}g_{ap}) \,\ .
}
\end{equation}
Notice that a similar dependence on the coupling is found also using the naive OPE emissivity. 
 Imposing that this luminosity does not exceed the SN 1987A bound,
 the allowed parameter space in the plane $g_{an}$-$g_{ap}$ is shown in 
 Fig.~\ref{fig:gangap}. 
 Imposing the condition $L_{a} \lesssim L_\nu\simeq 2\times 10^{52}\textrm{erg} \,\  \textrm{s}^{-1} $ one finds the constraint on the axion-nucleon couplings
 \begin{equation} 
 {
g_{an}^2+ 0.61 g_{ap}^2 + 0.53 g_{an}g_{ap} \lesssim 8.26 \,\times10^{-19}\,;
}
 \end{equation}
 or, equivalently, the bound on the axion mass
\begin{equation} 
{
{m_a} \lesssim 5.67 \,\ \textrm{meV} \times  (C_{an}^2+ 0.61 C_{ap}^2 + 0.53 C_{an}C_{ap})^{-1/2} \,\ .
}
\label{eq:massbound}
 \end{equation}
 
\begin{figure}[t!]
	\vspace{0.3cm}
	\hspace{2.2cm}
	\includegraphics[width=0.6\textwidth]{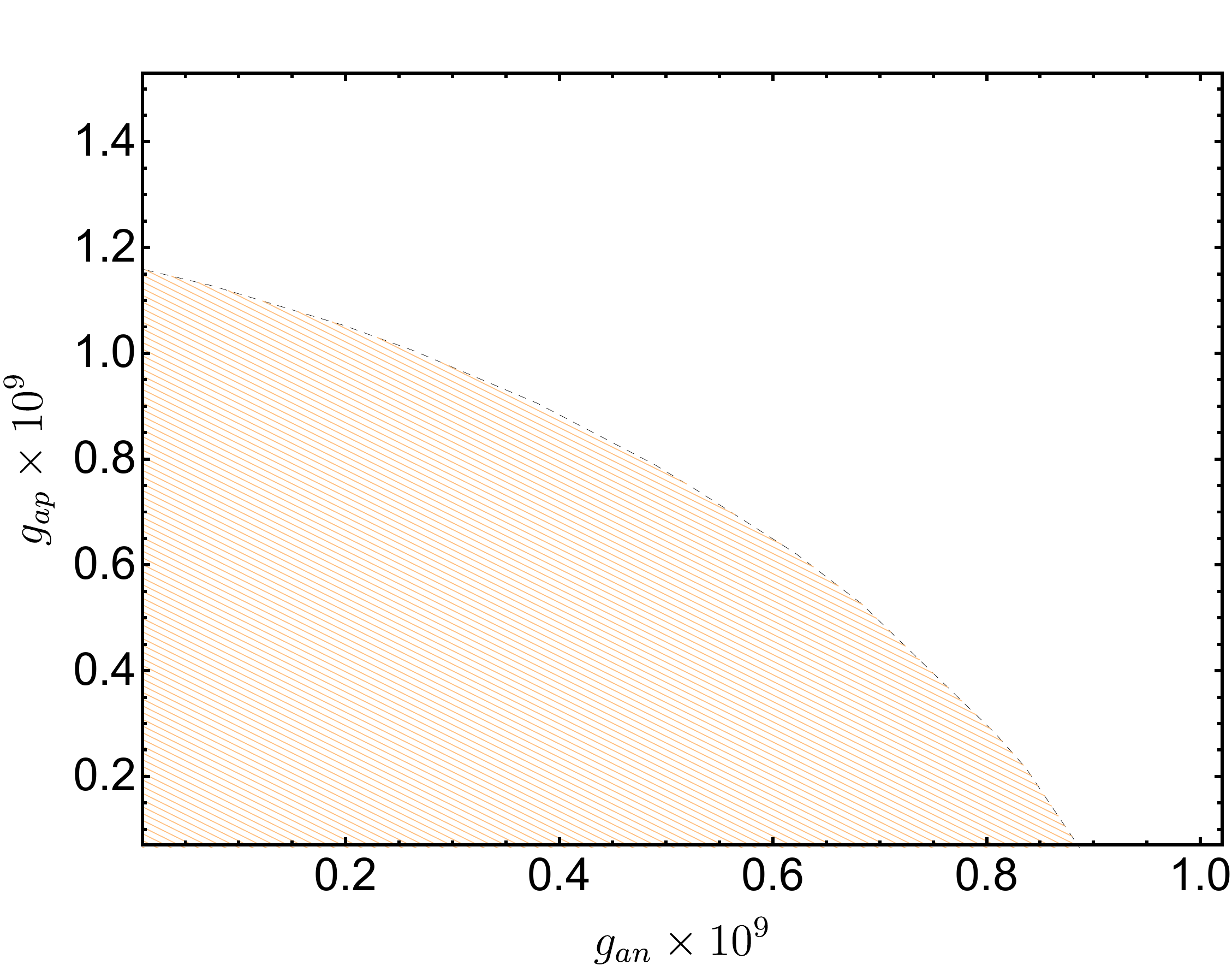}
	\caption{Allowed parameter space in the plane  $g_{an}$-$g_{ap}$ according 
	to $L_a \lesssim 2 \times 10^{52}$~erg~s$^{-1}$.}
	\label{fig:gangap}
\end{figure}

\begin{figure}[t!]
	\vspace{0.cm}
	\hspace{2.2cm}
	\includegraphics[width=0.6\textwidth]{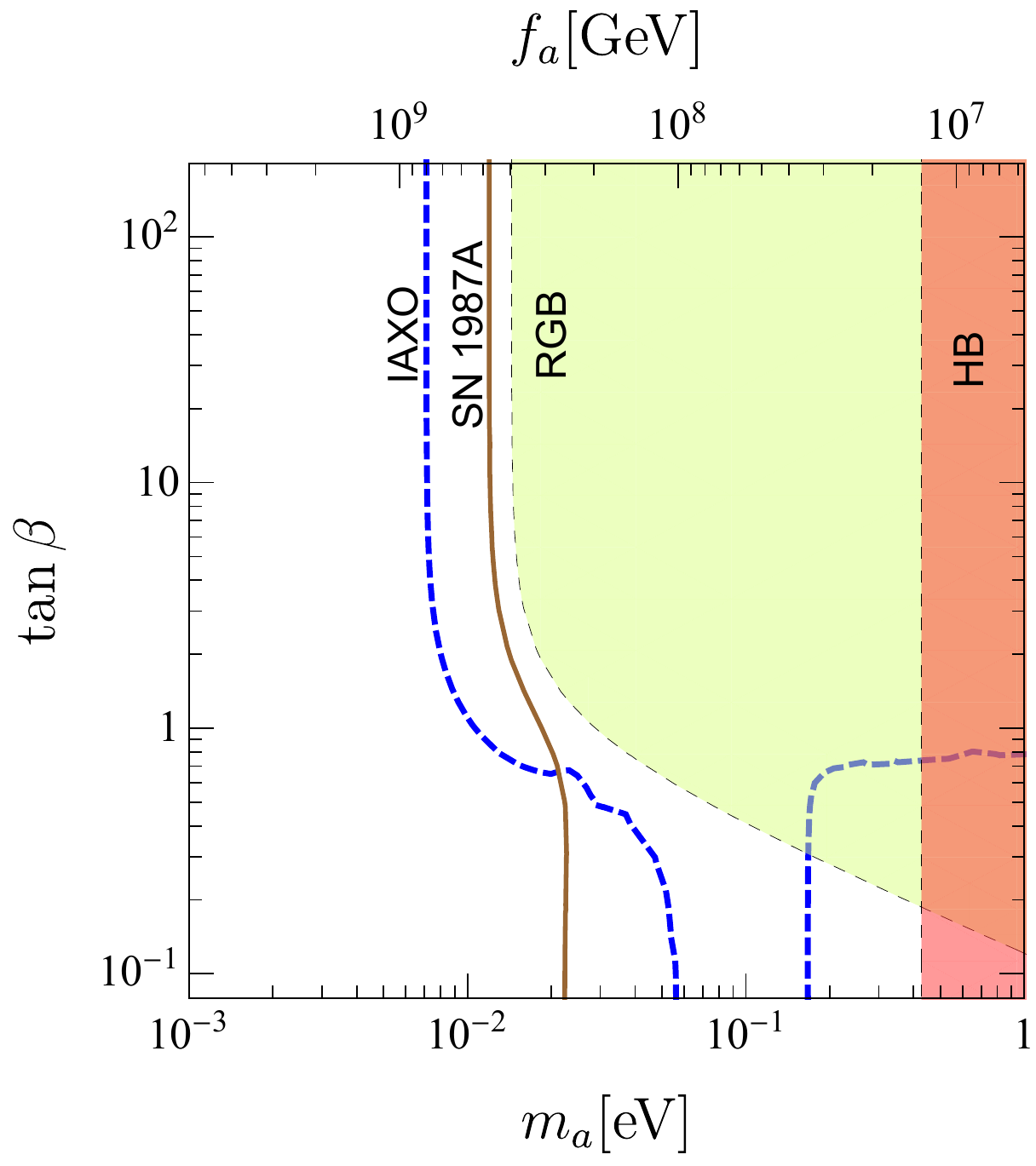}
	\caption{
The brown line is the SN 1987A bound for DFSZ axions, according to Eq.~(\ref{eq:massbound}). 
The RGB and the HB limits are also included. 
The blue dashed line is the IAXO potential~\cite{Armengaud:2019uso}.
}
	\label{fig:DFSZ}
\end{figure}
 
{In particular, for the particularly interesting case  of  DFSZ axions     our analysis
 indicates $m_a \lesssim 15-20$~meV, the exact value depending on $ \tan \beta $.
This bound is shown in Fig.~\ref{fig:DFSZ}.
As evident from the figure, for most parameters the SN bound is slightly more  stringent than the RGB constraint and it dominates  on it in  regions of small $ \tan\beta $.
It is also evident that the SN and RGB bounds leave a part of the axion parameter space accessible to the next generation of axion experiments such as 
IAXO.
~\cite{Giannotti:2017hny,Armengaud:2019uso}.}

We finally mention that a stronger bound on the axion-neutron coupling, namely
$g_{an} < 2.8 \times 10^{-10}$,  has been placed in~\cite{Beznogov:2018fda} from the
thermal evolution of the hot young neutron star in the supernova remnant HESS J1731-347.
However, just like the SN bound, also this one is affected by uncertainties associated 
 with the phase structure and linear response properties of matter at supra-nuclear density.
 Therefore, it is always important to use more than one argument to constrain axion-nucleon couplings.

\section{Axion opacity}

In the case of axions with mass exceeding $10^{-1}$~eV or so, the axion mean free-path for absorption (at the temperatures
and densities of a proto-neutron star) becomes less than the radius of a proto-neutron star. Therefore, such axions would not be free-streaming but 
``trapped'' and emitted from an axion-sphere in analogy with neutrinos~\cite{Burrows:1990pk}. 
We closely follow the treatment of the axion trapping given in~\cite{Raffelt:1996wa}.
To calculate the axion luminosity, one has to compute the
reduced Rosseland mean opacity $\kappa_a$, defined as (see, e.g., Sec. 4.4 of~\cite{Raffelt:1996wa})
\begin{equation}
\frac{1}{\rho \kappa_a}= \frac{15}{4 \pi^2 T^3}\int_{0}^\infty d \omega \lambda_{\omega} (1-e^{-\omega/T})^{-1}\partial_T B_{\omega}(T) \,\ ,
\end{equation}
where $\rho$ is the mass density of the medium, $\lambda_{\omega}$ is the axion mean free-path against absorption and
\begin{equation}
B_{\omega} (T) = \frac{1}{2 \pi^2}\frac{\omega^3}{e^{-\omega/T}-1} \,\ ,
\end{equation}
is the axion spectral density for one degree of freedom. Applying the operator $\partial_T= \partial/\partial T$ one gets
\begin{equation}
\frac{1}{\rho \kappa_a}= \frac{15}{8 \pi^4 }\int_{0}^\infty d x \lambda_{x} 
\frac{x^4 e^{2x}}{(e^{x}-1)^3}
 \,\ ,
\end{equation}
where $x\equiv \omega/T$.
 
One can determine the axion mean free-path, starting from the axion emissivity $Q_a$ of Eq.~(\ref{eq:emissivity}) 
writing it as
\begin{equation}
Q_{a} =\frac{T^{4}}{2\pi^{2}}\int dx\,x^{3}e^{-x}\lambda_{x}^{-1} \,\ .
\end{equation}
In Appendix B we give details of the calculation of the mean free-path starting from our emissivity.
Finally the optical depth is defined as
\begin{equation}
\tau(r)= \int_{r}^{+\infty} dr \kappa_a \rho \,\ .
\end{equation}
Setting $\tau  (r_{\textrm ax}) = 2/3$ we determine the radius of the axion-sphere
$r_{\textrm ax}$.
Using the temperature and density profiles shown in Fig.~\ref{fig:Temp} and \ref{fig:dens}, we obtain the 
radius and temperature  of the axion-sphere in function
of the post-bounce time $t_{\rm pb}$  as shown in
Fig.~\ref{fig:raxionsp} in the left and right panels, respectively, for axions with $g_{an}=g_{ap}=9\times10^{-7}$ 
for the complete emissivity ({dashed }curve) and for OPE (dotted curve), compared with the analogous
 quantities for non-electron neutrinos $\nu_x$ ({continuous }curve).
For the complete recipe for axion emissivity, one realizes that the axion-sphere radius is $\sim 10$~km after the core-bounce and 
{it is rather flat }
with respect to time. This radius corresponds to a temperature 
$T \sim 40$~MeV (see also Fig.~\ref{fig:Temp}). 
{ We see that the naive OPE prescription, having a larger axion absorption rate, would have given a larger axion-sphere radius
 $r \simeq 20$~km, and a smaller decoupling temperature $T \sim 5$~MeV (Fig.~\ref{fig:Temp}), comparable to the ones of 
the $\nu_x$'s whose mean free path is determined by neutral currents.}

\begin{figure}[t!]
\vspace{0.cm}
\includegraphics[width=0.5\textwidth]{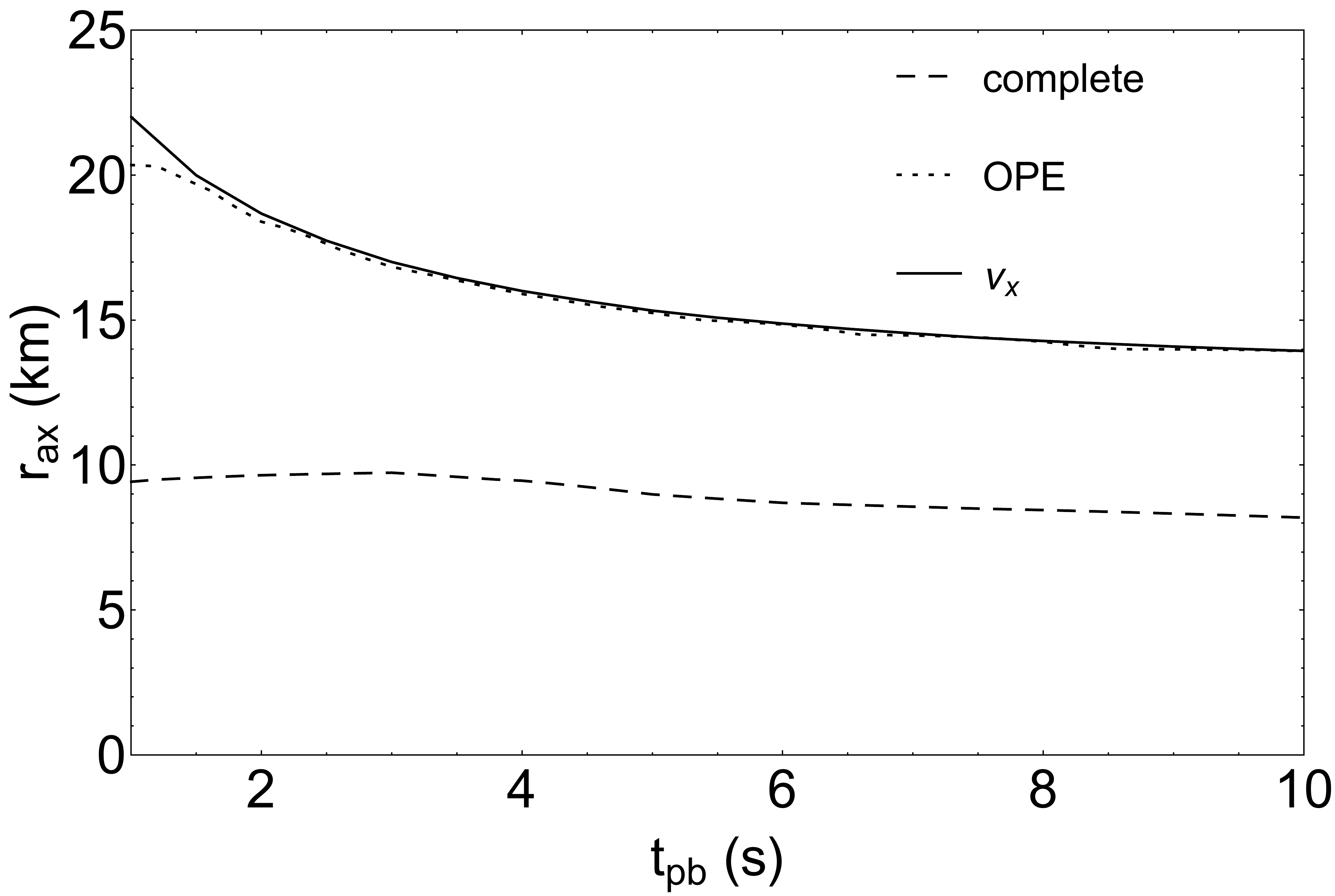}
\includegraphics[width=0.5\textwidth]{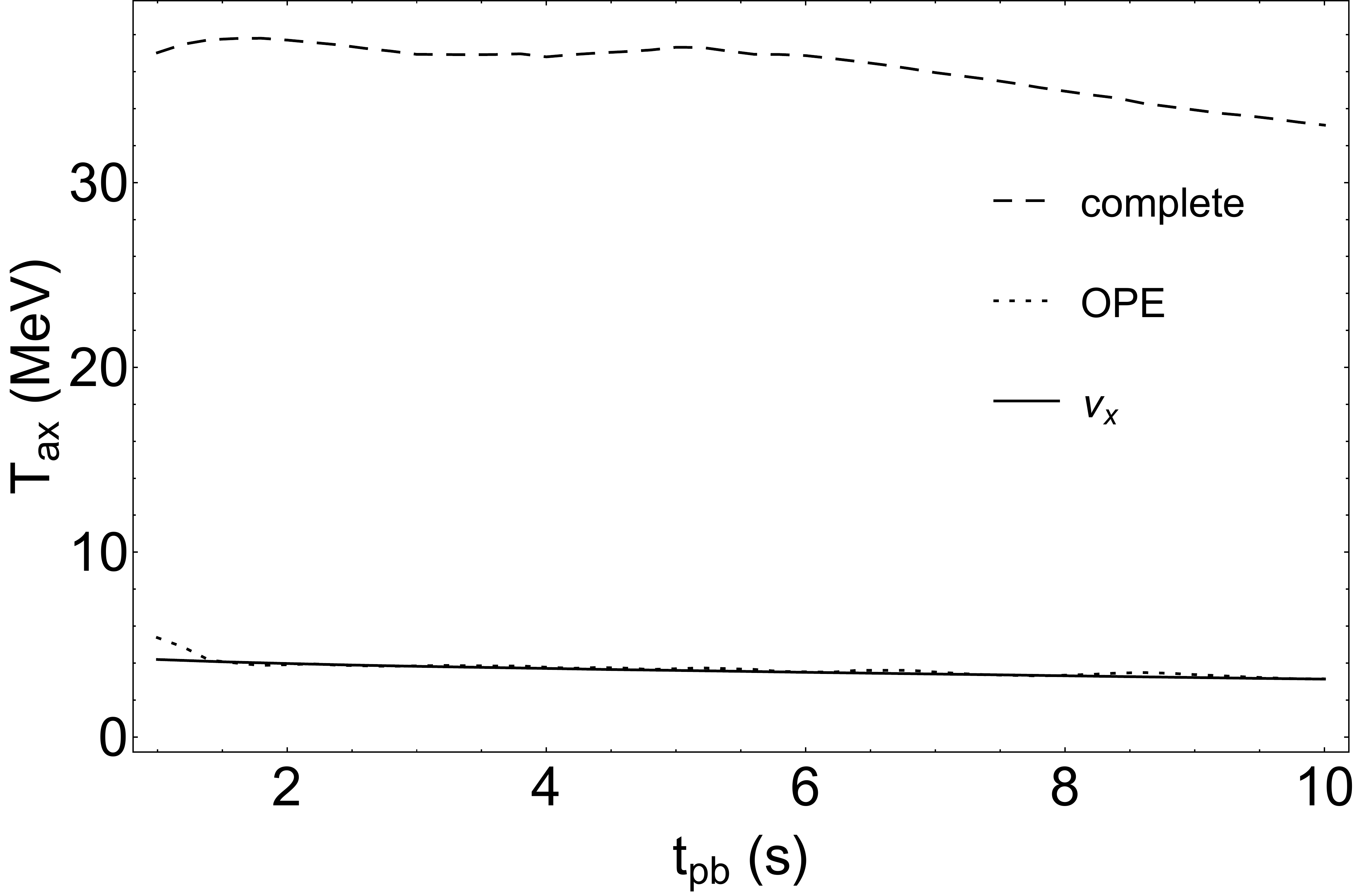}
\caption{Trapping regime. Axion-sphere radius (left panel) and temperature (right panel) as a function
of the post-bounce time $t_{\rm pb}$ for an axion 
with $g_{an}=g_{ap}=9\times10^{-7}$ for the complete emissivity (dashed curve) and for OPE (dotted curve),
compared with the analogous quantities for non-electron neutrinos $\nu_x$ (continuous curve).}
\label{fig:raxionsp}
\end{figure}

\begin{figure}[t!]
\vspace{1.cm}
\hspace{1.cm}
\includegraphics[width=0.8\textwidth]{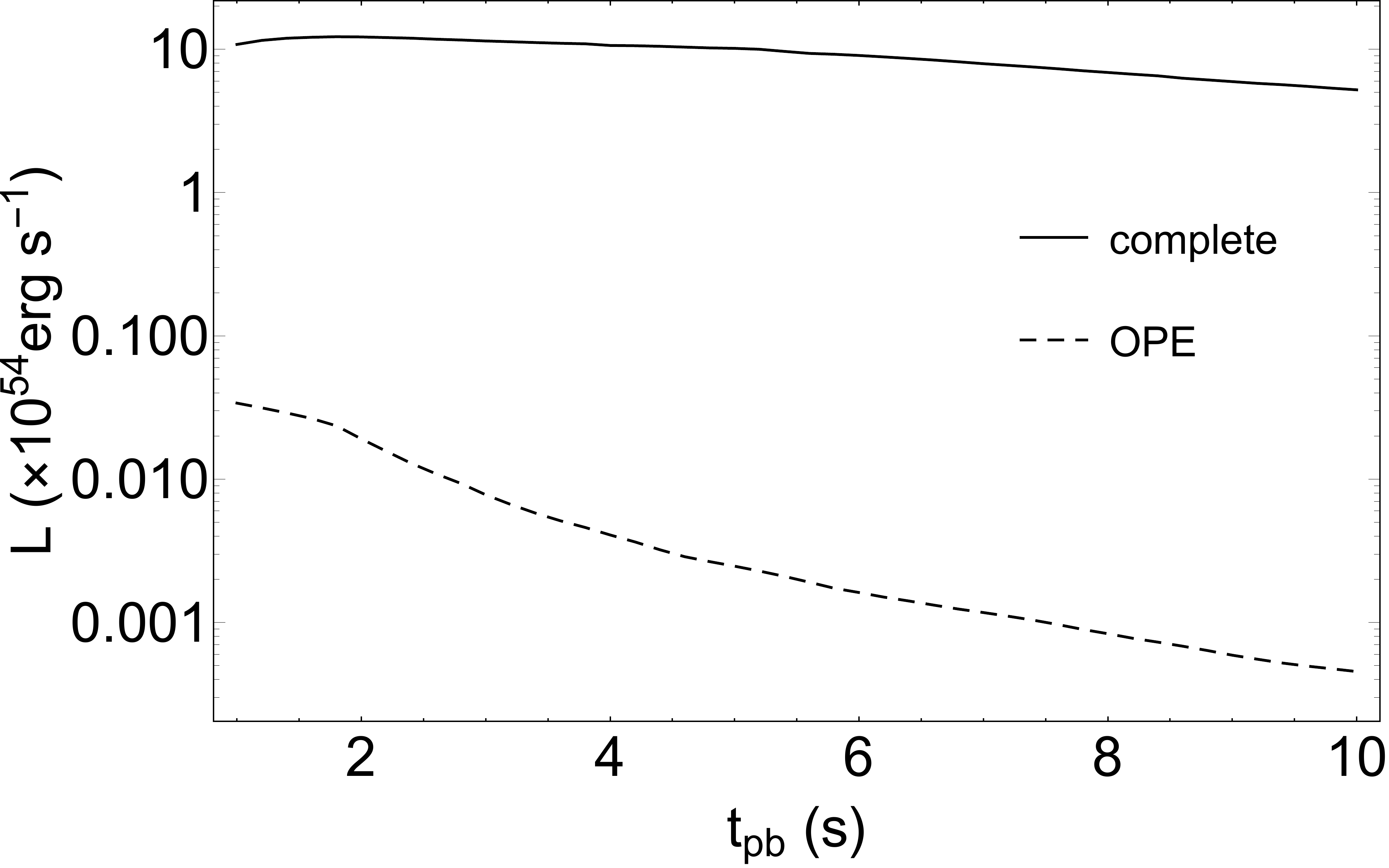}
\caption{Trapping regime. Axion luminosity in function
of post-bounce time $t_{\rm pb}$  for $g_{an}=g_{ap}=9\times10^{-7}$ 
for the complete emissivity (continuous curve) and for OPE (dotted curve).}
\label{fig:Lumtrapp}
\end{figure}

Finally, the axion luminosity can be determined by the Stefan's law
\begin{equation}
L_a= \frac{\pi^2}{120} 4 \pi r_{\rm ax}^2 T^4(r_{\rm ax}) \,\ ,
\end{equation}
and shown in Fig.~\ref{fig:Lumtrapp} for the complete emissivity (continuous curve) and for OPE (dotted curve).
Notice that, contrarily to the free-streaming case, the complete calculation in the trapping regimes gives a larger
axion luminosity than the naive OPE. This is easily explained since in the complete calculation we have seen that the axion-sphere
has a larger temperature.
 Note also that for the chosen axion-nucleon coupling, the axion luminosity would
be much larger than the unperturbed neutrino one (see Fig.~\ref{fig:LaLnu}).
In Fig.~\ref{fig:LaLnu}, which shows the ratio $L_a/L_a$,  the trapping case corresponds to $g_{an}=g_{ap} \gtrsim 10^{-6}$.
Notice that increasing the nucleon couplings the axion luminosity decreases.
This should not be surprising. 
In fact, the corrections to OPE reduce the axion interaction rate implying a larger axion-sphere radius, where the temperature is higher.  
The luminosity being proportional to the square of the axion-sphere radius and the fourth power of the axion-sphere temperature, gets contribution from both effects but the lowering of the temperature dominates for essentially the full range of couplings. 
One realizes that in the standard OPE scenario, the axion luminosity curve intersects the neutrino luminosity at $ g_{an}=g_{ap}\simeq 2\times 10^{-5} $.
In that case, larger couplings are allowed since the axion luminosity would be reduced below $ L_{\nu} $.
The OPE corrections lead to $L_a \simeq L_{\nu}$ at $ g_{an}=g_{ap}\simeq 6\times 10^{-5}$, i.e. a relaxation by a factor of 3 with respect to the OPE result.

\section{Discussions and Conclusions}

Axions can be produced in a SN core by nucleon-nucleon bremsstrahlung, providing
 an additional source of energy loss or transfer with respect to the standard neutrino emission.
 This process has been often modeled  at the level of the vacuum OPE approximation.
 Starting from this naive recipe, we have refined the calculation of the axion emissivity, systematically including different effects: 
 a non-zero pion mass, the contribution of the two-pions exchange, effective in-medium nucleon masses and 
 multiple nucleon scatterings. Moreover, we have  obtained axion emission rates valid for generic nucleon degeneracy. 
 We find that the axion emissivity is significantly reduced (by about an order of magnitude) with respect to the naive OPE.
From the SN 1987A neutrino observations (more specifically, requiring that the axion energy loss does not exceed the energy loss in neutrinos), 
we obtain a constraint on
the axion-nucleon coupling
 {$g_{an}=g_{ap} \lesssim  6 \times 10^{-10}$, 
to be compared with $g_{an}=g_{ap} \lesssim 2 \times 10^{-10}$, found in the simple  OPE approximation.}
{
In the case of  KSVZ model, we find $g_{ap} \lesssim 1.2 \times  10^{-9}$,  corresponding to $m_a \lesssim 15 $~meV.
This corresponds to a relaxation by a factor of 3 with respect to the OPE result, and it is comparable to the one 
quoted in~\cite{Raffelt:2006cw}.}

{In the case of DFSZ axions  our analysis indicates $m_a \lesssim 15-20$~meV, the exact value depending on $ \tan \beta $.
This bound is shown in Fig. \ref{fig:DFSZ}.
As evident from the figure, the SN bound is slightly stringent than the RGB bound,  and it dominates on it for low values of $ \tan\beta $.}
When compared with the expected IAXO potential~\cite{Giannotti:2017hny,Armengaud:2019uso},  
it is evident that helioscopes of the next generation have the capability to explore large regions of the axion parameter space that are not affected by the SN analysis.
As stressed above, the results of our analysis should not be considered as a robust bound as we are not including the axion feedback on the star, which at these couplings is most likely non-negligible. 
In fact, since axions are assumed to be transporting energy away from the SN at a rate comparable with that of neutrinos, it is likely that the stellar model with axion would be colder and therefore the axion rate reduced. 
Based on this, we can expect the axion bound to be further reduced when the axion feedback is taken into account in a self-consistent simulation of the SN evolution. 

Passing to the case of trapped axions, we find that $L_a \gg L_\nu$ for $g_{an}=g_{ap} \simeq 10^{-6}-10^{-7}$, closing the so called  ``hadronic axion window",
left opened by previous analyses.
This result is in agreement with the findings in~\cite{Chang:2018rso} and strengthens the cosmological argument for the exclusion of eV axions 
in the so-called   ``hadronic axion window"~\cite{Hannestad:2005df,Giusarma:2014zza}.

We stress that our improvement of the OPE case gives results similar to those based on the vacuum $T$-matrix, and therefore is well justified for our and similar studies.
Though we have considered all relevant effects on the axion emissivity, it is important to clarify that the present study is not fully self-consistent since the nuclear potential in nuclear medium, as well as the nucleon effective masses, have not been studied self-consistently. A possible improvement would require to calculate the in-medium $T$-matrix and the nucleon effective masses within the self-consistent Green's function method, based on a realistic two- plus three-body nuclear potential \cite{Dickhoff:2004xx}.
However, such studies, for a broad range of conditions relevant for SN matter, are still lacking. Our study, though not fully self-consistent, provides a reliable estimation of axion emissivity considering all important effects. 

Finally, and perhaps most importantly, our estimation of the axion luminosity is based on an unperturbed SN model. 
Neglecting the axion production feedback on the SN is not fully justified and we expect a further relaxation of the free streaming axion bound when such effects are taken into account.
Moreover, though the requirement that the axion luminosity does not exceed the neutrino one is certainly very reasonable, it is not obvious that this should be an exact criterion to fix the SN axion bound. 
A better, though much lengthier, approach requires a study of the axion induced modifications of the neutrino signal, which need then to be compared with the available data.  
%
We do plan to improve the current result, implementing what discussed above, in a forthcoming work. 
In particular, we plan to include the axion emissivity in a fully self-consistent SN simulation, as done in~\cite{Fischer:2016cyd}, in order to
characterize the feedback on the neutrino signal. 

In any case, we believe that our calculation represent a step forward in the effort to characterize the axion bounds from SN and a starting point for future investigations.

\section*{Acknowledgements}

 We warmly thank Georg Raffelt and Andreas Ringwald for useful discussions and suggestions during the development of this work.
The work of P.C. and 
A.M. is supported by the Italian Istituto Nazionale di Fisica Nucleare (INFN) through the ``Theoretical Astroparticle Physics'' project and by Ministero dell'Istruzione, Universit\`a e Ricerca (MIUR). P.C. thanks the ChETEC COST Action (CA16117) funded by COST (European Cooperation in Science and
Technology) for financial support.
T.F. acknowledges support from the Polish National Science Center (NCN) under grant number UMO-2016/23/B/ST2/00720.
G.G. and G.M.P. are partly supported by the German Research Foundation
(DFG, German Research Foundation) - Projektnummer
279384907 - SFB 1245.


\section*{Appendix A. Calculation of the emission rate}

We show here the details of the calculation of the axion emissivity including all the corrections described before.
The axion emissivity can be calculated as~\cite{Brinkmann:1988vi} 
\begin{equation}
Q_a = \int 
 \frac{d^3p_a}{2\omega_a (2\pi)^3} \prod_{i=1,4} \frac{g_i d^3p_i}{2 E_i(2\pi)^3}~\omega_a f_1 f_2 (1-f_3) (1-f_4)
 \sum_{\rm spins} S \times| {\mathcal M}|^2
\delta^4(p_1+p_2-p_3-p_4-p_a)  \,\ ,
\label{eq:emissivityap}
\end{equation}
where the nucleon degeneracy factor $g_i=2$, $\omega_a$ is the axion energy.
The distribution functions for the  different {non-interacting} species obey the Fermi-Dirac statistics,
\begin{equation}
f_N\left(p;\{\mu_N,T\}\right) = \left[\exp\{\beta(E(p)-\mu_N)\}+1\right]^{-1} \,\ ,
\label{eq:fddistrib}
\end{equation}
with inverse temperature $\beta=1/T$ and chemical potential $\mu_N$. Commonly used modern equations of state for supernova studies consider the strongly interacting nucleons at the mean field level~\cite{Hempel:2009mc,Hempel:2011mk}, which are based on the single-particle self energy that can be separated into scalar ($S$) and the vector parts ($V$). This leads to the nucleons's energy dispersion relation, $E(p)=\sqrt{p^2 + m^{*2}_{N}}+\Sigma_V$, with the nucleon effective mass, $m^*_N=m_N+\Sigma_S$.  

In our calculations of the axion emission rate we assume non-relativistic nucleons and the dispersion relation becomes \cite{Hempel:2014ssa},
\begin{equation}
E(p) \approx m_N + \frac{p^2}{2 m^*_N} + U\;,\quad\text{with}\;\; U=\Sigma_S+\Sigma_V \,\ ,
\end{equation}
that allows us to rewrite the Fermi-Dirac distribution,
\begin{equation}
f_N\left(p;\{\mu_N,T\}\right) \approx \left[\exp\left\{\frac{p^2}{2 m^*_N T}-\eta_N)\right\}+1\right]^{-1}\,\ ,
\end{equation}
where we have introduced the nucleon degeneracy parameter, ${\eta}_N$, as follows,
\begin{equation}
{\eta}_N = \frac{\mu_N -m_N - U}{T}\,\ .
\label{eq:eta}
\end{equation}
In our previous analysis we neglected the vector contribution, $\Sigma_V$, to the self-energy $U$ in the calculation of the degeneracy parameter.
This caused  inconsistencies in the calculation of the nuclear densities. 
Including the attractive $\Sigma_V$ contribution leads to a larger $ \eta_N $ (thus, to a slightly more degenerate plasma).

The factors
$(1-f_i)$ in Eq.~(\ref{eq:emissivityap}) take into account the nucleon Pauli blocking effect, that we inserted
since we considered  situations of arbitrary nucleon degeneracy.
The squared matrix element
 $| {\mathcal M}|^2$ is summed over initial and final
spins, and $S$  is the usual symmetry factor for identical
particles in the initial and final states.
Assuming one-pion-exchange (OPE) one would get
{
\begin{equation}
S\times\sum|\mathcal{M}|^{2} =\frac{1}{4}\times\frac{g_{a}^{2}}{4m_{N}^{2}} \omega_{a}^{2}{\overline M}  \,\ ,
\label{eq:inimatrix}
\end{equation}
with

\begin{equation}
 {\overline M}= 4\times\frac{256}{3}m_{N}^{4} \omega_{a}^{-2} \left(\frac{g_{A}}{2f_{\pi}}\right)^{4}\left(\mathcal{A}_{nn}+\mathcal{A}_{pp}+4 \mathcal{A}_{np}\right) \,\ ,
\label{eq:matrix}
\end{equation}
where   $g_{A}=1.26$ is the  axial coupling, 
 $f_\pi= 92.4$~MeV, the pion decay constant  and
\begin{eqnarray}
\mathcal{A}_{NN}&=& C_{aN}^{2}
\left[\left(\frac{|\bk|^{2}}{|\bk|^{2}+m_{\pi}^{2}}\right)^{2}+\left(\frac{|\bl|^{2}}{|\bl|^{2}+m_{\pi}^{2}}\right)^{2}+(1-\xi)\left(\frac{|\bk|^{2}}{|\bk|^{2}+m_{\pi}^{2}}\right)\left(\frac{|\bl|^{2}}{|\bl|^{2}+m_{\pi}^{2}}\right)\right] \nonumber \\
\mathcal{A}_{np}&=&\left(C_{+}^{2}+C_{-}^{2}\right)\left(\frac{|\bk|^{2}}{|\bk|^{2}+m_{\pi}^{2}}\right)^{2}+\left(4C_{+}^{2}+2C_{-}^{2}\right)\left(\frac{|\bl|^{2}}{|\bl|^{2}+m_{\pi}^{2}}\right)^{2}+\nonumber \\
&-& 2\left[\left(C_{+}^{2}+C_{-}^{2}\right)-\left(3C_{+}^{2}+C_{-}^{2}\right)\frac{\xi}{3}\right]\left(\frac{|\bk|^{2}}{|\bk|^{2}+m_{\pi}^{2}}\right)\left(\frac{|\bl|^{2}}{|\bl|^{2}+m_{\pi}^{2}}\right) \,\ ,
\label{eq:amplitude}
\end{eqnarray}

with
\begin{eqnarray}
\xi&=&3(\hat{\bk}\cdot\hat{\bl})^{2} \,\ , \\
C_{\pm}&=& \frac{1}{2}\left(C_{an}\pm C_{ap}\right) \,\ ,
\end{eqnarray}
where $\bk=\bp_{1}-\bp_{3}$ and $\bl=\bp_{1}-\bp_{4}$.  
We introduce the variables~\cite{Raffelt:1993ix}
\beq
\bp_{1/2}=\bp_{0}\pm\bp ; \quad\bp_{3/4}=\bp_{0}\pm\bq\\
\eeq
with 
\beq
\begin{split}
&\bp_{0}=|\bp_{0}|\left(\sin\delta\cos\phi,\sin\delta\sin\phi,\cos\delta\right)\\
&\bp=|\bp|\left(0,0,1\right)\\
&\bq=|\bq|\left(\sin\theta,0,\cos\theta\right) \,\ .
\end{split}
\eeq
Using these variables we get
\beq
\begin{split}
&\hat{\bp}\cdot\hat{\bq}=\cos\theta\quad \hat{\bp}\cdot\hat{\bp}_{0}=\cos\delta\quad \hat{\bq}\cdot\hat{\bp}_{0}=\sin\delta\sin\theta\cos\phi+\cos\delta\cos\theta \,\ .\\
\end{split}
\eeq
Finally we introduce
\beq
\begin{split}
&u=\frac{|\bp|^{2}}{m_{N}T}\quad v=\frac{|\bq|^{2}}{m_{N}T}\quad w=\frac{|\bp_{0}|^{2}}{m_{N}T} \\
&  y=\frac{m_{\pi}^{2}}{m_{N}T}\quad z=\cos\theta \,\ . \\
\end{split}
\eeq
{Moreover, in the non-relativistic limit for the nucleon, one may neglect the axion radiation in the law of conservation of moments, 
so that Eq.~(\ref{eq:emissivityap}) is simplified as 
\begin{equation}
\delta^4(p_1+p_2-p_3-p_4-p_a) \to \delta (E_1+E_2-E_3-E_4-\omega)\delta^3({\bf p}_1+{\bf p}_2-{\bf p}_3-{\bf p}_4) \,\ .
\end{equation}

 }

We discuss now the inclusion of the different corrections beyond OPE.

\emph{Pion mass exchange.}
	In terms of the new kinematical variables Eq.~(\ref{eq:amplitude}) reads

\begin{eqnarray}
{\mathcal A}_{nn}+{\mathcal A}_{pp}&=& \left(C_{an}^{2}+C_{ap}^{2}\right)\left[F_{-}^{2}+F_{+}^{2}+(1-\xi)F_{+}F_{-}\right] \,\ , \nonumber \\
{\mathcal A}_{np}&=& \left(C_{+}^{2}+C_{-}^{2}\right)F_{-}^{2}+\left(4C_{+}^{2}+2C_{-}^{2}\right)F_{+}^{2}-2\left[\left(C_{+}^{2}+C_{-}^{2}\right)-\left(3C_{+}^{2}+C_{-}^{2}\right)\frac{\xi}{3}\right]F_{+}F_{-}  \,\ ,  \nonumber \\
\end{eqnarray}

where 
\begin{eqnarray}
F_+ &=&    \frac{|\bl|^{2}}{|\bl|^{2}+m_{\pi}^{2}} =     \frac{u+v+2\sqrt{uv}z}{u+v+2\sqrt{uv}z+y}  \,\ ,    \nonumber \\
F_- &=& 	\frac{|\bk|^{2}}{|\bk|^{2}+m_{\pi}^{2}}=\frac{u+v-2\sqrt{uv}z}{u+v-2\sqrt{uv}z+y} \,\ ,	      \nonumber  \\
\xi&=& 3(\hat{\bk}\cdot\hat{\bl})^{2}=3\frac{(u-v)^{2}}{(u+v)^{2}-4uvz^{2}} \,\ ,
\label{eq:ffactor}
\end{eqnarray}
having  used 
\begin{eqnarray}
|\bk|^{2}&=& m_{N}T(u+v-2z\sqrt{uv}) \,\ , \nonumber \\
|\bl|^{2}&=& m_{N}T(u+v+2z\sqrt{uv}) \,\ .\\
\end{eqnarray}

\emph{$\varrho$ exchange.}
In order to take into account the $\varrho$-meson exchange, as in Eq.~(\ref{eq:rhomass}), 
we shift the quantities in Eq.~(\ref{eq:ffactor}) as
\beq
\begin{split}
&F_{\pm}\rightarrow {\tilde F}_{\pm} =F_{\pm}-C_{\varrho}G_{\pm} \,\ ,\\
&G_{\pm}=\frac{u+v\pm2\sqrt{uv}z}{u+v\pm2\sqrt{uv}z+r} \,\ ,\\
&r=\frac{m_{\rho}^{2}}{m_{N}
T} \,\ ,
\end{split}
\eeq
where $C_{\varrho}=1.67$.

{\emph{Structure function.}
It is useful to express the axion emissivity $Q_a$ of Eq.~(\ref{eq:emissivityap}) in terms of the structure function  as~\cite{Raffelt:1993ix}
{
\begin{equation}
Q_a = \frac{g_a^2}{16 \pi^2}\frac{n_B}{m_N^2} \int_{0}^{+\infty} d\omega e^{-\omega/T} \omega^4 S_\sigma(\omega) \,\ ,
\end{equation}
where $S_{\sigma}(\omega)$ may be written as
\begin{equation}
S_{\sigma}(\omega)= \frac{\Gamma_\sigma}{\omega^2} s(\omega/T) \,\ ,
\end{equation}
where $\Gamma_\sigma$ is called the  nucleon ``spin fluctuation rate'', while $s(x)$ is a dimensionless function of $x = \omega/T$. 
The factorization between $\Gamma_\sigma$ and $s(x)$ is not unique.
Following~\cite{Hannestad:1997gc,Raffelt:1993ix}
we  take $\Gamma_\sigma$ such that $s(0)=1$ for a medium of only one non-degenerate nucleon species and when 
the $\pi$ and the $\varrho$ mass exchange is neglected, leading to~\cite{Hannestad:1997gc,Raffelt:1993ix}
\begin{equation}
\Gamma_{\sigma}=4\pi^{-1.5}\rho\left(\frac{g_{A}}{2f_{\pi}}\right)^{4}T^{0.5}m_N^{0.5}=21.6 \,\ \textrm{MeV} \rho_{14}T_{\rm MeV}^{0.5}m_{938}^{0.5} \,\ ,
\end{equation}
where $ \rho_{14}= \rho/10^{14} \textrm{g} \,\ \textrm{cm}^{-3}$, and $T_{\rm MeV}= T/1\,\ \textrm{MeV}$ and
$m_{938}= m_N/ 938 \,\ \textrm{MeV}$.

For a medium composed of neutron and protons one would get 
{
\begin{equation}
s(x) =s^{nn}(x)+s^{pp}(x)+s^{np}(x)\,\ ,
\end{equation}
with 
\begin{equation}
\begin{split}
s^{nn}(x)&=\frac{1}{3}C_{an}^{2}Y_{n}^{2}(s_{\bk}+s_{\bl}+s_{\bk\bl}-3s_{\bk\cdot\bl})\;,\\
s^{pp}(x)&=\frac{1}{3}C_{ap}^{2}Y_{p}^{2}(s_{\bk}+s_{\bl}+s_{\bk\bl}-3s_{\bk\cdot\bl})\;,\\
s^{np}(x)&=\frac{4}{3}Y_{n}Y_{p}\left(C_{+}^{2}+C_{-}^{2}\right)s_{\bk}+\frac{4}{3}Y_{n}Y_{p}\left(4C_{+}^{2}+2C_{-}^{2}\right)s_{\bl}+\\
&-\frac{8}{3}Y_{n}Y_{p}\left[\left(C_{+}^{2}+C_{-}^{2}\right)s_{\bk\bl}-\left(3C_{+}^{2}+C_{-}^{2}\right)s_{\bk\cdot\bl}\right] \,\ ,
\label{eq:sfunction}
\end{split}
\end{equation}
where including the all the correction described above and considering arbitrary degeneracy one obtains
\beq
\begin{split}
s_{\bk}(x)&=\int \frac{d\cos\delta}{2}\, \frac{d\phi}{2\pi} \,\frac{\sqrt{w}dw}{\sqrt{\pi}/2}\,du\,\frac{d\cos\theta}{2}\left[\frac{\rho Y_{1}}{2m_{N}}\left(\frac{2\pi}{m_{N}T}\right)^{1.5}\right]^{-1}\left[\frac{\rho Y_{2}}{2m_{N}}\left(\frac{2\pi}{m_{N}T}\right)^{1.5}\right]^{-1}\\
&\sqrt{u(u-x)}e^{w-{\eta}_{3}}e^{u-{\eta}_{4}}H^{+}_{u}({\eta}_{1})H^{-}_{u}({\eta}_{2})H^{+}_{v}({\eta}_{3})H^{-}_{v}({\eta}_{4})\tilde{F}_{-}^{2}\bigr|_{v=u-x\ge0}\\
s_{\bl}(x)&=\int \frac{d\cos\delta}{2}\, \frac{d\phi}{2\pi} \,\frac{\sqrt{w}dw}{\sqrt{\pi}/2}\,du\,\frac{d\cos\theta}{2}\left[\frac{\rho Y_{1}}{2m_{N}}\left(\frac{2\pi}{m_{N}T}\right)^{1.5}\right]^{-1}\left[\frac{\rho Y_{2}}{2m_{N}}\left(\frac{2\pi}{m_{N}T}\right)^{1.5}\right]^{-1}\\
&\sqrt{u(u-x)}e^{w-{\eta}_{3}}e^{u-{\eta}_{4}}H^{+}_{u}({\eta}_{1})H^{-}_{u}({\eta}_{2})H^{+}_{v}({\eta}_{3})H^{-}_{v}({\eta}_{4})\tilde{F}_{+}^{2}\bigr|_{v=u-x\ge0}\\
s_{\bk\bl}(x)&=\int \frac{d\cos\delta}{2}\, \frac{d\phi}{2\pi} \,\frac{\sqrt{w}dw}{\sqrt{\pi}/2}\,du\,\frac{d\cos\theta}{2}\left[\frac{\rho Y_{1}}{2m_{N}}\left(\frac{2\pi}{m_{N}T}\right)^{1.5}\right]^{-1}\left[\frac{\rho Y_{2}}{2m_{N}}\left(\frac{2\pi}{m_{N}T}\right)^{1.5}\right]^{-1}\\
&\sqrt{u(u-x)}e^{w-{\eta}_{3}}e^{u-{\eta}_{4}}H^{+}_{u}({\eta}_{1})H^{-}_{u}({\eta}_{2})H^{+}_{v}({\eta}_{3})H^{-}_{v}({\eta}_{4})\tilde{F}_{+}\tilde{F}_{-}\bigr|_{v=u-x\ge0}\\
s_{\bk\cdot\bl}(x)&=\int \frac{d\cos\delta}{2}\, \frac{d\phi}{2\pi} \,\frac{\sqrt{w}dw}{\sqrt{\pi}/2}\,du\,\frac{d\cos\theta}{2}\left[\frac{\rho Y_{1}}{2m_{N}}\left(\frac{2\pi}{m_{N}T}\right)^{1.5}\right]^{-1}\left[\frac{\rho Y_{2}}{2m_{N}}\left(\frac{2\pi}{m_{N}T}\right)^{1.5}\right]^{-1}\\
&\sqrt{u(u-x)}e^{w-{\eta}_{3}}e^{u-{\eta}_{4}}H^{+}_{u}({\eta}_{1})H^{-}_{u}({\eta}_{2})H^{+}_{v}({\eta}_{3})H^{-}_{v}({\eta}_{4})\frac{\xi}{3}\tilde{F}_{+}\tilde{F}_{-}\bigr|_{v=u-x\ge0} \, \\
\end{split}
\eeq
where $Y_i$ are the numbers of nucleons $i$ per baryons.
In the previous expressions, one introduces the functions
\begin{eqnarray}
H^{\pm}_{u}({\eta})&=& (e^{\frac{w+u}{2}\pm\sqrt{uw}\cos\delta-{\eta}}+1)^{-1} \,\ , \nonumber \\
H^{\pm}_{v}({\eta})&=& (e^{\frac{w+v}{2}\pm\sqrt{vw}(\sin\delta\sin\theta\cos\phi+\cos\delta\cos\theta)-{\eta}}+1)^{-1} \,\ .
\end{eqnarray}
In the OPE and non-degenerate limit we recover the results shown in \cite{Raffelt:1993ix}.
}
}

{
 As discussed before, many-body effects caused by multiple nucleon scatterings can also reduce the axion emissivity. 
 One can take these effects into account by the ansatz
 \begin{equation}
 S_{\sigma}(\omega)= \frac{\Gamma_\sigma}{\omega^2+ \Gamma^2}s(\omega/T) \,\ .
 \end{equation}
}

The calculation of the axion emissivity requires the solution of six-dimensional integrals. For this purpose we use the subroutine D01GDF for multidimensional
Gaussian quadrature from the Numerical Algorithms Group 
(NAG) \cite{nag}.

\section*{Appendix B. Axion opacity}

One can determine the axion mean free-path, starting from the axion emissivity $Q_a$, writing this latter as 
\begin{equation}
Q_{a} =\frac{T^{4}}{2\pi^{2}}\int dx\,x^{3}e^{-x}\lambda_{x}^{-1} \,\ .
\end{equation}

{
For each channel $pp$, $nn$, $np$, from the emissivity one can extract a mean free-path
\beq
Q^{ij}_{a}=\frac{T^{4}}{2\pi^{2}}\int dx\,x^{3}e^{-x}(\lambda^{ij}_{x})^{-1}\rightarrow (\lambda^{ij}_{x})^{-1}=\frac{g_{a}^{2}}{8}\frac{\rho}{m_{N}^{3}}\frac{\Gamma_{\sigma}/T}{x^{2}}xs^{ij}(x)\;;
\eeq
where $i,j=n,p$.
In conclusion, the opacity is
\beq
\begin{split}
\frac{1}{k_{a}\rho}&=\frac{15}{8\pi^{4}}\int dx\,\frac{x^{4}e^{2x}}{(e^{x}-1)^{3}}\left(\lambda^{nn}_{x}+\lambda^{pp}_{x}+\lambda^{np}_{x}\right)=\\
&=\frac{15}{8\pi^{4}}\int dx\,\frac{x^{4}e^{2x}}{(e^{x}-1)^{3}}\left[\frac{g_{a}^{2}}{8}\frac{\rho}{m_{N}^{3}}\frac{\Gamma_{\sigma}/T}{x^{2}}x\right]^{-1}\left[\frac{1}{s^{nn}(x,{\eta})}+\frac{1}{s^{pp}(x,{\eta})}+\frac{1}{s^{np}(x,{\eta})}\right]\; .
\end{split}
\label{eq:opacity}
\eeq
Including also the MS effect, the opacity becomes
\beq
\begin{split}
\frac{1}{k_{a}\rho}&=\frac{15}{8\pi^{4}}\int dx\,\frac{x^{4}e^{2x}}{(e^{x}-1)^{3}}\left[\frac{g_{a}^{2}}{8}\frac{\rho}{m_{N}^{3}}\frac{\Gamma_{\sigma}/T}{x^{2}+\Gamma^{2}/T^{2}}x\right]^{-1}\left[\frac{1}{s^{nn}(x,{\eta})}+\frac{1}{s^{pp}(x,{\eta})}+\frac{1}{s^{np}(x,{\eta})}\right]\; .
\end{split}
\label{eq:opacityMS}
\eeq

}


\end{document}